\documentclass[twocolumn,showpacs,preprintnumbers,amsmath,amssymb,superscriptaddress]{revtex4-1}

\usepackage[english]{babel}
\RequirePackage[T1]{fontenc}
\RequirePackage{times} 

\usepackage{amsfonts}
\usepackage{subfigure}
\usepackage{amsmath}
\usepackage{amssymb}
\usepackage{graphicx,epstopdf}
\usepackage{array}
\usepackage{color}
\usepackage{my_symbols}
\usepackage{bm}
\usepackage[normalem]{ulem}
\usepackage{hyperref}
\usepackage{color}
\usepackage{mathrsfs}
\usepackage{tocvsec2}
\usepackage{enumitem}
\usepackage{amssymb}
\usepackage[utf8]{inputenc}
\usepackage{xeCJK}

\hypersetup{
	colorlinks=true,
	linkcolor=blue,
	filecolor=magenta,
	urlcolor=cyan,
}

{\par}

\date{\today}

\begin{document}

\title{Moir\'e spintronics: Emergent phenomena, material realization and machine learning accelerating discovery}

\author{Fengjun Zhuo (卓冯骏)}
\email[~]{fengjun.zhuo@zju.edu.cn}
\affiliation{Center for Quantum Matter, School of Physics, Zhejiang University, Hangzhou 310058, China}

\author{Zhenyu Dai (戴震宇)}
\affiliation{Department of Physics, University of Houston, Houston, Texas 77204, USA}

\author{Kai Chang (常凯)}
\affiliation{Center for Quantum Matter, School of Physics, Zhejiang University, Hangzhou 310058, China}

\author{Hongxin Yang (杨洪新)}
\email[~]{hongxin.yang@zju.edu.cn}
\affiliation{Center for Quantum Matter, School of Physics, Zhejiang University, Hangzhou 310058, China}

\author{Zhenxiang Cheng (程振祥)}
\email[~]{cheng@uow.edu.au}
\affiliation{ Institute for Superconducting and Electronic Materials (ISEM), University of Wollongong, Wollongong 2500, Australia}

\begin{abstract}
Twisted van der Waals (vdW) materials have emerged as a promising platform for exploring exotic quantum phenomena and engineering novel material properties in two dimensions, potentially revolutionizing developments in spintronics. This Review provides an overview of recent progress on emerging moir\'e spintronics in twisted vdW materials, with a particular focus on two-dimensional magnetic materials. Following a brief introduction to the general features of twisted vdW materials, we discuss recent theoretical and experimental studies on stacking-dependent interlayer magnetism, non-collinear spin textures, moir\'e magnetic exchange interactions, moir\'e skyrmions and moir\'e magnons. We further highlight the potential of machine learning to accelerate the discovery and design of multifunctional materials for moir\'e spintronics. Finally, we conclude by addressing the most pressing challenges and potential opportunities in this rapidly expanding field.
\end{abstract}

\maketitle

\tableofcontents

\section{Introduction: Spintronics with a twist}
Stimulated by the discovery of the giant magnetoresistance (GMR) effect \cite{Baibich1988,Binasch1989} in magnetic metallic multilayers nearly 40 years ago, scientists have extensively explored the interplay between charge, spin, and orbital degrees of freedom in spintronic materials \cite{Prinz1998,Wolf2001}. Theoretically, diverse new concepts in spintronics have emerged, including spin transfer torque (STT) \cite{Slonczewski1996,Berger1996}, spin-orbit torque (SOT) \cite{Manchon2015,Manchon2019}, Dzyaloshinskii–-Moriya interaction (DMI) \cite{Dzyaloshinskii1958,Moriya1960,Yangh2023}, topological spin texture \cite{Bogdanov2001,Robler2006,Muhlbauer2009,Yu2010,Yu2018,Jani2021}, topological magnon insulator \cite{Zhuof2021,Zhuof2022,Zhuof2024}, spin Hall effect (SHE) \cite{Dyakonov1971,Hirsch1999,Sinova2015}, topological Hall effect \cite{Shindou2001,Tokura2017}, magnon thermal Hall effect \cite{Onose2010,Hirschberger2015}, etc. Technologically, spintronics offers a distinct advantage over conventional silicon-CMOS by utilizing the electron's intrinsic angular momentum (spin) rather than charge as the information carrier. This shift enables novel functional devices with superior performance, including ultrafast processing, low power consumption, nonvolatility, and high circuit integration density \cite{Zutic2004,Chappert2007,Jungwirth2016,Baltz2018,Manipatruni2018,Lin2019}. Over the past three decades, great efforts have been devoted to realize the applications of spintronic techniques \cite{Fert2008,Mao2006,Dave2006,Fan2015,Zahedinejad2020,Kurenkov2019}, such as magnetic hard disk drives, magnetic random access memory (MRAM), magnetic tunnel junction (MTJ) device, magnetic nonvolatile logic device, spin Hall nano-oscillator, and artificial neuromorphic networks. While spintronic devices have many outstanding characteristics in the industrial application, conventional magnetic materials face scalability limits at atomic thicknesses. This bottleneck has propelled spintronic research into two-dimensional (2D) magnetic materials, where reduced dimensionality enables unprecedented control over magnetic states. 
\begin{figure*}[htp]
	\centering
	\includegraphics[width=0.98\textwidth]{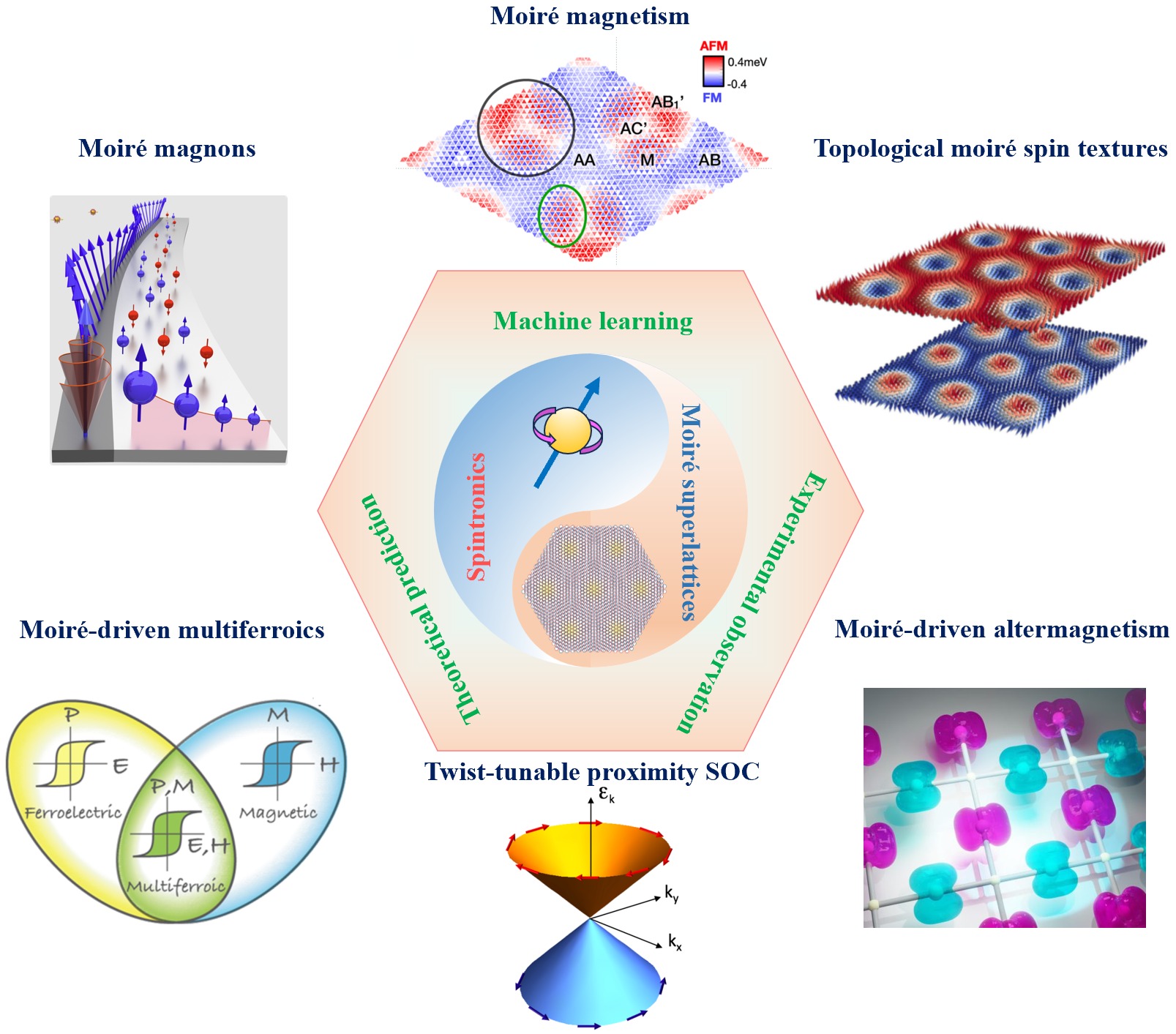}
	\caption{Overview of moir\'e spintronics with emergent phenomena.} 
	\label{fig0}
\end{figure*}

Since the successful exfoliation of layered graphene utilizing an adhesive tape by Geim and Novoselov in 2004 \cite{Novoselov2004}, there has been immense advancement in various extensions of the list of 2D materials to almost all functionalities of condensed matter systems including metals \cite{Cheny2018,Sarkar2020,Wangt2021}, semiconductors \cite{Zhaob2021,Zhaos2021,Zhouk2023}, topological insulators \cite{Koul2017,Rachel2018,Krishnamoorthy2023}, semimetals \cite{Maj2019,Lix2025,Yuh2025}, superconductors (SC) \cite{Qiu2021,Maggiora2024}, and magnetism \cite{Gibertini2019,Mak2019,Jiang2021}. Notably, 2D materials are thermostatically stable even with atomically thin layered structures, while exhibiting unique electrical, thermal, optical, acoustical, mechanical, and magnetic properties distinct from their bulk counterparts \cite{Pesin2012,Fiori2014,Chhowalla2016,Balandin2011,Xiaf2014}. To date, over 2000 stable atomically or molecularly thin 2D materials have been discovered, driven by rapid development in experimental preparation methods \cite{Mannix2017,Dongr2018,Caix2018,Xiaox2018}, including micromechanical exfoliation, liquid-phase exfoliation, chemical vapor deposition (CVD), molecular beam epitaxy (MBE), wet-chemical synthesis, reduction-oxidation, etc. Prominent examples include graphene \cite{Geim2007,Geim2009}, black phosphorus \cite{Liuh2015,Gusmao2017,Abate2018}, hexagonal boron nitride (hBN) \cite{Wengq2016,Zhangk2017,Caldwell2019,Roy2021}, transition metal dichalcogenides (TMDs) \cite{Wangq2012,Voiry2015,Manzeli2017,Choi2017,Chowdhury2020}, and metal carbides and nitrides (MXenes) \cite{Naguib2014,VahidMohammadi,Weip2021}. In recent years, alongside the advancements in monolayer 2D materials, novel research efforts have intensified toward heterostructures fabricated by vertically stacking various atomically thin crystals, which are connected by the van der Waals (vdW) interactions \cite{Geim2013,Novoselov2016,Liuy2016}. The fundamental concept is straightforward: sequentially stacking two or more 2D atomic layers (e.g., a monolayer atop another) create a precisely assembled artificial material. The stability of these structures relies on strong intralayer covalent bonds within each layer, while the stacked layers are held together by relatively weak vdW forces instead of direct chemical bonding  without the constraint of lattice matching \cite{Haigh2012,Butler2013}. Nevertheless, the intrinsic characteristics of each monolayer remain largely unaltered with negligible modulation or interlayer interactions due to stacking in the case of very weak interlayer coupling. Such multilayer vdW heterostructures, now experimentally realized, have surpassed performance expectations, demonstrating remarkable functionality \cite{Ponomarenko2011,Georgiou2013,Wangj2024}. 

The recent discovery of intrinsic magnetism in 2D vdW materials provides a thrilling arena for studying spin-related physics and achieving low-power spintronic devices, which has sparked significant interest within both the spintronics and 2D materials communities \cite{Burch2018,Gong2019,Lih2019,Cortie2020,Wangq2022}. Therefore, great efforts have been devoted to realizing magnetism in existing 2D materials (e.g., graphene and TMDs) \cite{Gmitra2015,Hanw2014,Guguchia2018} via defects, band engineering (e.g., exploiting van Hove singularities (vHs)), or the magnetic proximity effect \cite{Yangh2011,Yangh2013,Krasheninnikov2009,Castro2008,Caot2015}. However, the first experimental observation of intrinsic 2D ferromagnetism was independently reported in 2D vdW crystals CrI$_3$ down to the monolayer limits \cite{Huangb2017} and in Cr$_2$Ge$_2$Te$_6$ down to bilayers \cite{Gongc2017} with low phase transition temperatures ($T_c \approx 45$ K for monolayer CrI$_3$ and $T_c \approx 30$ K for bilayer Cr$_2$Ge$_2$Te$_6$) in 2017. This breakthrough overcomes the limitation of the Mermin--Wagner theorem predicted that the long-range magnetic order is unstable in low dimensional systems with sufficiently short-range isotropic interactions \cite{Mermin1966,Halperin2019,Jenkins2022}, as the continuous symmetries at finite temperatures cannot be spontaneously broken in the case. Subsequently, intense investigations identified a wide variety of emerging 2D intrinsic ferromagnetism or antiferromagnetism. These intrinsic 2D magnets offer enhanced efficiency in controlling or switching magnetic states by external perturbations (e.g., electric fields, carrier doping and strain) other than magnetic fields \cite{Yangh2018,Hellman2017,Kuepferling2023,Yangh2025}, since interface-based mechanisms become more effective in ultrathin layers, such as STT, SOT, and voltage-controlled magnetic anisotropy \cite{Amiri2012,Lix2017}. Moreover, 2D magnetic insulators also allow the exploration of low-damping, long spin-wave propagation lengths, and reduced Ohmic losses in designs for magnetic logic devices \cite{Dattas2012,Weid2018,Hanj2019}. Strong magnetic proximity effects in vdW heterostructures may provide new possibilities for manipulating spintronic, superconducting, excitonic, valleytronic and topological phenomena \cite{Lum2015,Morales2019,Saito2017,Stier2018,Bae2022,Xux2014,Mak2018,Liuj2023}. Beyond ferromagnetism, 2D magnets can also host a diverse range of magnetic states, including antiferromagnets \cite{Xingw2019,Jiangs2020,Rahman2021}, altermagnets \cite{Smejkal2022,Songc2025}, quantum spin liquids \cite{Banerjee2016,Banerjee2017,Takagi2019}, and topological spin textures \cite{Zhangh2022,Tangj2021,Fragkos2022}. These states can be seamlessly integrated into functional devices, offering novel avenues for investigating and manipulating such phenomena \cite{Songt2024}. As a result, 2D materials offer a prospective platform for discovering emergent quantum phenomena and realizing diverse next-generation quantum devices for practical applications \cite{Mellnik2014,Liux2019,Liw2020,Jinc2018}. 

Beyond monolayer properties of 2D vdW materials, stacking 2D layers with a twist angle provides novel new insights into spintronic research. The twist angle---a fundamental yet long-overlooked structural parameter--has recently emerged as a uniquely critical degree of freedom, leading to the development of a subfield known as "Twistronics" \cite{Carr2017,Andrei2020,Hennighausen2021,Yangy2020}. Notably, a moir\'e superlaatice forms due to a small rotational twist and/or lattice mismatch in the stacked layers, creating a unique platform for discovering novel physical phenomena and precisely engineering material properties \cite{Huangd2022,Adak2024,Christos2022,Liuy2021}. Therefore, twistronics investigates how twist angles between stacked 2D materials influence their properties and functionalities, enabling unprecedented control over material characteristics \cite{Carr2020,Jorio2022,Cui2024}. By adjusting twist angles, researchers can systematically tune electronic, optical, mechanical, and magnetic properties of 2D vdW heterostructures, with potential breakthroughs in electronics, superconductivity, spintronics, and optoelectronics \cite{Caoy2016,Ribeiro2018,Torma2022,Dul2023,Kim2022,Balents2020}. Specially, weak vdW coupling permits lattice-mismatch-independent vdW heterostructures \cite{Haigh2012,Butler2013}, where the relative twist angle generates moir\'e superlattices that spatially modulate electronic and magnetic interactions \cite{Huangd2022,Adak2024,Christos2022,Liuy2021}. Manipulating twist angle in magnetic layered structures (bilayers/multilayers) presents a potential means to engineer interlayer magnetic exchange interactions, which can potentially generate magnetic moir\'e patterns capable of stabilizing spatially nonuniform spin textures \cite{Hejazi2020,Songt2021}. Consequently, twistronics thus represents a transformative approach to manipulating 2D materials, opening new avenues for advanced material design and innovative applications in nanotechnology and quantum devices \cite{Kennes2021,Confalone2025,Caiy2021,Wangj2019,Sunx2024}. 

In parallel with these materials science advances, the field of artificial intelligence (AI) has been revolutionized by deep learning, a class of machine learning (ML) algorithms based on artificial neural networks with many layers \cite{LeCun2015}. A particularly impactful breakthrough within deep learning has been the development of the transformer architecture, which introduced the attention mechanism \cite{Vaswani2017}. This innovation has enabled the creation of powerful large-scale models and has led to remarkable scientific achievements, such as the transformer-based model AlphaFold2 solving the long-standing protein folding problem with high accuracy \cite{Jumper2021}. Inspired by these successes, the application of ML has expanded rapidly across the physical sciences to analyze complex data and accelerate discovery~\cite{Carleo2019}. In the domain of magnetism and spintronics, for instance, ML models are now used to predict key material properties, such as the Curie temperature, directly from chemical composition \cite{Nelson2019}, and to guide the autonomous discovery of novel rare-earth-free permanent magnets \cite{Xia2022}. A particularly promising approach is the development of physics-informed neural networks (PINNs), which embed known physical laws directly into the learning process to improve accuracy and data efficiency in simulations of complex physical systems \cite{Raissi2019, Karniadakis2021}. Specifically for 2D materials where the combinatorial design space is vast, with the assistance of high-throughput screening and ML, theoretical scientists have efficiently predicted abundant new classes of 2D materials \cite{Torelli2019,Lus2020,Lub2024,Bhawsar2025,Gouveia2025}. As a result, ML has become an ideal tool for creating large materials databases and predicting structural and electronic properties, thereby greatly reducing the time and cost of discovery \cite{Mounet2018, Lu2024_review}.

The goal of this review is to presents the state of the art and future prospects for emerging \textit{moir\'e spintronics} [\Figure{fig0}] in twisted vdW materials \cite{Lado2021}, with a particular focus on 2D magnetic materials. But we do not wish to deliver an exhaustive overview of the vast field of twistronics in magic angle graphene moir\'e materials, nevertheless, there are several excellent reviews related to this topic \cite{Andrei2020,Balents2020,Nimbalkar2020,Lium2022,Bhowmik2024}. Instead, we have tended to cover recent relevant literature primarily after 2018, when studies on physical properties of twisted 2D magnets began to emerge as a crucial research frontier. The review is organized as follows. Following the Introduction, in Section \uppercase\expandafter{\romannumeral2}, we first provide a brief introduction to fundamental theory of moir\'e superlattices. Then in Section \uppercase\expandafter{\romannumeral3}, we turn to present an overview of the recent progress on emerging moir\'e spintronics, including stacking-dependent interlayer magnetism, non-collinear spin textures, moir\'e magnetic exchange interactions, moir\'e skyrmions and moir\'e magnons. Section \uppercase\expandafter{\romannumeral4} focuses on recent advances in the study of these characteristics in moir\'e spintronics with the assistance of ML. Finally, we conclude with a discussion of vital challenges and promising future opportunities based on the current research status on this field. 

\section{Moir\'e geometry in twisted vdW materials}
\begin{figure*}[htp]
	\centering
	\includegraphics[width=0.9\textwidth]{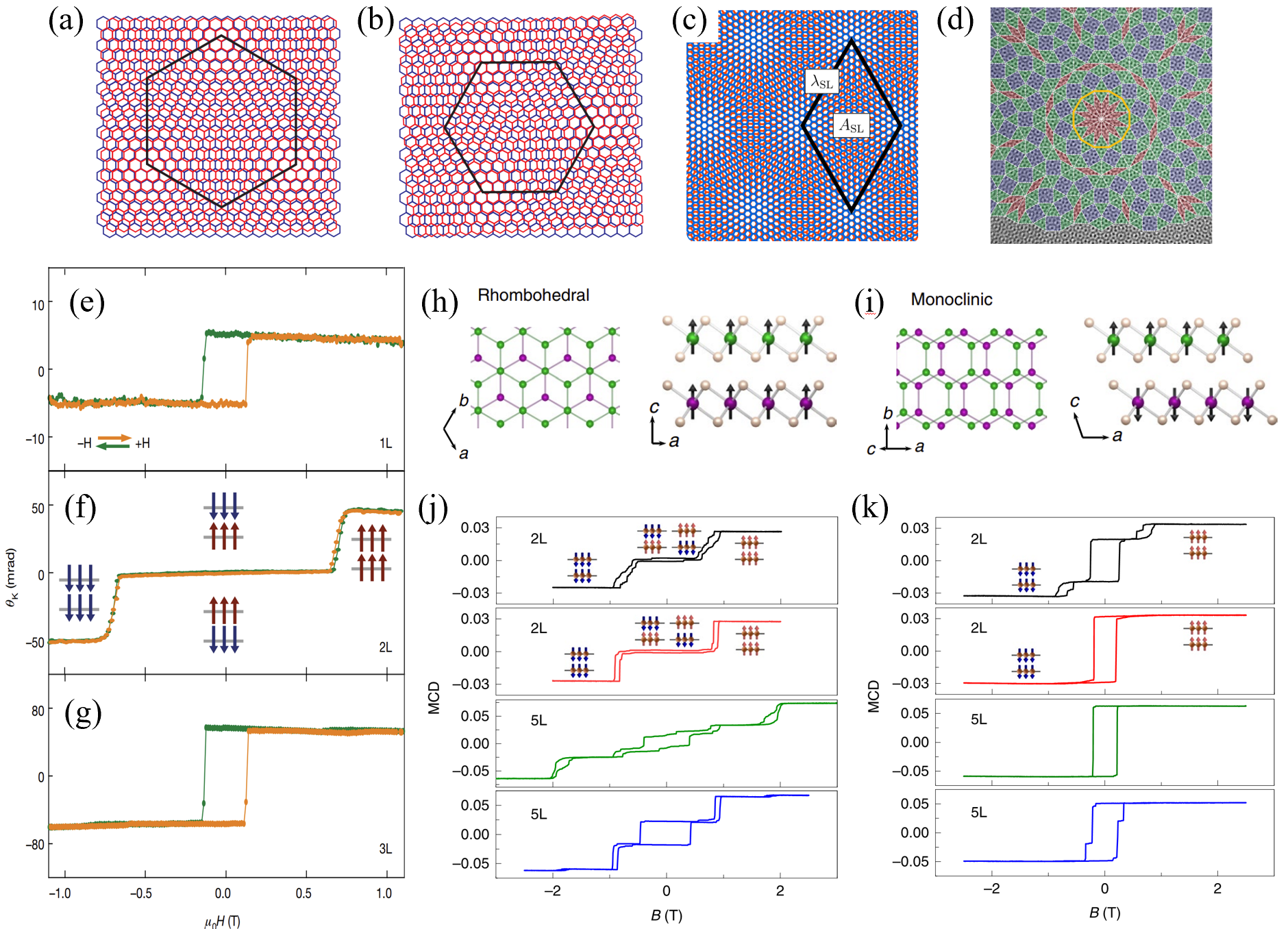}
	\caption{Schematic illustration of the moir\'e pattern of graphene (red) on hBN (blue) with a relative rotation angle between the crystals of (a) $0^{\circ}$ and (b) $3^{\circ}$. Black hexagons mark the moiré plaquette. (c) Schematic of TBG and its superlattice unit cell. $\lambda_{\mathrm{SL}}$ is the moir\'e period and $A_{\mathrm{SL}}$ is the unit cell area. (d) A false-colored TEM image of graphene quasicrystal in TBG rotated exactly 30$^{\circ}$, mapped with 12-fold Stampfli-inflation tiling. Magneto-optical Kerr effect for (e) monolayer (1L), (f) bilayer (2L) and (g) trilayer (3L) of CrI$_{3}$. By measuring the Kerr rotation angle $\theta_{\mathrm{K}}$, the evolution of magnetization with the application of an external magnetic field in the out-of-plane configuration are traced. (h) and (i) Schematic of rhombohedral (monoclinic) stacking with top (left) and side view (right) indicating the ferromagnetic (antiferromagnetic) interlayer coupling. The green (purple) atoms represent the Cr atoms in the top (bottom) layer while the brown ones represent the I atoms. (j) and (k) Magnetic field dependence of MCD at 3.5K for two 2L and two 5-layer (5L) regions of the CrI$_{3}$ flake before (j) and after (k) applying a pressure of 1.8Gpa, indicating stacking order transition, as the rhombohedral (monoclinic) phase prefers ferromagnetic (antiferromagnetic) interlayer coupling. Panels (a) and (b) reproduced with permission from Woods \textit{et al}., Nat. Phys. \textbf{10}, 451 (2014). Copyright 2014 Springer Nature. \cite{Woodsc2014} Panel (c) reproduced with permission from Cao \textit{et al}., Phys. Rev. Lett. {\bf 117}, 116804 (2016). Copyright 2016 American Physical Society. \cite{Caoy2016} Panel (d) reproduced with permission from Ahn \textit{et al}., Science  {\bf 361}, 782 (2018). Copyright 2018 American Association for the Advancement of Science. \cite{Ahn2018} Panels (e)-(g) reproduced with permission from Huang \textit{et al}., Nature \textbf{546}, 270 (2017). Copyright 2017 Springer Nature. \cite{Huangb2017} Panels (h) and (i) reproduced with permission from Song \textit{et al}., Nat. Mater. \textbf{18}, 1298 (2019). Copyright 2019 Springer Nature. \cite{Songt2019} Panels (j) and (k) reproduced with permission from Li \textit{et al}., Nat. Mater. \textbf{18}, 1303 (2019). Copyright 2019 Springer Nature. \cite{Lit2019}} 
	\label{fig1}
\end{figure*}

The weak vdW interlayer coupling allows two individual 2D materials with identical or misaligned periodic lattices to stack on top of each other with a relative twist angle, providing a powerful tool for exploring novel quantum phenomena \cite{Mak2022,Fox2024}. Interestingly, a large-scale geometrical pattern, known as a moir\'e superlattice, emerges due to the interference induced by a lattice mismatch [\Figure{fig1}(a)] and/or a rotational twist [\Figure{fig1}(b)] between the constituted layers \cite{Guoh2021,Hef2021}. These patterns are classified into two types based on symmetry \cite{Woodsc2014,Liy2024}: commensurate superlattices [\Figure{fig1}(c)] with long-range periodicity (e.g., moir\'e patterns or superlattice) \cite{Caoy2016} and incommensurate superlattices lacking translational symmetry, sometimes exhibiting quasicrystalline order [\Figure{fig1}(d)] \cite{Yaow2018,Ahn2018}. This review focus on the commensurate superlattices in twisted vdW materials arising from a twist angle and/or a mismatch in the lattice constants. To date, moir\'e superlattices have been extensively observed in both vdW homostructures and heterostructures formed by diverse layered 2D materials \cite{Andrei2021}. Additionally, the periodicity of moir\'e superlattice can be tuned by adjusting lattice constants difference, twist angles, and translational sliding, providing an effective approach to control material properties \cite{Bistritzer2011,Rosenberger2020,Kim2017}. The moir\'e supperlattice wavelength (moir\'e wavelength) $\lambda$  [\Figure{fig1}(c)] for twisted bilayer structures with a twist angle $\theta$ can be described as follows \cite{Ribeiro2018}:
\begin{equation}
	\lambda = \frac{\left( 1+\delta \right) a }{\sqrt{2\left( 1+\delta \right)\left( 1-\cos \theta \right)+\delta^{2}  }},
	\label{eq1}
\end{equation} 
where $\delta$ is the fractional lattice mismatch of the two 2D monolayers, which can be expressed as: $\delta = | a-a'|/a$; $a$ and $a'$ are the lattice constants of the two 2D monolayers. The typical moiré period is $\lambda = 10$nm $\gg a\approx 1 $\AA{}. When the vdW moir\'e materials form by either stacking two identical monolayers (homobilayer) with a small twist angle or two different monolayers (heterobilayer) with a slight lattice mismatch and a small twist angle, the above equation simplifies into $\lambda \approx a/\theta $ for the former and  $\lambda \approx a/ \sqrt{\delta^{2}+\theta^{2}}$ for the latter. 

For vdW homobilayer moiré superlattices, where the identical monolayers are vertically stacked, the most prominent examples are twisted bilayer graphene (TBG) \cite{Mele2010,Yankowitz2019}, twisted bilayer black phosphorus \cite{Kangp2017,Srivastava2021,Huangs2024}, twisted bilayer hBN \cite{Xianl2019,Woods2021,Taboada2023,Bennett2023}, twisted TMD homobilayer (e.g., MX$_2$, in which the transition metal atom M = Mo or W and the chalcogen atom X = S, Se or Te) \cite{Zhangz2020,Naik2018,Wangl2020,Weston2020,Guoy2025}, and twisted 2D magnets (e.g., CrI$_3$, CrBr$_3$, RuCl$_3$, and Fe$_3$GeTe$_2$ (FGT)) \cite{Xuy2022,Xieh2022,Cheng2021,Chend2022}. In homostructures, the moir\'e period is determined solely by the twist angle. For instance, the interlayer vdW interactions and band structures of TBG can be easily modulated by the twist angle \cite{Tarnopolsky2019,Choi2021}. Therefore, decreasing the twist angle brings low-energy vHs gradually moves closer, accompanied by a significant suppression of Fermi velocity due to the strong interlayer coupling \cite{Kerelsky2019}. Near the magic angle ($\sim 1.1^{\circ}$), the Fermi velocity nearly vanishes, resulting in two highly non-dispersive flat moiré bands at the charge neutrality point \cite{Brihuega2012,Sherkunov2018,Yinj2016}. Consequently, TBG near the magic angle exhibits an exceptionally wide range of exotic quantum phenomena, including correlated insulating states, unconventional superconductivity, ferromagnetism, and topological states \cite{Wus2021,Caoy2020,Saito2020,Shenc2020}. In contrast, the moir\'e superlattice arises primarily from both the lattice mismatch and rotational misalignment in a vdW heterostructure composed of constituent layers with different lattice constants, crystal structures or compositions, such as TBG/hBN \cite{Yoo2019,Shij2021,Nuckolls2020,Oh2021,Lisi2021}, graphene/TMD \cite{David2019,Naimer2021,Arora2020}, TMD/TMD (e.g., MoSe$_2$/WSe$_2$, MoSe$_2$/WS$_2$, and WS$_2$/WSe$_2$.) \cite{Seyler2019,Wangx2021,Zhangy2021,Polovnikov2024,Rafizadeh2025}. The newly introduced additional degrees of freedom bring about more methods and approaches to control the periodicity of a moir\'e superlattice, which give rise to unique features of moir\'e superlattices in diverse material systems. As a result, a variety of electronic, optical, magnetic, ferroelectric, and topological functionalities can be integrated into a vdW heterostructure, leading to a revolution in fundamental study and technological innovation.   

\section{Moir\'e spintronics with twisted vdW magnets}
So far, most experimental and theoretical investigations in twistronics has concentrated on TBG and its extensions to triple- and quadruple-layer homostructures \cite{Caoy2018,Caoy2018_2,Cheng2019,Burg2019,Hem2021}. Expanding the catalogue of engineered lattice structures to ultrathin vdW materials and their heterostructures, 
as twistronics meet spin, they provide a powerful platform for discovering emergent spin-related phenomena and implementing spintronic devices in the 2D limit. By tuning the new degree of freedom--twisting angle, moir\'e superlattices in ultrathin vdW materials and their heterostructures offer possibilities for engineering the magnetism and various quasiparticle excitations. 
\begin{figure*}[htp]
	\centering
	\includegraphics[width=0.95\textwidth]{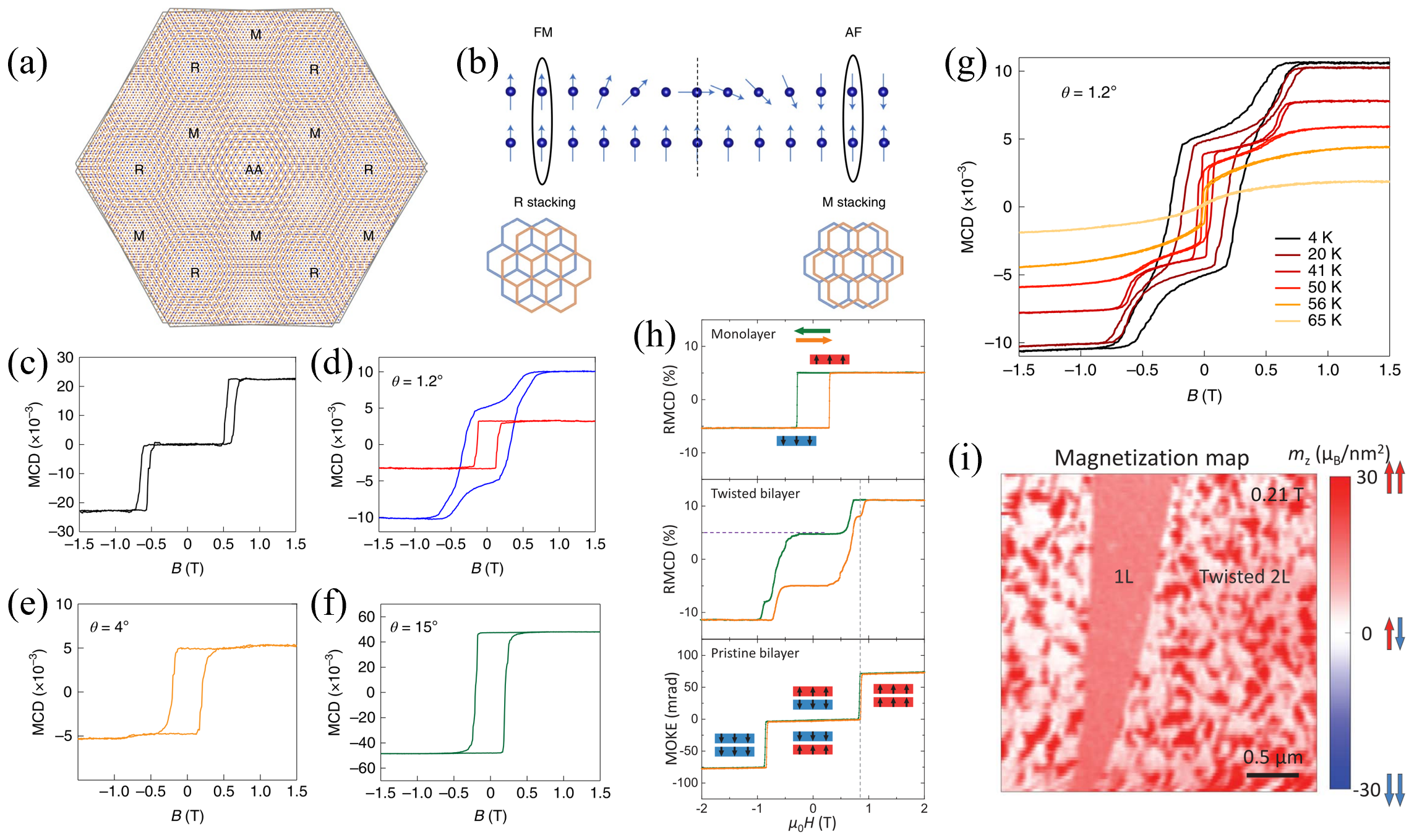}
	\caption{Moir\'e superlattice structure and moiré magnetism in twisted bilayer CrI$_{3}$. (a) Schematics of a moire\'e superlattice structure in bilayer CrI$_{3}$ with a small-twist-angle. R, M and AA represent rhombohedral, monoclinic and AA stacking, respectively. (b) Schematic illustration of a magnetic domain wall between the R- and M-stacking domain regions. (c)-(f) Magnetic-field dependence of MCD of a bilayer CrI$_{3}$ by natural stacking (c) and twisted stacking with twist angle 1.2$^{\circ}$ (d), 4$^{\circ}$ (e) and 15$^{\circ}$ (f), respectively. (g) MCD as a function of magnetic field at selective temperatures for a bilayer CrI$_{3}$ with twist angle 1.2$^{\circ}$. (h) Polar reflective MCD and MOKE signals as a function of magnetic field for the monolayer (top panel), the twisted bilayer (middle panel), and a pristine bilayer (bottom panel) CrI$_{3}$, respectively. (i) Scanning NV magnetometry magnetization map of twisted bilayer CrI$_{3}$ at 0.21T. Nanoscale antiferromagnetic and ferromagnetic domains are clearly resolved. Panels (a)-(g) reproduced with permission from Xu \textit{et al}., Nat. Nanotechnol. {\bf 17}, 143 (2022). Copyright 2022 Springer Nature. \cite{Xuy2022} Panels (h) and (i) reproduced with permission from Song \textit{et al}., Science \textbf{374}, 1140 (2021). Copyright 2021 American Association for the Advancement of Science. \cite{Songt2021}} 
	\label{fig2}
\end{figure*}

In the past few years, researchers have shown an increased interest in the study of moir\'e spintronics for its fundamental interest and potential impacts in logic operations and data storage devices. Early studies on magnetism in twistronics mostly concentrated on the emergent ferromagnetism  induced by strong electron–electron interactions at fractional fillings of twisted bilayer systems with a specific twist angle composed of nonmagnetic materials, such as twisted bilayer/trilayer/double bilayer graphene \cite{Sharpe2019,Liux2020,Cheng2020}, and twisted homobilayer/heterobilayer TMDs \cite{Caij2023,Xuy2020,Devakul2021}. Additionally, the spin and valley-spin in nonmagnetic TMDs monolayer have been demonstrated to be manipulated via the magnetic proximity effect by stacking on 2D magnets (e.g., CrI$_3$ and CrBr$_3$) \cite{Qiaoz2014,Zhaoc2017,Scharf2017,Zhongd2017,Zhongd2020}. In this section, we will highlight some of the new aspects and opportunities in moir\'e spintronics offered by twisted 2D vdW materials. 

\subsection{Stacking-dependent interlayer magnetism}
Whereas, the stacking dependence of magnetism is more remarkable and widespread in 2D vdW magnets, since the interlayer magnetic interaction depends strongly on both relative distance and orientation between the magnetic moments. As shown in figures 1(e)-(g), Huang \textit{et al.} \cite{Huangb2017} unravel the thickness-dependent behaviour of magnetic order in few-layer CrI$_3$. The experiment shows that samples with an odd number of layers (1L and 3L) indicate a ferromagnetic (FM) state, displaying a hysteresis loop with a coercive field of about 0.13T. In addition, the bilayer CrI$_{3}$  [\Figure{fig1}(f)] is characterized by FM order within the individual layers, with antiferromagnetic (AFM) coupling between the layers when $|\mu_0H|<0.65$T. A spin-flip process occurs at a field of about $\pm 0.65$T , so that magnetization in one layer flips to align with the external magnetic field as the field increases. Interestingly, it possible to control the interlayer magnetic interaction between 2D magnetic layers by electric gating or magnetic fields due to the weak vdW interlayer coupling \cite{Dengy2018,Wangz2018,Huangb2018,Jiangs2018}, which enables the experimental fabrication of ultrathin magnetic tunnel junction (MTJ) with giant tunneling magnetoresistance (TMR) \cite{Songt2018,Wangz2018_2}. 

Furthermore, early first-principles studies demonstrated the arrangement of interlayer magnetic interaction is directly related to the stacking order in bilayer CrI$_3$ \cite{Sivadas2018,Jiangp2019,Jangs2019}. Specifically, there is a one-to-one correspondence between the stacking structure and interlayer magnetic states: the monoclinic (M) phase supports an interlayer AFM magnetic state [\Figure{fig1}(h)]; and the rhombohedral (R) phase supports an interlayer FM magnetic state [\Figure{fig1}(i)]. Subsequently, this stacking order dependence of interlayer magnetic interaction was confirmed experimentally in CrI$_3$ and CrBr$_3$ \cite{Songt2019,Lit2019,Chenw2019}, leading to a efficient way for accurate designing 2D magnetism by controlling stacking order. For instance, Li \textit{et al.} \cite{Lit2019} measured the hysteresis curves of bilayer CrI$_3$ before [\Figure{fig1}(j)] and after [\Figure{fig1}(k)] application of hydrostatic pressure, which demonstrates the change from AFM to FM interlayer coupling, clearly indicating a stacking order phase transition from monoclinic to rhombohedral phase. While stacking-dependent interlayer magnetism is intrinsic to untwisted layered magnets such as CrI$_3$, this "one-to-one correspondence" forms the foundation for understanding the spatially varying interlayer exchange interactions that arise within the moir\'e superlattice when a twist is introduced. A small twist angle ($\theta < 5^{\circ }$) creates a moir\'e superlattice with spatially varying R/M stacking domains [\Figure{fig2}(a)], which results in competing FM/AFM interactions \cite{Jiangp2019} and emergent non-collinear spin textures via domain walls \cite{Xuy2022}.

\begin{figure*}[htp]
	\centering
	\includegraphics[width=0.98\textwidth]{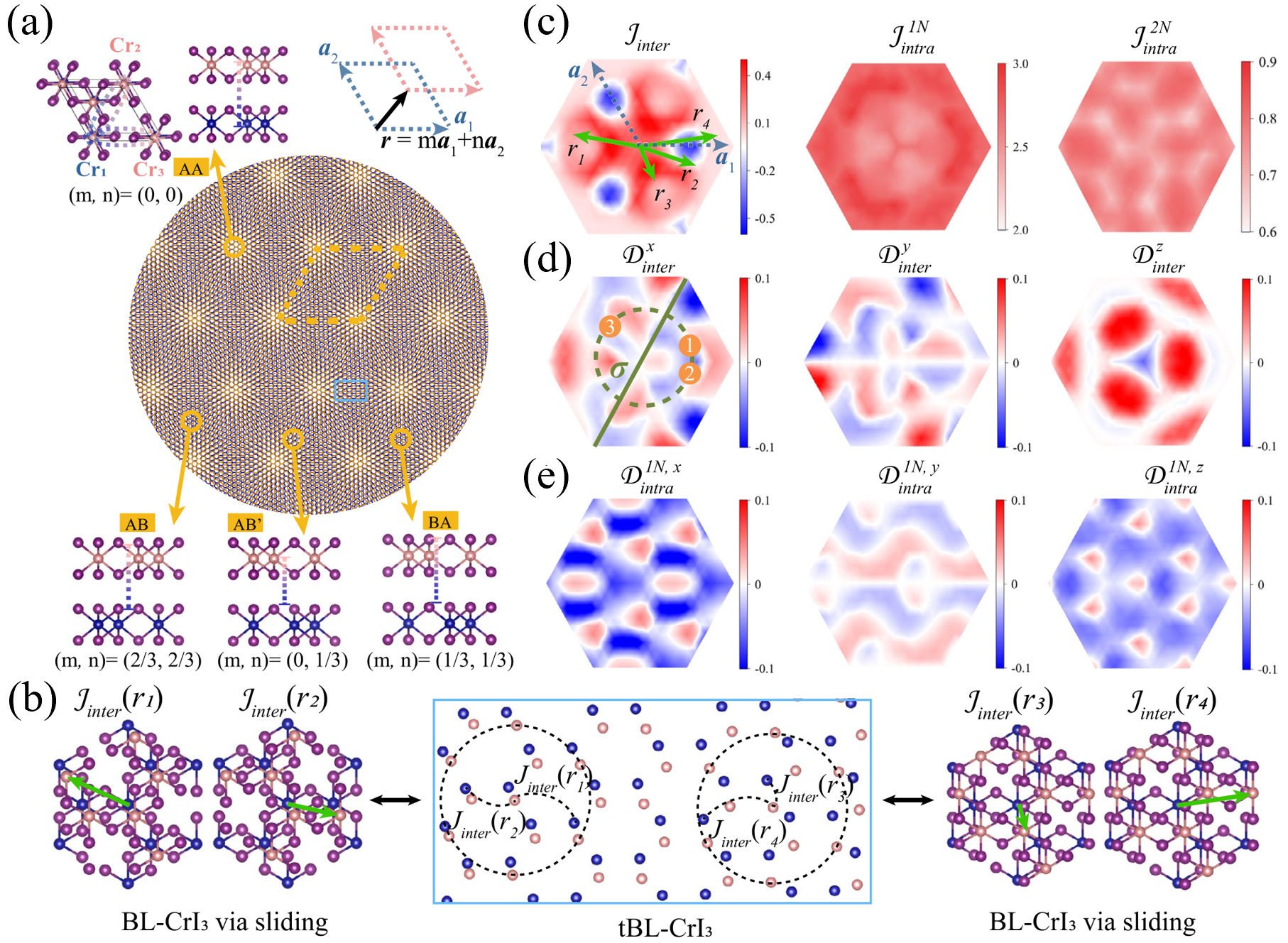}
	\caption{Schematics of generating MMEIs via Sliding-mapping Approach. (a) Moir\'e superlattice of the tBL-CrI$_3$. Insets are four domain regions with different stacking orders: AA, AB, AB' and BA stacking. Blue frame indicates the local site in (b). Dashed-yellow line represents the unit cell of moir\'e superlattice. (b) Examples of one-to-one correspondence between $J_{\mathrm{inter}}$ in tBL-CrI$_3$ and $\mathcal{J}_{\mathrm{inter}}(\textbf{\textit{r}})$ in sliding BL-CrI$_3$. (c)-(e) Color mapping of interlayer exchange interaction (left panel in c), intralayer nearest-neighboring exchange interaction (middle panel in c), intralayer next-nearest-neighboring exchange interaction (right panel in c), interlayer DMI (d), and intralayer DMI (e) between two Cr atoms as a function of sliding vector. Here, all the magnetic parameters are in the unit of meV. Panels (a)-(e) reproduced with permission from Yang \textit{et al}., Nat. Comput. Sci. {\bf 3}, 314 (2023). Copyright 2023 Springer Nature. \cite{Yangb2023}} 
	\label{fig3}
\end{figure*}
\subsection{Moir\'e magnetism}
\subsubsection{Non-collinear spin textures: Coexisting FM--AFM states}
Recently, researchers have shown an increased interest in the long-periodic moir\'e superlattice of twisted magnetic homostructures with small twist angles, where the stacking order changes smoothly over a long range \cite{Songt2021}. As shown in \Figure{fig2}(a), the triangular moir\'e superlattice formed by both R and M stacking regions can emerge in small-twist-angle CrI$_3$ bilayers \cite{Xuy2022}, suggesting the coexistence of FM and AFM domains as the interlayer magnetic exchange interactions strongly depends on the stacking order. In addition, magnetic moments are forced to follow non-collinear configurations between FM and AFM domains, forming FM–-AFM domain walls (DWs) as illustrated in \Figure{fig2}(b). Accordingly, this competing interlayer AFM and FM interactions in moir\'e superlattice can induce non-collinear magnetic textures with coexisting AFM and FM domains. 

Indeed, this non-collinear magnetic textures in twisted bilayer CrI$_3$ (tBL-CrI$_3$) have been observed by using magnetic circular dichroism (MCD) measurements in experiments \cite{Songt2021,Xuy2022}, as the out-of-plane magnetization of samples is linearly proportional to the value of MCD. Figures 2(c)-(f) shows the MCD as a function of out-of-plane magnetic field for four bilayer samples with different twist angle. Unlike a simple switching between FM and AFM response, the magnetic response in the sample with 1.2$^{\circ}$ twist angle shows a non-linear behaviour [\Figure{fig2}(d)]. \Figure{fig2}(g) shows the field dependence of the magnetic response at different temperatures. Additionally, this coexisting FM--AFM state in tBL-CrI$_3$ disappears at a critical angle $\theta_{\mathrm{c}}\approx 3^{\circ}$, above which only FM response exists. Utilizing single-spin quantum magnetometry, Song \textit{et al.} \cite{Songt2021} visualized nanoscale magnetic domains and moir\'e patterns, and measured domain size and moir\'e magnetism in tBL-CrI$_3$. In comparison with the reflective MCD signals of a monolayer and the magneto-optical Kerr effect (MOKE) signals of a pristine bilayer CrI$_3$ under the same experimental conditions, the MCD signal in a $\sim 0.2^{\circ}$ twisted bilayer sample dispalys a FM hysteresis loop with a nonzero remanent MCD signal [\Figure{fig2}(h)]. These observations indicate the coexisting FM--AFM state, which can be certainly verified in \Figure{fig2}(i).  

Besides, Xie \textit{et al.} \cite{Xieh2023} fabricated twisted double bilayers CrI$_3$ (tDB-CrI$_3$)  with successful twist engineering, and then reported direct evidence of non-collinear magnetic textures. Magneto-optical measurements identified signatures of moir\'e magnetism, demonstrating that both net magnetization and non-collinear magnetic texture exist at small twist angles from 0.5$^{\circ}$ to 5$^{\circ}$. Notably, at intermediate twist angles near 1.1$^{\circ}$, magnetization canted as noncollinear magnetic textures between FM and AFM domains only in the second layer, which can be adjusted with external magnetic fields. Overall, for a small twist angle, the emergent moir\'e magnetism can be viewed as a weighted linear superposition of the magnetism of two- and four-layer CrI$_3$. Whereas for a large twist angle, its magnetic ground state bears a strong resemblance to the independent two-layer CrI$_3$. However, at an intermediate twist angle, a finite net magnetization was observed due to spin frustrations caused by competition between FM and AFM interlayer exchange couplings within individual moir\'e superlattice. Li \textit{et al.} \cite{Lis2024} report observation of coexisting FM-AFM phases within individual moir\'e supercells in twisted double trilayer CrI$_3$ by utilizing scanning nitrogen-vacancy (NV) magnetometry techniques. Cheng \textit{et al.} \cite{Chengg2023} fabricated a dual-gate device based on twisted double bilayers CrI$_3$ sandwiched by top and bottom h-BN flakes. The coexistence of FM and AFM domains with non-zero net magnetization in a wide range of twist angles (0 $\sim$ 20$^{\circ}$) are observed, indicating non-collinear magnetic textures. Moreover, they report the high electrical tunability of moir\'e magnetism, which can be interpreted by a phase diagram associated with the wavelength of moir\'e superlattice and magnetic exchange parameters. 

Significantly, CrI$_3$ serves as a primary model system in moir\'e spintronics due to its robust ferromagnetism in monolayer limits, experimental accessibility, and well-characterized stacking-dependent interlayer magnetism \cite{Huangb2017}. Similar effects has also been observed in CrBr$_3$ \cite{Chenw2019} and predicted in other systems, such as RuCl$_3$ \cite{Cheng2021} and NiI$_2$ \cite{Zhuh2025}, though critical fields and stacking energies vary between materials. Similarly, Liang \textit{et al.} \cite{Liang2023} discovered a homostructure of two layered vdW magnet FGT with different thickness, which exhibited an interesting two-step magnetic hysteresis loop. The two coercivities in this specific hysteresis loop differ from that of the thinner and thicker constituent FGT layer, and meanwhile, the Curie temperature of the homostructure is in the middle of theirs, suggesting an interaction between the two constituents without merging into a naturally stacking thicker layer. However, further studies beyond CrI$_3$ remain essential for discovering various moir\'e phenomena.

\subsubsection{Moir\'e magnetic exchange interactions}
From the perspective of theoretical model, the magnetization of twisted vdW magnets can be modeled with the effective spin Hamiltonian modulated by the moir\'e potential
\begin{equation}
	\begin{aligned}
	\mathcal{H} =& -\sum_{i,j,l} J_{ij}^{l} \mathbf{S}_{i}^{l}\cdot \mathbf{S}_{j}^{l}
	+ \sum_{i,j,l} \mathbf{D}_{ij}^{l} \cdot \left ( \mathbf{S}_{i}^{l} \times \mathbf{S}_{j}^{l} \right ) - \sum_{i,l} \mathbf{B}\cdot \mathbf{S}_{i}^{l} \\
	& - \sum_{i,l} K_{e}^{l} \left ( \mathbf{S}_{i}^{l} \cdot \hat{\mathbf{e}}  \right ) ^{2} - \sum_{i,j,l,l'} J_{\perp}(\mathbf{r}_{ij}) \mathbf{S}_{i}^{l}\cdot \mathbf{S}_{j}^{l'}\\ 
	& + \sum_{i,j,l,l'} \mathbf{D}_{\perp} ( \mathbf{r}_{ij}) \cdot \left ( \mathbf{S}_{i}^{l} \times \mathbf{S}_{j}^{l'} \right )
	+ U   
	\end{aligned}
	\label{eq2}
\end{equation} 
where $\mathbf{S}_{i}^{l}$ denotes the spin operators at site $i$ in layer $l$; $J_{ij}^{l}$ and $J_{\perp} $ represent the interlayer and intralayer Heisenberg exchange interactions; $ \mathbf{D}_{ij}^{l}$ and $\mathbf{D}_{\perp} $ represent the interlayer and intralayer DMI; $K_{e}^{l}$ is the easy axis magnetic anisotropy in layer $l$; $\mathbf{B} $ is an external magnetic field. The last term $U$ contains additional interlayer or intralayer contributions, such as high-order Heisenberg exchange interaction \cite{Kartsev2020,Paul2020,Zhangs2023}, Kitaev exchange interaction \cite{Xuc2018,Takagi2019,Kims2022}, DMI from the substrate effect \cite{Yangh2023}, dipole-dipole interaction (DDI) \cite{Wuhrer2024,DeBell2000,Lux2020} and so on. In contrast with conventional magnetism (e.g., FM and AFM states) described by a few simple magnetic exchange interactions, moir\'e superlattices with large-scale periodicity always contain more than tens of thousands of nonequivalent magnetic atoms, forming complicated interlayer magnetic exchange interactions ($ J_{\perp}$ and $ \mathbf{D}_{\perp}$) --- moir\'e magnetic exchange interactions (MMEIs). The strength and sign of MMEIs in moiré superlattices are primarily governed by the local atomic registry (stacking geometry), which varies periodically with the moir\'e wavelength modulated by twist angle ($\lambda \propto 1/\theta $) \cite{Weston2020,Zhangc2017,Huangm2023}. Besides, intrinsic material properties (e.g., orbital overlap, magnetic anisotropy) and external factors such as pressure, strain or electric field modulate the coupling strength at specific stacking geometries \cite{Dengy2018,Wangz2018,Huangb2018,Jiangs2018,Lit2019,Kapfer2023}. Substrate effects can also introduce additional interlayer DMI and modify anisotropy \cite{Yangh2023,Yangb2023}. As a result, directly solving the MMEIs in moir\'e superlattices with samll twist angles from the first-principles or density functional theory (DFT) calculations is very unlikely. To date, a perfect theoretical formalism has not yet been discovered to make an accurate comparison of the results observed in experiments, owing to extremely complex interlayer interactions and twist-angle sensitivity. However, developing an effective spin Hamiltonian comprising reasonable MMEI parameters in moir\'e superlattices plays a crucial role to unveil the mystery of moir\'e magnetism in twisted 2D magnets. 
\begin{figure*}[htp]
	\centering
	\includegraphics[width=0.95\textwidth]{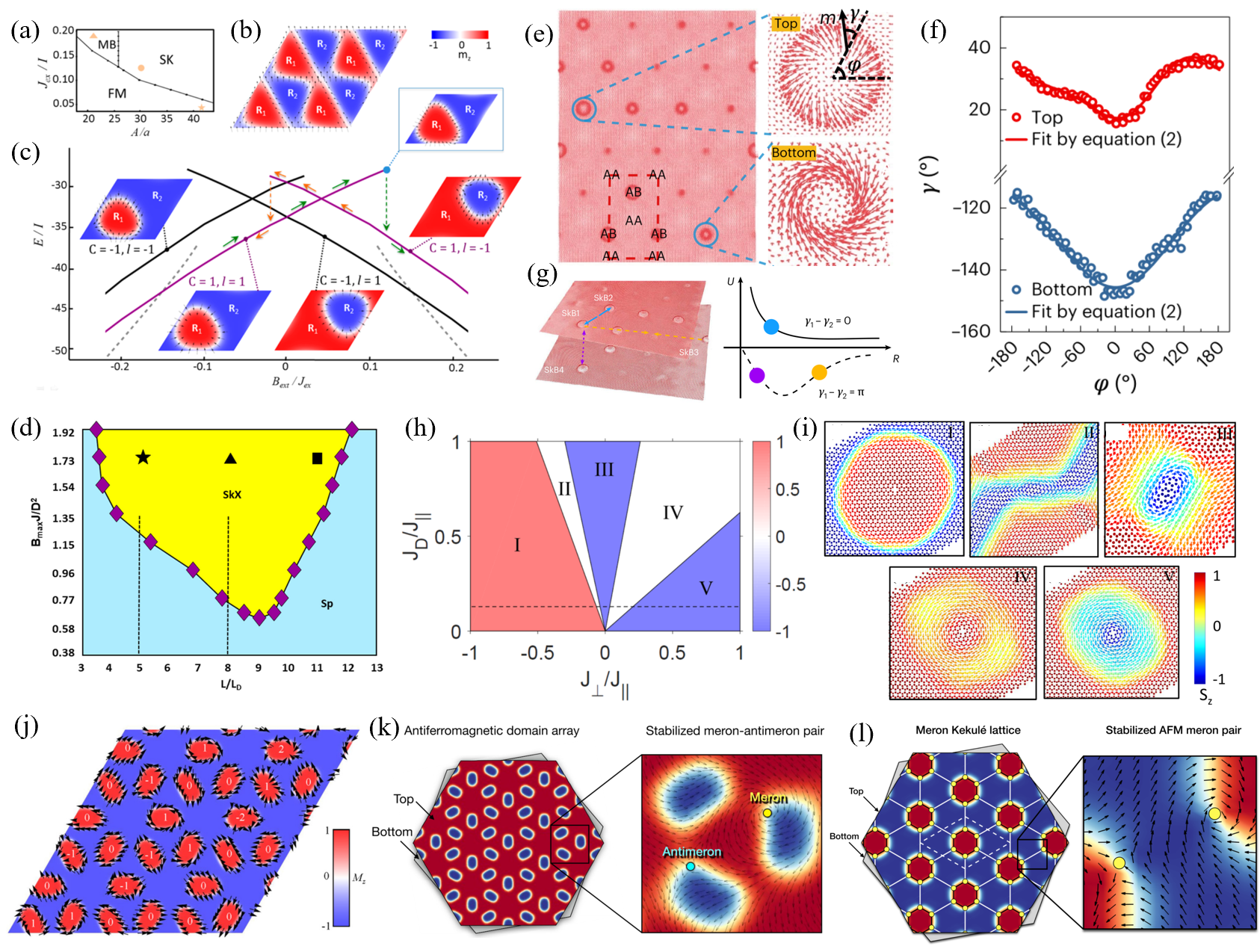}
	\caption{(a) Phase diagram as a function of the maximum MMEI and the moir\'e periodicity at zero external magnetic field. $J_{ex}$ denotes the maximum of MMEI in the  moir\'e superlattice. The triangle in the magnetic bubble lattice phase marks parameters for panel (b), and the dot in the skyrmion lattice phase is for panel (c). (b) Spin texture of the magnetic bubble lattice, shown in 2$\times$2 moir\'e supercells. The arrows show the in-plane orientation at DWs. (c) Energies per supercell of the lowest lying stable magnetic configurations as a function of external magnetic field. Solid curves are four skyrmion lattices of distinct topological skyrmion number C and vorticity $l$. Dashed curves are FM states. Green (orange) arrows indicate the evolution by sweeping the magnetic field up (down), where vorticity and location of skyrmion can switch under the conservation of topological charge. (d) Moir\'e periodicity $L$ vs maximum interlayer exchange field $B_{\mathrm{max}}$ phase diagram with spiral (Sp) and skyrmion crystal (SkX) phases. (e) Spin texture in tBL-CrI$_3$ at twist angle 1.41$^\circ$. Insets show the SkBs in the top and bottom layers. (f) Helicity $\gamma$ as a function of azimuth angle $\varphi$. (g) Schematic diagram of SkB-SkB interactions in the same layer or different layers. (h) Phase diagram of five distinct topological phases in the  $J_{\perp}-J_{D}$ parameter space. $J_{\perp}$ and $J_{D}$ represent the interlayer exchange coupling and DDI, respectively. Red and blue color distinguishes skyrmion number of $\pm$1. (i) Schematic diagram of five distinct spin textures as shown in (h). (i)  Steady-state spin textures evolved from a paramagnetic state for a 3$\times$3 moir\'e supercell in tBL-CrI$_3$. The local topological skyrmion number is indicated at each domain. (k) Schematic diagram of a stable meron-antimeron pair in a twisted magnet. Twist-induced AFM domain array in a FM order background (left panel). Schematic diagram of a stable Meron-antimeron pair (right panel). Red and blue indicate parallel and antiparallel spin alignments between the top and bottom layers respectively. Arrows and circles depict their in-plane winding textures and core positions, respectively. (l) Schematic illustration of a Meron Kekul\'e lattice in a twisted bilayer antiferromagnet. Yellow dots denote the cores of AFM merons that form a Kekul\'e lattice structure. Black solid lines denote intracell bonds between meron cores within the same moiré supercell, while white solid lines denote intercell bonds between meron cores across different supercells. The white dashed line depicts a single moir\'e supercell. Blue color indicates parallel alignment, while red indicates antiparallel alignment between N\'eel vectors across the top and bottom layers. Arrows represent the in-plane components of N\'eel vectors (right panel). Panels (a)-(c) reproduced with permission from Tong \textit{et al}., Nano Lett. {\bf 18}, 7194 (2018). Copyright 2018 American Chemical Society. \cite{Tongq2018} Panel (d) reproduced with permission from Akram \textit{et al}., Phys. Rev. B {\bf 103}, L140406 (2021). Copyright 2021 American Physical Society. \cite{Akram2021} Panels (e)-(g) reproduced with permission from Yang \textit{et al}., Nat. Comput. Sci. {\bf 3}, 314 (2023). Copyright 2023 Springer Nature. \cite{Yangb2023} Panels (h) and (i) reproduced with permission from Ray \textit{et al}., Phys. Rev. B {\bf 104}, 014410 (2021). Copyright 2021 American Physical Society. \cite{Roy2021} Panel (j) reproduced with permission from Xiao \textit{et al}., Phys. Rev. Research {\bf 3}, 013027 (2021). Copyright 2021 American Physical Society. \cite{Xiaof2021} Panel (k) reproduced with permission from Kim \textit{et al}., Nano Lett. {\bf 24}, 74 (2024). Copyright 2024 American Chemical Society. \cite{Kim2024} Panel (l) reproduced with permission from Kim \textit{et al}., npj Quantum Mater. {\bf 10}, 68 (2025). Copyright 2025 Springer Nature. \cite{Kim2025}} 
	\label{fig4}
\end{figure*}

Apart from the above experiments, several theoretical studies have attempted to demonstrate MMEIs and moir\'e magnetism in twisted 2D magnets \cite{Hejazi2020,Wangc2020,Xiaof2021,Kim2023,Kim2023_2,Yangb2023}. Hejazi \textit{et al.} \cite{Hejazi2020} have first developed a continuum field theory to analyze the non-collinear magnetic configurations of the magnetic moir\'e patterns in detail three different examples of twisted bilayers: FM, AFM, and zigzag AFM. Through this approach, they presented a complex phase diagram of non-collinear magnetic textures with respect to moir\'e wavelength and interlayer magnetic exchange. Based on the idea that the stacking order in each local region is similar to the lattice-matched stacking configuration but changes very smoothly over long range in a long-periodic moir\'e pattern, Xiao \textit{et al.} \cite{Xiaof2021} developed an indirect computing method: they first obtained the interlayer magnetic interaction from first-principles calculations in each local region and then extend to the whole moir\'e pattern by introducing vdW corrections (opt-vdW density functionals) in DFT calculations as an effective magnetic field. Subsequently, Kim \textit{et al.} \cite{Kim2023} developed a generic ab initio spin Hamiltonian for tBL-CrI$_3$ based on DFT combined with the magnetic force theorem (MFT). They characterized a functional form of the interlayer exchange coupling, which satisfies the underlying point group symmetry of the atomic structure in the long-periodic moir\'e pattern. Then the unspecified parameters of the coupling function are determined by fitting the DFT data, as a result, the fitting results of the coupling function agree well with the DFT calculations. Additionally, they investigated moir\'e magnetism in  twisted trilayer CrI$_3$ with their coupling function \cite{Kim2023_2}. 

In contrast, a more complete and precise method to obtain approximate MMEI parameters instead of directly solving in tBL-CrI$_3$ was proposed by Yang \textit{et al.} \cite{Yangb2023}, namely the sliding-mapping method as depicted in \Figure{fig3}. The method is based on the fact that local stacking orders in the moir\'e pattern are nearly equivalent to a series of AA-stacking bilayer CrI$_3$ with smoothly sliding by a vector $\textbf{\textit{r}}=m\textbf{\textit{a}}_1+n\textbf{\textit{a}}_2$, as shown in \Figure{fig3}(b). For instance, as shown in \Figure{fig3}(a), the four typical stacking orders AA, AB, BA, and AB' corresponds to interlayer sliding vectors $(\textbf{m},\textbf{n})=$(0,0),(2/3,2/3),(1/3,1/3) and (0,1/3), respectively. \Figure{fig3}(b) presents an illustration of using the sliding-mapping method to calculate four typical interlayer Heisenberg exchange interactions $J_{\mathrm{inter}}$ with different distance $\textbf{\textit{r}}$ ($\textbf{\textit{r}}_1,\textbf{\textit{r}}_2,\textbf{\textit{r}}_3$ and $\textbf{\textit{r}}_4$) in tBL-CrI$_3$. It could be found that the local atomic position around $J_{\mathrm{inter}}(\textbf{\textit{r}})$ in tBL-CrI$_3$ is almost identical with  $\mathcal{J}_{\mathrm{inter}}(\textbf{\textit{r}})$ in bilayer CrI$_3$ (BL-CrI$_3$), after sliding the top layer CrI$_3$ with respect to the bottom layer with same vector $\textbf{\textit{r}}$. That is to say, all interlayer or intralayer magnetic parameters (e.g., Heisenberg exchange interaction \textit{J} and DMI \textit{D}) between two Cr atoms in tBL-CrI$_3$ can be effectively obtained through mapping these magnetic parameters ($\mathcal{J}$ and $\mathcal{D}$) of BL-CrI$_3$ via smooth sliding, as the local structures in the two cases are equivalent. Importantly, an approximate one-to-one maping of the magnetic parameters between tBL-CrI$_3$ and sliding BL-CrI$_3$ could always be found, on condition that the sliding process is sufficiently smooth. Specially, this approach is quite suitable for tBL-CrI$_3$ with small twist angles, as a smaller twist angle leads to more accurate mapping parameters. As a result, direct one-to-one correspondences of the magnetic parameters in tBL-CrI$_3$ from counterparts in BL-CrI$_3$ vis sliding were presented in Figures 3(c)-(e), including interlayer and intralayer Heisenberg exchange and DM interactions.  

\subsection{Topological aspects of moir\'e spintronics}%
Over the past decades, topological magnetic materials, topological quantum effects, and topological excitations have become a booming research frontier in spintronics \cite{Smejkal2018,Bonbien2021,Heq2022}. In real space, topologically protected unconventional magnetic configurations such as (anti)skyrmions, skyrmionium, (anti)merons, bimerons, (anti)hedgehogs, and hopfions have been stabilized in magnetic systems \cite{Tokura2021,Nagaosa2013,Gobel2021,Fert2017}. On the other hand, manifestations of topological spintronics in momentum space include topological magnons \cite{McClarty2021,Zhuo2025}, magnetic topological insulators and semimetals \cite{Smejkal2017,Tokura2019,Changc2020}. In the present section, we concentrate on the topological spin textures and magnons in twisted 2D magnets. 

\subsubsection{Topological moir\'e spin textures in twisted magnets: Moir\'e skyrmions}
Magnetic skyrmion is a particle-like magnetization texture with non-collinear and non-coplanar spin configurations characterized by a topological integer number (i.e., skyrmion number). Due to their topologically protected stability and observation in noncentrosymmetric chiral magnets at room-temperature \cite{Muhlbauer2009,Yu2010}, skyrmions have been proposed to serve as candidate information carriers for next-generation memory devices \cite{Fert2017}. In addition, it was shown that the N\'eel type skyrmions can be stabilized in 2D magnetic systems by the interfacial DMI due to broken inversion symmetry at the interface of ferromagnet/heavy-metal heterostructures \cite{Romming2013,Luchaire2016,Wangl2018}. However, the DMI induced by broken inversion symmetry and strong spin-orbital coupling has greatly limited the material realizations and device applications of skyrmions. By twist angles and translational sliding, the tunable periodicity of moir\'e superlattice in vdW heterostructure of 2D magnets can play crucial roles in skyrmion formation as a new and general mechanism. Recent theoretical studies have predicted the emergence of topological spin textures in twisted 2D magnets \cite{Yangb2023,Xiaof2021,Tongq2018,Akram2021,Akram2021_NL,Hejazi2021,Fumega2023,Ray2021,Ghader2022,Shaban2023,Akram2024,Kim2024,Kim2025}. 

Tong \textit{et al.} \cite{Tongq2018} first explored the possibility of stabilizing and controlling magnetic skyrmion lattice in the long-periodic moir\'e superlattices, produced by twisting a FM monolayer in an angle $\theta$ on a layered AFM substrate both of a hexagonal lattice structure. In regard to the possible candidate materials, CrX$_3$ (X=Cl, Br, I) for the FM monolayer and the layered manganese chalcogenophosphates MnPX$_3$ (X=S, Se, Te) for AFM substrate are good choices, respectively. \Figure{fig4}(a) shows the phase diagram as a function of the maximum MMEI and the moir\'e periodicity at zero external magnetic field. The MMEI and DDI between the top monolayer and the substrate jointly induce an effective alternating exchange field, that the competition between them determines the FM phase in the top monolayer transforms to the magnetic bubble (MB) lattice phase [\Figure{fig4}(b)] or skyrmion lattice phase [\Figure{fig4}(c)]. Furthermore, sweeping the external magnetic field can tune the skyrmion size and switch the vorticity and location of skyrmion under the conservation of topological charge as shown in \Figure{fig4}(c). Unlike DDI stabilizing a skyrmion lattice in Ref. \cite{Tongq2018}, Akram \textit{et al.} \cite{Akram2021} show that the skyrmion lattice can also be stabilized in the top monolayer by a DMI in the absence of an external magnetic field. \Figure{fig4}(d) shows the moir\'e periodicity vs maximum interlayer exchange field phase diagram without the MD phase. In addition, they found the easy-axis anisotropy is essential to stabilize skyrmions. 
\begin{figure*}[htp]
	\centering
	\includegraphics[width=0.98\textwidth]{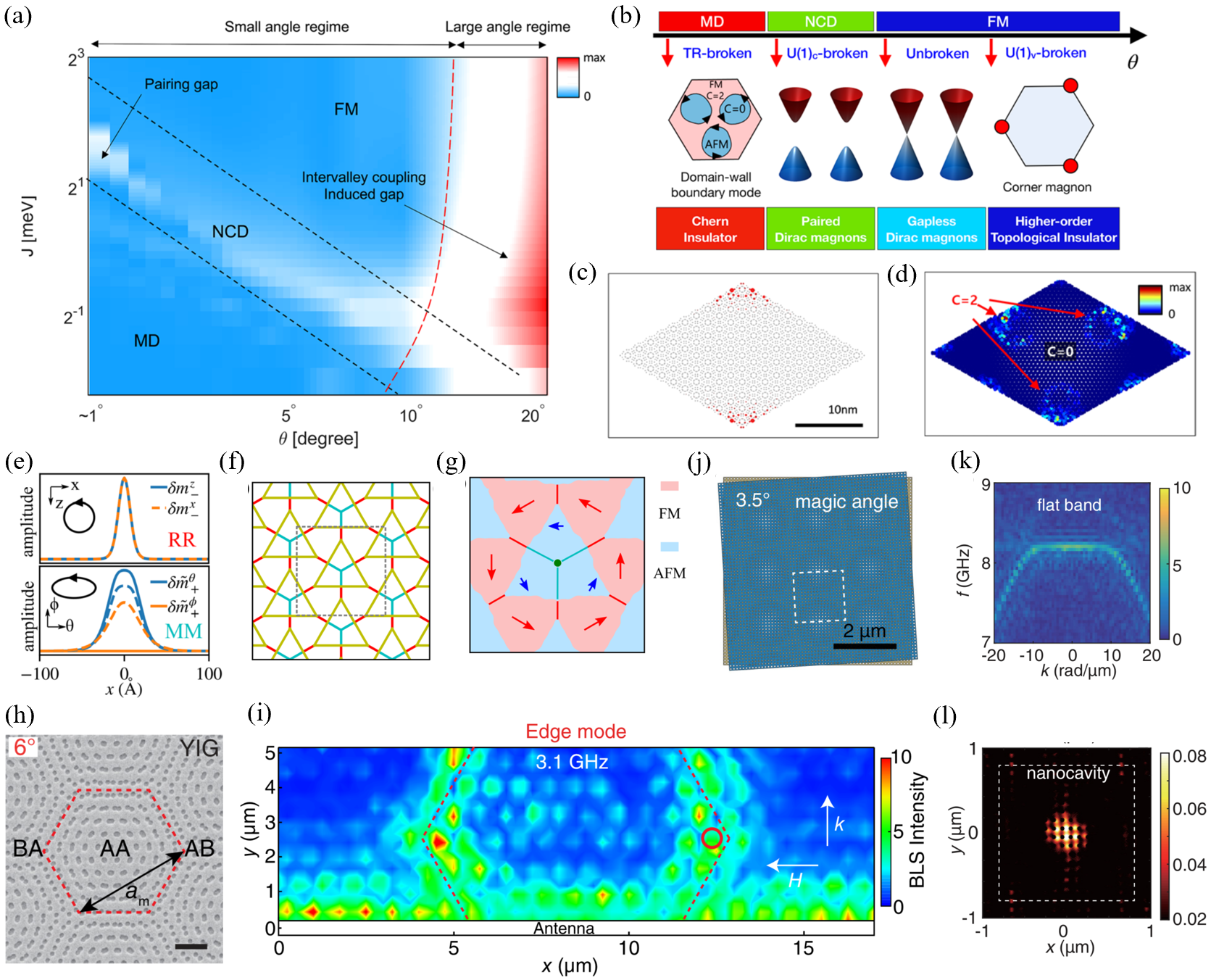}
	\caption{(a) The gap of Dirac magnons as a function of twist angle and intralayer exchange coupling. (b) Summary of the broken symmetries in the magnetic phases and the corresponding topological magnon phases. (c) Corner mode wave function of the higher-order topological insulator. (d) DW boundary mode of the Chern insulator. Here, C denotes the Chern number in each domain. (e) Bounded magnons in the RR (top) and MM (bottom) DWs. (f) Sketch of the magnon network in tBL-CrI$_3$ with twist angle 0.1$^\circ$. (g) Enlargement into the region marked by the gray rectangle in (f) and shows the stacking and magnetic domain patterns. Red (blue) arrows represent stacking order for rhombohedral (monoclinic) stackings. Red (cyan) lines represent the RR (MM) stacking DWs. (h) Scanning electron microscope image of a moir\'e magnonic lattice based on YIG grown on a GGG substrate with a twist angle of 6$^\circ$ (magic angle). (i) Microfocused Brillouin light scattering ($\mu$BLS) signals detected as a function offrequency at the AB region with an applied field of 50 mT. (j) Schematic illustrations of twisted bilayer magnonic crystals with twist angles of 3.5$^\circ$ (magic angle). (k) The spin wave dispersion near the flat-band region simulated at the center of the moir\'e spuercell. (l) Magnetization $m_x$ snapshots of a 6$\mu$m $\times$ 2$\mu$m area at $t = 4.3$ns after the start of continuous microwave excitation.
		Panels (a)-(d) reproduced with permission from Kim \textit{et al}., Nano Lett. {\bf 23}, 6088 (2023). Copyright 2023 American Chemical Society. \cite{Kim2023} Panels (e)-(g) reproduced with permission from Wang \textit{et al}., Phys. Rev. Lett. {\bf 125}, 247201 (2020). Copyright 2020 American Physical Society. \cite{Wangc2020} Panels (h) and (i) reproduced with permission from Wang \textit{et al}., Phys. Rev. X {\bf 13}, 021016 (2023). Copyright 2023 American Physical Society. \cite{Wangh2023} Panels (j)-(l) reproduced with permission from Chen \textit{et al}., Phys. Rev. B {\bf 105}, 094445 (2022). Copyright 2022 American Physical Society. \cite{Chenj2022}} 
	\label{fig5}
\end{figure*}

Another candidate moir\'e system for skyrmion formation is twisted homobilayer of vdW magnetic layers with ferromagnetic order, for instance, several theoretical studies reported skyrmions in twisted bilayer CrX$_3$. Most of them adopted the effective spin Hamiltonian with the DMI to explore the moir\'e magnets \cite{Yangb2023,Akram2021_NL,Hejazi2021,Fumega2023}. For instance, as shown in \Figure{fig4}(e), Yang \textit{et al.} \cite{Yangb2023} found a new type of magnetic skyrmion bubble (SkB) due to the emergence of MMEIs in tBL-CrI$_3$ with a twist angle 1.41$^{\circ}$, which has non-conserved helicity $\gamma$ [\Figure{fig4}(f)] and has never been predicted pr observed in previous works. \Figure{fig4}(f) shows that SkBs located in same (different) layers have same (opposite) sign of $\gamma$. Additionally, the topological number and vorticity for SkBs in top and bottom layer are -1 and 1, respectively. \Figure{fig4}(g) shows the interactions between SkBs as a function of in the same layer or different layers the distance between them. Interestingly, there is an attractive interaction between them two SkBs when they are far from each other [see SkB1 and SkB3 in \Figure{fig4}(g)]. Conversely, a much stronger repulsion between them emerges if two SkBs are close to each other [see SkB1 and SkB4 in \Figure{fig4}(g)]. As a result, SkBs in different layers cannot exist at the same stacking site. By contrast, Ray \textit{et al.} \cite{Ray2021} determined the spin textures in twisted magnetic bilayers with interlayer DDI at low temperature by carrying out Monte Carlo (MC) simulations. \Figure{fig4}(h) presents a phase diagram of five distinct phases of spin textures in parameter space of the interlayer Heisenberg exchange coupling and DDI with a twist angle 1.61$^{\circ}$. The spin textures in five different phases are demonstrated in \Figure{fig4}(h). They found three distinct skyrmion phases (phase \uppercase\expandafter{\romannumeral1}, \uppercase\expandafter{\romannumeral3}, and \uppercase\expandafter{\romannumeral5}), which have different topological charge distributions (point, rod, and ring shapes) as shown in \Figure{fig4}(i). Xiao \textit{et al.} \cite{Xiaof2021} found the formation of nonuniform magnetization textures by applying a uniform external magnetic field in tBL-CrX$_3$ in the absence of DMI and DDI,  including magnetic skyrmion with different topological numbers and topologically trivial magnetic bubble as shown in \Figure{fig4}(j). Moreover, moir\'e skyrmions are discovered in tBL-CrI$_3$ \cite{Ghader2022} and twisted bilayer $\alpha$-RuCl$_3$ \cite{Akram2024} by introducing Kitaev interactions in a generic spin Hamiltonian. Overall, the stabilization mechanisms of skyrmion (lattice) in magnetic moir\'e system can be classified as follows: The substrate-induced interfacial DMI including the interlayer and intralayer DMI with a uniform external magnetic field dominates skyrmion stabilization in tBL-CrI$_3$ with layered BN substrate \cite{Yangb2023,Akram2021}. The coexistence of FM and AFM interlayer couplings from competing FM/AFM domains in twisted homobilayers like tBL-CrI$_3$ may allow for a further reduction in magnetic energy to form an in-plane winding texture instead of uniform magnetization even without asymmetric interactions such as the DMI \cite{Xiaof2021,Kim2023}. When twisting the FM monolayer on an AFM substrate, the MMEIs combined with DDI or interfacial DMI stabilizes skyrmions in FM layer \cite{Akram2021,Tongq2018}. However, DMI-induced skyrmions offer potential for device applications due to convenient current-driven motion. Conversely, moir\'e skyrmion lattices induced by interlayer interaction frustration benefit from inherent electrical tunability of the moir\'e pattern itself. Furthermore, high scalability and room-temperature realization remain key challenges for moir\'e skyrmions. The vdW heterostructure combining high-$T_c$ magnets with strong SOC layers (e.g., TMDs) may offer promising solutions \cite{Huangk2022}.

Beyond moir\'e skyrmions, another kind of topological spin texture--stable magnetic meron is predicted in twisted 2D magnets \cite{Kim2024,Kim2025}. In Ref\cite{Kim2024}, Kim \textit{et al.} presented a novel magnetic texture dubbed the “Meron Quartet” in tBL-CrI$_3$, which is composed of four merons (two for each layer) as depicted in \Figure{fig4}(k). Besides, twist-induced AFM domains (blue areas) arranged into a Kagome lattice in a FM background (red areas), in which stable meron-antimeron pairs appear. Since then, they proposed a new magnetic state "Meron Kekul\'e lattice" in twisted bilayer easy-plane antiferromagnets, that a Kekul\'e lattice are composed of antiferromagnetic merons as shown in \Figure{fig4}(l) \cite{Kim2025}.     

\subsubsection{Low energy excitations in twisted magnets: Moir\'e magnons}
As low-energy excitations in magnetic systems, magnons are promising information carriers due to their low dissipation and long coherence length. In the past decade, topological phases in magnonic systems have attracted intensive attention, such as topological magnon insulators, magnon Chern insulators, and topological magnon semimetals. Recently, significant efforts have been directed toward the exploration of magnons in twisted vdW magnets--moir\'e magnons \cite{Liy2020,Wangh2023,Chenj2022,Chenj2024,Liuj2025,Huac2023,Ganguli2023,Ghader2020}, unveiling a vast landscape of opportunities for both fundamental research and practical applications of magnonic devices.

The study presented in Ref. \cite{Liy2020} was centered on moir\'e magnons in twisted bilayer magnets with four different magnetic ground states, in which the out-of-plane collinear magnetic order of each layer is preserved in the presence of next-nearest-neighbor DMI and weak interlayer Heisenberg exchange interaction. The authors unveil that the twist angle has a powerful influence on the magnon dispersion, leading to the appearance of topological flat magnon bands and intricate magnon thermal Hall effect. Another intriguing studies of moir\'e magnons is the significance of magnons associated with non-collinear magnetic textures in coexisting FM--AFM states in twisted vdW magnets \cite{Kim2023,Wangc2020}. The topological band theory of moir\'e magnons was well established in different magnetic ground state phases in Ref. \cite{Kim2023}, including FM, non-collinear domain, and magnetic domain. \Figure{fig5}(a) shows the gap of Dirac magnons as a function of twist angle and intralayer Heisenberg exchange interaction. However, the mass gap opening of the Dirac magnons with increasing twist angle originates from the breaking of the symmetries as depicted in \Figure{fig5}(b). A second-order topological magnon insulator with magnon corner states [\Figure{fig5}(c)] was predicted in tBL-CrI$_3$ with a large commensurate angle 21.78$^{\circ}$ in Refs. \cite{Kim2023,Huac2023}. Moreover, a pair of the topological edge mode of magnons flowing around the DW was observed because of the Chern number difference 2 inside and outside the DW, as shown in \Figure{fig5}(d). Similarly, Wang \textit{et al.} \cite{Wangc2020} also demonstrate 1D magnon channels confined in the stacking DWs hosting interconnected moir\'e magnons network in small-angle tBL-CrI$_3$ [Figures \Figure{fig5}(e)-(g)], which are more stable against external perturbations in comparison with confined 1D magnons in magnetic DWs proposed in previous works \cite{Ferrer2003,Sanchez2015}. 

Experimental studies in Ref. \cite{Wangh2023} demonstrate the observation of spin-wave moir\'e edge and cavity modes using Brillouin light scattering spectromicroscopy in an artificial moir\'e lattice, consisting of two twisted triangle antidot lattices made on a yttrium iron garnet thin film [\Figure{fig5}(h)]. By precisely tuning the twist angle, the emergence of spin-wave moir\'e edge mode localized at the edges of a moir\'e unit cell are detected around the optimal twist angle or the "magic angle" 6$^{\circ}$ with a selective excitation frequency [\Figure{fig5}(i)]. Furthermore, Chen \textit{et al.} \cite{Chenj2022,Chenj2024} report the formation of a magnonic flat band at the center of the Brillouin zone [\Figure{fig5}(k)] in twisted bilayer magnonic crystals by stacking two square antidot lattices with the magic angle 3.5$^{\circ}$ [\Figure{fig5}(j)]. As demonstrated in \Figure{fig5}(l), the magnon DOS remarkably increases at the center of a moir\'e unit cell suggesting the strong confinement of magnons as a result of magnon excitation at the flat band. Interestingly, this efficient accumulation of magnon density in a confined region presents exciting opportunities for potential applications, such as Bose-Einstein condensation of magnons \cite{Demokritov2006,Giamarchi2008}. 

\section{Machine learning accelerating discovery in moir\'e spintronics}
The study of moir\'e materials presents a dual challenge of complexity: the large number of possible material combinations and the immense computational cost of simulating any single configuration. ML and AI are emerging as powerful tools to address both of these challenges, offering new capabilities for accelerating simulation, analyzing complex experimental data, and performing materials inverse design. This section reviews the recent advances in applying ML methods to moir\'e spintronics, categorized into two main areas: overcoming computational barriers in simulation and enabling higher-level data analysis and design \cite{Tritsaris2021,Li2023,Liu2024}. In this section, we clarify how ML is evolving from a mere data analysis tool into a driver for material discovery and twist-angle optimization, providing concrete examples where it has altered our physical understanding of moir\'e systems.

\subsection{Overcoming computational barriers: ML-accelerated simulation}
\begin{figure*}[htp]
	\centering
	\includegraphics[width=0.95\textwidth]{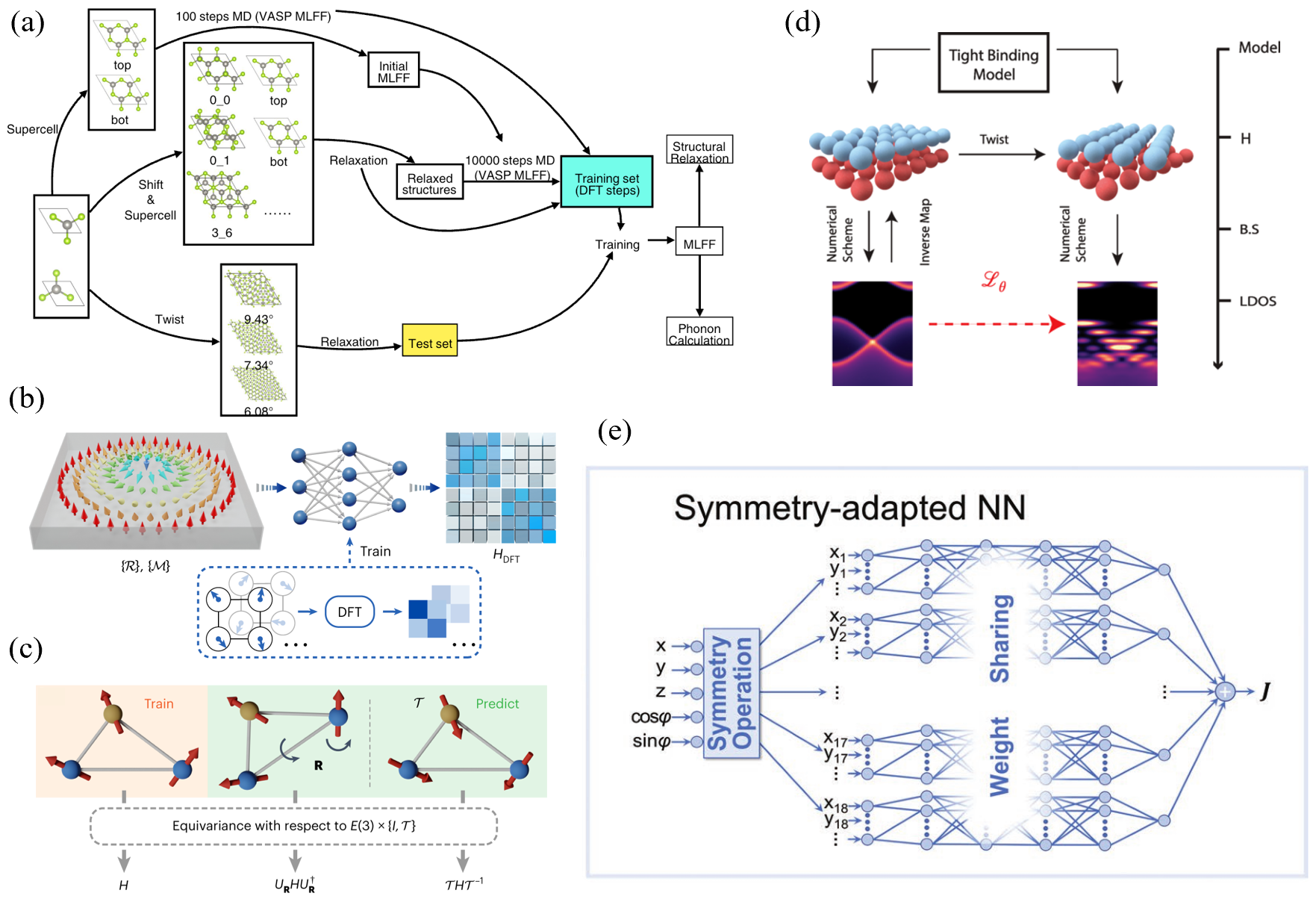}
	\caption{Schematic workflows for ML-accelerated simulation of moir\'e systems. (a) Workflow for generating a Machine Learning Force Field (MLFF) for structural relaxation using the \texttt{DPmoire} methodology. The model is trained on a dataset constructed from small, non-twisted bilayer structures with various lateral shifts and molecular dynamics steps. (b) Workflow for a Deep Learning Hamiltonian using the \texttt{xDeepH} framework. A neural network is trained on DFT data from small structures to learn the mapping from a magnetic structure (e.g., a skyrmion) to its DFT Hamiltonian ($H_\text{DFT}$). (c) A key principle for the \texttt{xDeepH} model is respecting physical symmetries. The neural network is constructed to be equivariant under symmetry operations such as rotation (R) and time-reversal ($\mathcal{T}$). (\textbf{d}) Workflow for learning a generalized "twist operator". A network learns the transformation from the electronic structure of an aligned bilayer to that of a twisted moir\'e system, enabling material-agnostic predictions. (e) Architecture of a symmetry-adapted neural network (SANN) used to learn the parameters of a classical spin model, such as the exchange interaction $J$. Panel (a) reproduced with permission from Liu \textit{et al.} npj Comput. Mater. {\bf 11}, 248 (2025). Copyright 2025 Springer Nature. \cite{Liu2024}. Panels (b) and (c) reproduced with permission from Li \textit{et al.}, Nat. Comput. Sci. {\bf 3}, 321 (2023). Copyright 2023 Springer Nature. \cite{Li2023} Panel (d) reproduced from Liu \textit{et al.}, Phys. Rev. Research {\bf 4}, 043224 (2022). Copyright 2022 American Physical Society. \cite{Liu2022_seeing} Panel (e) reproduced from Zheng, Adv. Funct. Mater. {\bf 33}, 2206923 (2023). Copyright 2023 John Wiley and Sons.  \cite{Zheng2022_magnetic} }
	\label{fig6}
\end{figure*}

The primary obstacle to the theoretical investigation of moir\'e systems, particularly at small twist angles, is the computational scaling problem. The moir\'e supercells can contain thousands or even tens of thousands of atoms, making first-principles calculations using DFT prohibitively expensive, as the computational cost scales cubically with the number of atoms \cite{Li2023,Liu2024}. This computational scaling problem severely limits the scope of structures that can be studied \cite{Soriano2023}. ML offers a path to bypass this limitation by learning the fundamental relationships between atomic structure and physical properties from a smaller, computationally tractable set of DFT data. The primary strategies for this, using ML to learn either the atomic forces, the electronic Hamiltonian, or the electronic structure directly, are illustrated in Figure \ref{fig6}.

\subsubsection{Machine learning force fields for structural relaxation}

The precise atomic structure of a moir\'e superlattice is not rigid; the atoms relax to minimize the total energy, and this structural relaxation significantly influences the electronic and magnetic properties \cite{Liu2024}. Accurately calculating these relaxed structures is therefore a critical first step. Machine Learning Force Fields (MLFFs) provide a solution by learning the potential energy surface from DFT data. A notable methodology in this area is embodied by the open-source software package \texttt{DPmoire}, which is specifically tailored for constructing MLFFs for moir\'e systems. The core strategy of \texttt{DPmoire} is to construct a training dataset not from large, expensive twisted structures, but from computationally cheaper, small, non-twisted bilayer structures with various lateral stacking shifts, as shown in Figure \ref{fig6}(a). This approach has been successfully demonstrated for TMD systems, including various MX$_{2}$ materials (M=Mo, W; X=S, Se, Te).

To augment the training data, \texttt{DPmoire} employs an on-the-fly learning strategy during molecular dynamics (MD) simulations, utilizing the MLFF module within the Vienna Ab initio Simulation Package (VASP). This module operates on a Bayesian linear regression framework, which has a significant advantage: it can estimate the uncertainty of its own predictions in real-time. During an MD run, if the model estimates its prediction error for a configuration to be low, it uses the fast MLFF-calculated forces. If the estimated error is high, the model triggers a full, computationally expensive DFT calculation. This accurate DFT data is then added to the training set, iteratively refining the MLFF in a continuous active learning loop. This process allows for extensive sampling of the configuration space while minimizing the number of costly DFT calculations, thereby efficiently generating a high-quality dataset. Once trained on this comprehensive dataset, the resulting MLFF can predict forces with near-DFT accuracy but at a fraction of the cost. The performance of these MLFFs is robust; for example, in twisted bilayer WSe$_{2}$ and MoS$_{2}$, the root mean square errors for atomic forces were as low as $0.007~$eV/\AA and $0.014~$eV/\AA, respectively. Crucially, this ML-driven relaxation reveals that lattice reconstruction is a dominant factor in determining moir\'e potential depth and band flatness, capturing physics often missed by rigid-lattice approximations.

\subsubsection{Deep learning Hamiltonians for electronic and magnetic structure}
Beyond obtaining the correct atomic structure, understanding the spintronic properties requires knowledge of the system's electronic and magnetic Hamiltonian. Here too, ML models can learn the complex mapping from atomic and magnetic configurations to the DFT Hamiltonian matrix elements. The extended Deep Learning DFT Hamiltonian (\texttt{xDeepH}) framework is a prime example of this approach, specifically tailored for magnetic materials \cite{Li2023}. The general workflow, shown in Fig. \ref{fig6}(b), involves training a neural network on DFT data from small-scale structures to predict the Hamiltonian of large-scale magnetic superstructures.

The innovation of \texttt{xDeepH} lies in its use of a physics-informed neural network (PINN) architecture \cite{Karniadakis2021}. Instead of being a generic "black-box" model, it is explicitly designed to incorporate and respect the fundamental physics of the system. The network is constructed as an equivariant neural network (ENN) that respects the symmetry group E(3)$\times$\{I,$\mathcal{T}$\}, which includes Euclidean transformations (translations, rotations, inversions) and time-reversal symmetry \cite{Li2023}, a concept illustrated in Figure \ref{fig6}(c). This equivariance is critical for correctly capturing the behavior of magnetic moments and spin-orbit effects. The model also incorporates the "nearsightedness principle" of electronic matter, which posits that the electronic properties at a given point are primarily influenced by the local atomic environment. This is implemented using a message-passing neural network architecture where atoms and their interactions are represented as a graph, and information about the more localized magnetic structure is introduced with strict locality to improve training efficiency and accuracy. By embedding this physical knowledge directly into the model architecture, \texttt{xDeepH} achieves high accuracy and transferability, with reported mean absolute errors for Hamiltonian matrix elements as low as 0.36~meV for monolayer CrI$_3$ test sets. This level of performance has enabled the study of complex magnetic superstructures, such as the skyrmion phase in twisted CrI$_3$ at a twist angle of 63.48$^{\circ}$ (containing $\sim$4,500 atoms), a system far too large for direct DFT treatment \cite{Li2023,Soriano2023}. This application provides a concrete example of ML changing physical understanding. By enabling full-scale simulation, \texttt{xDeepH} revealed that the magnetic ground state involves complex non-conserved helicity in moir\'e skyrmions, a subtle texture feature that cannot be captured by simplified continuum models.

A complementary approach involves using ML to learn the parameters of a classical spin model rather than the full quantum mechanical Hamiltonian. In a study of twisted bilayer CrI$_3$, Zheng employed a symmetry-adapted artificial neural network (SANN) to learn the complex, stacking-dependent interlayer exchange interactions~\cite{Zheng2022_magnetic}. The SANN, whose architecture is shown in Fig. \ref{fig6}(e), was trained on data from first-principles calculations on small cluster models, thereby bypassing the need for direct DFT calculations on the full moir\'e supercell. Once trained, the network functions as a highly efficient surrogate model, providing the necessary exchange parameters to construct a full classical spin Hamiltonian for the large twisted system. Subsequent Landau-Lifshitz-Gilbert (LLG) simulations using this ML-parameterized Hamiltonian successfully explained the disorderly magnetic domains observed at small twist angles and predicted the formation of ordered skyrmion lattices at large twist angles near 60$^{\circ}$. This demonstrates the power of ML in bridging the gap between first-principles accuracy and the length scales required for micromagnetic simulations.

\subsubsection{Learning generalized transformations and other physical properties}
Beyond learning forces and Hamiltonians for specific material systems, another powerful ML strategy is to learn a generalized, material-agnostic transformation that maps the properties of an untwisted bilayer to its twisted moir\'e counterpart. Liu \textit{et al.} proposed learning a universal "twist operator", which they approximate with a CNN~\cite{Liu2022_seeing}, with the overall concept outlined in Figure \ref{fig6}(d). This approach frames the problem as one of image processing: the stacking-dependent local density of states (SD-LDOS) of an aligned bilayer is treated as an input image, and the network is trained to predict the corresponding SD-LDOS image of the twisted system. Because the network learns the general effect of the twist, it can make reasonable predictions for materials not included in its training set, offering a powerful tool for high-throughput screening of novel moir\'e materials without the need for material-specific parameterization.

This concept of using ML to learn local-to-global relationships extends to other physical properties as well. For example, Zheng \textit{et al.} used a "deep Wannier" model, a method originally developed to model the dielectric response of insulators by learning the environmental dependence of electronic charge centers~\cite{Zhang2020_deep}, to investigate moir\'e ferroelectricity in twisted hexagonal boron nitride (h-BN)~\cite{Zheng2025_machine}. The DW model is trained on a small set of DFT calculations of non-twisted, shifted bilayers to learn the mapping between local atomic environments and the centers of electronic charge. Once trained, the model can rapidly and accurately predict the full polarization pattern across a large moir\'e supercell, a task that would be computationally intractable with direct DFT. This demonstrates the broad applicability of ML surrogate models for exploring various moir\'e phenomena, including those closely related to spintronics like ferroelectricity and multiferroics.

\begin{figure*}[htp]
	\includegraphics[width=0.95\textwidth]{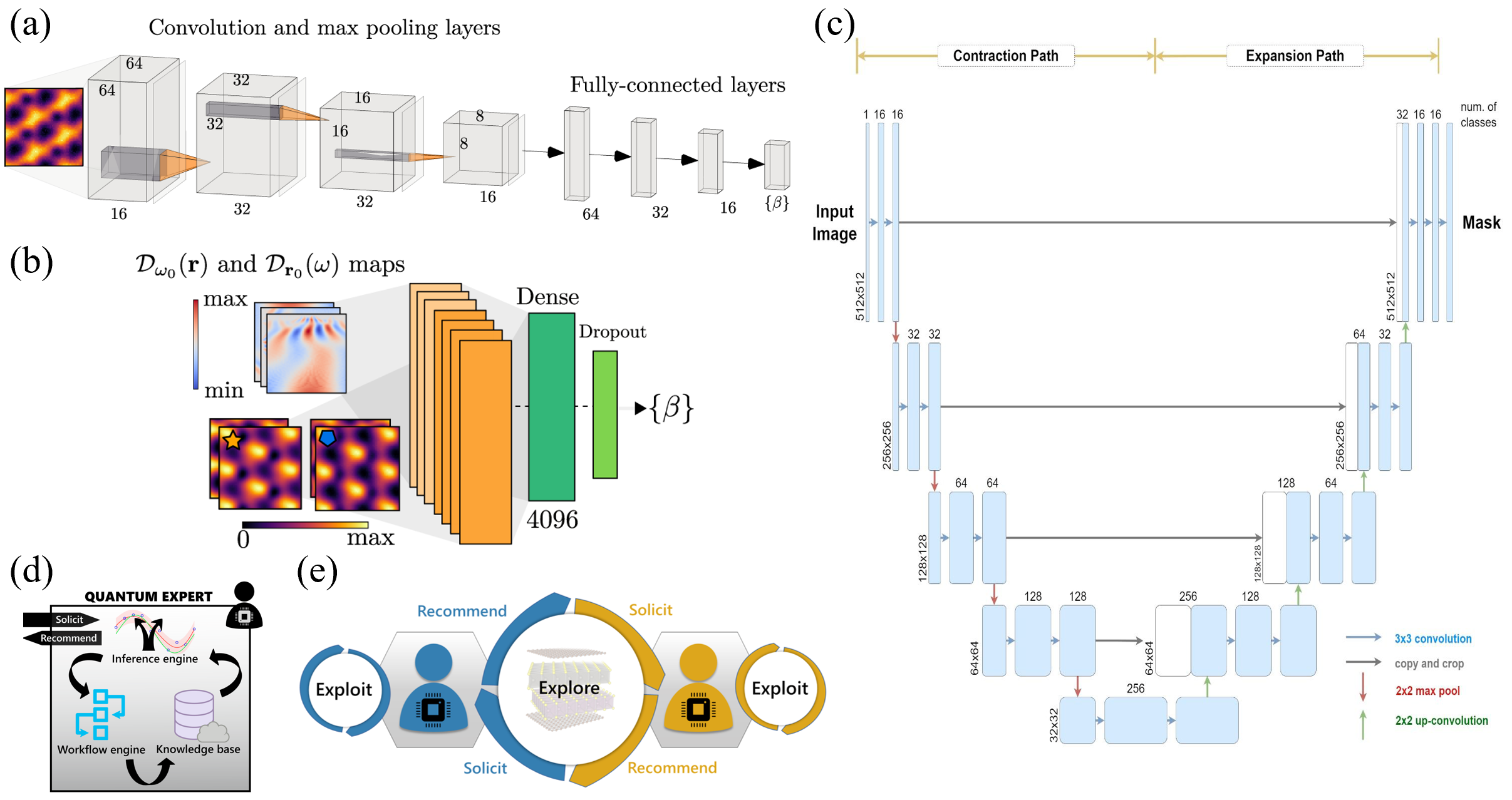}
	\caption{Machine learning for interpreting complex experimental data. (a) and (b) A multi-channel Convolutional Neural Network (CNN) architecture designed to solve the inverse problem of extracting microscopic physical parameters from experimental data. Spatially-resolved local density of states (LDOS) maps and point spectra are fed into parallel convolutional and fully-connected layers to distinguish between competing physical effects, such as electronic nematicity and lattice strain. (c) A U-Net derivative architecture, another type of CNN, used for image segmentation tasks. This architecture is effective for identifying features in noisy experimental images, such as detecting magnetic skyrmions. (d) and (e) Schematic of an AI-driven framework for autonomous materials discovery and inverse design. (d) The core components of an autonomous agent include an inference engine for planning, a workflow engine for executing virtual experiments (simulations), and a knowledge base for storing data. (e) The operational model for such a framework involves a closed loop where AI agents intelligently navigate the materials space by balancing \textit{exploitation} of known promising areas with \textit{exploration} of new configurations to efficiently find materials with desired target properties. Panels (a) and (b) reproduced with permission from Sobral \textit{et al.}, Nat. Commun. {\bf 14}, 5012 (2023). Copyright 2023 Springer Nature. \cite{Sobral2023} Panel (c) reproduced with permission from Labrie-Boulay \textit{et al.} Phys. Rev. Appl. {\bf 21}, 014014 (2024). Copyright 2024 American Physical Society. \cite{Labrie-Boulay2024_machine}. Panels (d) and (e) reproduced with permission from Tritsaris \textit{et al.}, Appl. Phys. Rev. {\bf 8}, 031401 (2021). Copyright 2021 AIP Publishing. \cite{Tritsaris2021}}
	\label{fig7}
\end{figure*}
\subsection{From data analysis to inverse design}
Beyond accelerating forward simulations, ML methods are increasingly applied to higher-level tasks such as interpreting complex experimental data and performing inverse design, where a material structure is sought based on a desired target property.

\subsubsection{Interpreting complex experimental data}
Modern experimental probes like STM generate vast, high-resolution, image-like datasets. These datasets are rich with information about the underlying physics but are often too complex for straightforward interpretation by humans. The inverse problem---extracting a microscopic physical model from the experimental data---is a significant challenge where ML excels \cite{Sobral2023}. The work of Sobral et al. demonstrates this power by using Convolutional Neural Networks (CNNs) to analyze STM data of nematic order in twisted double-bilayer graphene (TDBG). The architecture of such a network is shown in Figures \ref{fig7}(a) and (b). A key finding is that analyzing a local density of states (LDOS) map at a single energy is insufficient to uniquely determine the microscopic origin of the observed symmetry breaking. Different physical mechanisms, such as intrinsic electronic nematicity and extrinsic heterostrain, can produce visually similar patterns at one energy. However, by using multi-channel inputs---combining STM data from different bias voltages (energies) and different spatial locations---the CNN can learn the subtle correlations across the dataset. This allows it to successfully disambiguate the contributions of nematic order from those of strain, a task that is nearly impossible through manual inspection. This work showcases ML not just as a data analysis tool, but as a sophisticated interpretation engine that resolves competing physical explanations, directly impacting the understanding of the broken symmetries in moir\'e systems.

While the analysis of simulated STM data is powerful, applying ML to raw experimental images presents additional challenges, such as noise, low contrast, and physical defects. Labrie-Boulay \textit{et al.} addressed this by developing a CNN based on a U-Net architecture [Fig. \ref{fig7}(c)] to automatically detect magnetic skyrmions in MOKE microscopy images~\cite{Labrie-Boulay2024_machine}. They found that a simple binary classifier (skyrmion vs. background) performed poorly, as it learned to rely too heavily on pixel intensity and often confused bright defects with skyrmions. The key insight was to introduce a third class for defects, forcing the network to learn the distinct shapes and features of skyrmions rather than just their contrast. This 3-class model proved to be a robust and accurate method for automating the pre-processing of noisy experimental data, a critical step for enabling high-throughput analysis of magnetic textures.

\subsubsection{AI-driven discovery and inverse design}
The configuration space of moir\'e materials is combinatorially large, defined by the choice of constituent layers, stacking sequence, twist angle, strain, and external fields. An exhaustive search of this space is impossible \cite{Tritsaris2021}. AI-driven workflows offer a strategy to navigate this space intelligently. This concept is exemplified by autonomous discovery frameworks that employ agent-based simulations, as described in Figures \ref{fig7}(d) and (e). In such a framework, an autonomous AI agent intelligently navigates the vast material space by cyclically planning and executing virtual experiments (i.e., simulations) and analyzing the results to inform the next step. This approach shifts the paradigm from simple high-throughput screening to active learning and optimization. It enables the pursuit of inverse design, which shifts the objective from predicting the properties of a given structure to identifying a structure that exhibits a predefined target functionality. This principle applies directly to spintronics, where, for example, an AI agent could be tasked with finding a moir\'e heterostructure that exhibits a specific spin-polarized band gap, a robust topological Hall effect, or a desired magnetic ordering temperature.

These different ML methods can be viewed as components of a synergistic framework. For instance, an AI agent could propose a novel moir\'e structure for inverse design. \texttt{DPmoire} could then be used to efficiently obtain its relaxed atomic geometry, after which \texttt{xDeepH} could calculate its detailed electronic and magnetic properties. The simulated signatures from this pipeline could then be used to train a CNN to recognize the same physical phenomena in a future experiment, creating a powerful, end-to-end computational framework for the discovery, characterization, and experimental verification of novel moir\'e spintronic systems. 

\section{Conclusions and future perspectives}
In conclusion, we have reviewed recent experimental and theoretical progress on the novel quantum phenomena and topological phases in moir\'e spintronics enabled by twisting engineering of layered vdW magnetic materials, including stacking-dependent interlayer magnetism, non-collinear spin textures, moir\'e magnetic exchange interactions, moir\'e skyrmions and moir\'e magnons. We especially emphasize the important role of ML in deep understanding the mechanism of moir\'e magnetism and topological phase transitions in twisted magnets. Here, we summarize recent works and give a list of candidate materials that are currently being intensively investigated for moir\'e spintronics in Table 1. However, moir\'e systems requiring ultra-low temperatures ($<10$K) or exhibiting extreme sensitivity to disorder bring about greater challenges for near-term device integration. We look forward to the promising platforms for device applications include twisted bilayers of high-$T_c$ 2D vdW magnets with room-temperature magnetism (e.g., FGT \cite{Dengy2018}), FM/SC heterostructures for topological superconductivity \cite{Lum2015}, and moir\'e multiferroics enabling electric-field control \cite{Zhuh2025,Antao2024,Yeh2025}. Furthermore, we also provide a systematic summary of the relationship between twist angles and moir\'e  magnetism in Table 2. These results demonstrate the great potential of moiré spintronics in fundamental science and spintronic applications. 
\begin{table*}\label{tab1} 
 \caption{Possible materials for realizing moir\'e spintronics. (In this table, E and T represent experimental observation and theoretical prediction, respectively.)}
\centering
\tabcolsep=0.024\linewidth
\begin{tabular}{l c c c c}
   \hline
    Material  &  Moir\'e system & Moir\'e phenomena & E/T \\
    \hline
    CrX$_3$   &  Twisted bilayer & Non-collinear spin textures \cite{Songt2021,Xuy2022,Chenw2019} &  \\
     (X = I, Br, Cl) & Twisted trilayer &   Moir\'e skyrmions/merons \cite{Yangb2023,Akram2021_NL,Kim2024} & \\
      & Twisted double-bilayer &  Topological magnons \cite{Kim2023,Liy2020} & \\
      &   &   Stacking domain wall magnons \cite{Wangc2020} & E and T \\
      &   &    Magnon corner states \cite{Kim2023,Huac2023}& \\
      &   & Moir\'e-driven multiferroics \cite{Fumega2023, Yeh2025} & \\
      &   &   Altermagnetism \cite{Liuy2024,Her2023} & \\

     \hline 
    Fe$_3$GeTe$_2$ & Twisted few-layer & Non-collinear spin textures \cite{Liang2023} & E \\
    \hline
    CrSBr & Orthogonally twisted & Non-collinear spin textures \cite{Constant2024}  &   \\
    & bilayer & Moir\'e excitons \cite{Liuj2025} &  E and T \\
    &  &   Altermagnetism \cite{Liuy2024,Her2023} &  \\
    \hline
    MnPS$_3$ & Twisted 1L-MoSe$_2$/MnPS$_3$ & Moir\'e trion–magnon complexes \cite{Zhangy2022} & E \\
    \hline
    YIG & Twisted bilayer artificial&  Magnonic flat band \cite{Chenj2022,Chenj2024}  &  E and T \\
    &  magnonic crystals & Spin-wave moir\'e edge and cavity modes \cite{Wangh2023} &  \\
     \hline
    NiI$_2$ & Twisted bilayer & Moir\'e skyrmions \cite{Antao2024}& \\
    &  & Moir\'e-driven multiferroics \cite{Antao2024,Zhuh2025} &  T\\
    &  & Twist-tunable valley polarization \cite{Lil2024} &  \\
    \hline
    RuCl$_3$ & Twisted bilayer & Moir\'e skyrmions \cite{Akram2024}&  T\\
    &  & Quantum spin liquids \cite{Cheng2021} &  \\
    \hline
    NiZrI$_6$ & Twisted bilayer & Altermagnetism \cite{Panb2024,Zengs2024}&  T \\
    \hline
    NiCl$_2$ & Twisted bilayer & Altermagnetism \cite{Her2023,Zengs2024} & T \\
    \hline
  \end{tabular}
\end{table*}

Furthermore, beyond these exciting achievements, the integration of ML into the study of moir\'e materials is rapidly evolving from a tool for accelerating individual simulations into a cornerstone of a new paradigm for materials discovery. The progress in developing ML-accelerated simulation tools and AI-driven analysis methods points toward a future where the design of novel spintronic materials is a highly automated, collaborative, and efficient process. 

However, we believe it is only the beginning of moir\'e spintronics and no one can precisely predict what new discoveries will emerge, as these breakthroughs have been witnessed in the span of less than five years. In spite of the rapid development in the study of this intriguing field, it is far from mature before its practical applications, and there still remain plenty of appealing phenomena to be explored and key challenges to be overcome. However, based on the work done so far, we will further point out some of these challenges that require urgent attention and promising future opportunities related to the practical applications.  
\begin{table*}\label{tab2} 
 \caption{A summary of the relationship between twist angles and moir\'e magnetism.}
\centering
\tabcolsep=0.024\linewidth
\begin{tabular}{l c c c c}
   \hline
    Moir\'e magnetic system  & Twist angle range & Magnetic state/Remarkable feature\\
    \hline
    tBL-CrI$_3$ \cite{Songt2021,Yangb2023,Kim2023,Huac2023,Wangc2020} & $ \theta = 0.1^{\circ} $  &  Stacking domain wall magnons:  \\
    & & magnon network \\
    \cline{2-3}
    & $ 0.2^{\circ} <\theta <1.2^{\circ}$  &  Non-collinear spin textures:  \\
    & & coexisting FM/AFM domains \\
    \cline{2-3}
     &  $ 1.2^{\circ} < \theta  < 3^{\circ} $ &  Moir\'e skyrmions \\
     \cline{2-3}
     & $ \theta > 3^{\circ} $  &  FM \\ 
     \cline{2-3}
     & $ \theta = 21.78^{\circ} $  &  Magnon corner states  \\
     \hline
    tDB-CrI$_3$ \cite{Songt2021,Chengg2023,Xieh2022,Xieh2023}  & $ 0.14^{\circ} <\theta < 5^{\circ}\sim 10^{\circ}$  & Non-collinear spin textures: \\
    & & coexisting FM/AFM domains \\
    \cline{2-3}
    & $ \theta > 10^{\circ} $  &  FM \\
     \hline
    Twisted bilayer magnonic crystals  &  $\theta = 6^{\circ}$ & Magic angle for spin-wave moir\'e edge  \\
    based on YIG \cite{Wangh2023,Chenj2022} & & and cavity modes \\
    \cline{2-3}
      & $\theta = 3.5^{\circ}$  & Magic angle for moir\'e magnonic flat band \\
     \hline
  \end{tabular}
\end{table*}

From the experimental point of view, the development of observing these novel physical properties of moir\'e spintronics and realization of their device integration are still in their infancy. First, although a vast number of model studies and theoretical predictions have been made in recent years, the twisted 2D magnets explored in experiments are limited to transition metal trihalides CrI$_3$. So far, the fabrication of atomically thin vdW magnets down to a single monolayer or only few layers has been widely reported, such as Cr$_2$Ge$_2$Te$_6$ \cite{Gongc2017}, FGT \cite{Dengy2018}, CrBr$_3$ \cite{Chenw2019}, CrCl$_3$ \cite{McGuire2017}, MnBi$_2$Te$_4$ \cite{Dengy2020}, VSe$_3$ \cite{Bonilla2018}, MnSe$_x$ \cite{OHara2018}, 1T-CrTe$_2$ \cite{Mengl2021}, CrTe$_3$ \cite{Yaoj2022}, CrSBr \cite{Telford2020}, GdAlSi \cite{Parfenov2024}, NiPS$_3$ \cite{Hul2023} and so on. As a result, this large set of building blocks with different crystal and magnetic structures will allow us to further explore moir\'e spintronics. For instance, we surprisingly noticed the twist engineering of two CrSBr ferromagnetic monolayers, accompanied by a multistep magnetization switching \cite{Constant2024}. Moreover, a lot of 2D magnetic materials synthesized in experiments thus far are not isolated monolayer but instead grown on a substrate, such as single- and few-layer VSe$_3$ sheets grown on highly oriented pyrolytic graphite (HOPG) or MoS$_2$ substrates by MBE \cite{Bonilla2018}. It is still uncertain whether the intrinsic properties of isolated magnets can withstand the impact of substrate environment. In addition to broaden material choice for moir\'e spintronics, plenty of other key issues of experimental techniques still have to be resolved for this important yet nascent field, including but not limited to strategies for crystal stability of stacking layered 2D magnetic materials after twisting, large-area moir\'e superlattice growth, in situ control of twist angles at extreme conditions for measuring their properties. Besides, it will be valuable to develop advanced scanning probe techniques with an ultrahigh spatial resolution that might be used to detect and manipulate magnetic states on very small length scales at least less than the moir\'e superlattice in presence of very large demagnetization fields. So far in recent experiments, the measurements of magnetic configuration are unable to directly identify the non-collinear spin textures in twisted 2D vdW magnets. Developing new techniques that can locally probe the distribution of magnetization will undoubtedly promote our understanding of moir\'e magnetism and moir\'e skyrmions. Additionally, as the moir\'e magnons have not been observed, another pressing problem is to develop more advanced technology to detect excitations of magnons in twisted vdW magnets. Furthermore, the moir\'e pattern strongly depends on various factors except for twist angles, such as stacking orders, either intrinsic or extrinsic strain. The vast majority of studies always assume that the lattice structure is rigid and the moir\'e pattern only changes with the twisting, ignoring the lattice relaxation. However, the assumption has been questioned in recent experiments, which have shown that the lattice reconstructions widely exist in twisted layers and has an essential impact on their electronic properties \cite{Park2025,Lih2021,Lie2021}. As a result, lattice reconstruction of the twisted vdW magnets and relaxation after the twisting engineering seriously affect the study of moir\'e spintronics.   

On the theoretical side, one critical challenge revolves around the origin of the magnetic exchange interactions in 2D magnetic materials. Therefore, the exact quantitative determination of intralayer or interlayer magnetic exchange interaction between a pair of atoms for a specific magnetic material using the first-principles methods remains challenging. Different from previous studies, a recent work presented in Ref. \cite{Sabani2025} demonstrated a systematic method to quantify all possible mechanisms contributed to magnetic exchange interaction between any two transition-metal atoms in a given magnetic material, which was applied to the monolayers CrI$_3$ and NiI$_2$ as examples. In addition, the lattice reconstruction or structural relaxation also should be included into the calculations as interactions show strongly dependence, which has been preliminary discussed in recent studies \cite{Yangb2023,Wangd2021,Tilak2023,Agarwal2024}. Furthermore, the combinations of theoretical predictions and experimental verification are insufficient.  

In addition to the ML side, while the synergistic paradigm promises to dramatically accelerate the pace of materials discovery and optimization, several significant challenges remain on the path to this future. The generation of high-quality DFT data required to train these sophisticated ML models is still a major computational bottleneck, even if the models themselves are fast at inference \cite{Li2023}. Capturing the physics of strong electronic correlations, which often go beyond the capabilities of standard DFT, will require training ML models on data from more advanced, and even more computationally expensive, quantum many-body methods. Finally, the experimental realization of these precisely engineered moir\'e structures demands exquisite control over twist angle, strain, and interface quality, which continues to be a frontier of materials synthesis and fabrication \cite{Tritsaris2021}. Addressing these outstanding issues will require a close integration of theoretical, computational, and experimental efforts. Such a synergistic approach is essential for advancing from the discovery of emergent phenomena to the rational design of quantum materials with precisely engineered spintronic properties.

Looking forward, moir\'e spintronics based on multilayers of 2D vdW magnetic materials with twist degree of freedom is a new and promising research field with multitude of possibly unexplored properties and phenomena. To further promote progress of this field, a short perspective of moir\'e spintronics that should be studied in the future is discussed below:
\begin{enumerate}[leftmargin=*]
		\renewcommand{\labelenumi}{(\theenumi)}
		\item \textbf{Twisted 2D vdW antiferromagnets.} In comparison with FMs, AFMs have advantages including no stray magnetic fields, intrinsic terahertz (THz) frequency range of spin dynamics, and lower energy consumption \cite{Baltz2018,Rahman2021,Baih2022}. Currently, most monolayer 2D magnetic materials are FMs. However, several monolayer AFM materials are predicted in recent years, such as CrTe$_3$ \cite{Yaoj2022}, MPX$_3$ (M = Mn, Fe, Ni; X= S, Se) \cite{Hul2023,Chuh2020,Gaoy2023,Mavani2025}, V$_2$Se$_2$O \cite{Mah2021}, Cr$_2$SeO \cite{Khan2025}, and Cr$_2$Se$_2$O \cite{Gongj2024}. Therefore, comprehending the properties of moir\'e spintronics based on 2D vdW AFM homostructures or FMs/AFMs heterostructures with twist degree of freedom is of utmost importance. Recently, a nonrelativistic spin-momentum coupling was predicted in twisted Bilayer AFMs, that a momentum-dependent spin splitting can be induced by the twisting as the same order of magnitude as that arising from SOC \cite{Her2023}.    
		\item \textbf{Hybridization between moir\'e magnons and other excitations.} The coupling between magnons and other excitations in 2D magnets could provide a versatile playground to design new functional materials for spintronic devices. For example, it has been demonstrated that magnon-phonon hybrid excitations (i.e., magnon-polarons) with topological bands and their thermal Hall effect exist extensively in 2D magnets \cite{Takahashi2016,Zhangx2019,Lius2021,Pawbake2022}, where the symmetry is broken by magnon-phonon interactions arising from DMI, magnetoelastic coupling, or dipolar coupling. Recent studies also report the chirality selective magnon-phonon hybridization \cite{Cuij2023} and topological magnon-polaron excitations \cite{Linz2024} in a layered vdW magnet. The formation of moir\'e superlattices not only redefines the magnonic excitations in twisted 2D vdW magnets, but can also remarkably modify the phononic dynamics defined as moir\'e phonons in twisted TMDs (e.g., MoS$_2$ homo-bilayers, WSe$_2$/MoSe$_2$ and WSe$_2$/WS$_2$ hetero-bilayers) \cite{Linm2018,Parzefall2021,Chuangh2022}. However, the interplay between moir\'e magnons and moir\'e phonons  as well as the formation of magnon-phonon hybrid excitations in a twisted magnetic system have yet to be explored. Besides magnon-polarons, the interaction between an exciton, the bosonic quasiparticle of electron–hole pairs bound by the Coulomb interaction, and a magnon has been experimentally studied in the layered magnetic semiconductor CrSBr \cite{Bae2022,Diederich2023} and MnPS$_3$ \cite{Wangz2023}. Moreover, the exciton-magnon coupling has been further reported in twisted 2D systems, such as twisted CrSBr bilayers \cite{Liuj2025}, and twisted vdW heterostructures composed of monolayer MoSe$_2$ and AFM MnPS$_3$ \cite{Zhangy2022}. As a result, twisted magnetic vdW materials provide a versatile platform for realizing and controlling the hybrid moir\'e excitations, presenting significant prospects for exploring novel quantum phenomena in spintronics. 
		\item \textbf{Moir\'e-driven multiferroics.} Multiferroic materials \cite{Eerenstein2006,Luc2015,Fiebig2016}, that exhibit simultaneous magnetization and electric polarization, have recently provided a promising route to realize non-volatile electric field control of magnetism for low-power and ultra-high density information storage. In recent years, plenty of 2D multiferroic materials have been theoretically and/or experimentally proposed \cite{Wangc2023}, such as CuCrP$_2$S$_6$ \cite{Qij2018,Tangw2023,Wangx2023}, NiI$_2$ \cite{Songq2022,Fumega2022,Amini2024}, VOX$_2$(X = Cl, Br, and I) \cite{Tant2019}, electron-doped CrBr$_3$ \cite{Huangc2018}, ReWCl$_6$ \cite{Xum2020}, Cr$_2$SeO \cite{Khan2025}, and so on. However, the emergence of multiferroic order driven by the moir\'e pattern has been predicted in twisted homobilayers composed of these 2D multiferroic materials \cite{Antao2024}, as well as twisted bilayer CrX$_3$ (X = Br, I, Cl) \cite{Fumega2023,Yeh2025}and LaBr$_2$ \cite{Sunw2022}, which is absent in monolayer or aligned multilayers. More interesting, the twisting engineering allows stabilizing topological moir\'e spin textures and an out-of-plane ferroelectric polarization in these moir\'e-driven multiferroics, in which the strong magnetoelectric coupling enables the manipulation of phase transition between diverse skyrmion lattice phases via an external electric field \cite{Antao2024}. In addition, recent studies have predicted that magnetic skyrmions in a FM monolayer can be controlled by the ferroelectric polarization of an adjacent 2D vdW ferroelectric material due to the broken inverse symmetry at the interface of the vdW heterostructure \cite{Huangk2022,Sunw2023}. Overall, these findings highlight the potential of moir\'e-driven multiferroics for electric-field control of topological moir\'e spin textures, which establishes a foundation for designing novel information memory devices.   
		\item \textbf{Altermagnets in twisted bilayer vdW magnets.} Recently, an emerging category of magnetism called altermagnetism \cite{Smejkal2022,Songc2025,Smejkal2022_2,Smejkal2022_3,Jungwirth2024,Jungwirth2025} has attracted considerable attention in spintronics, due to its novel physical properties and potential device applications. Beyond FMs and conventional AFMs, altermagnets (AMs) have compensated magnetic order with a zero net magnetization in real space, while its energy band dispersion exhibits the non-relativistic spin-splitting without requiring relativistic SOC in reciprocal space. As a result, AMs have combined features of both FMs and AFMs, including the absence of stray field, ultrafast spin dynamics, sizable GMR and TMR effect, anomalous Hall and Kerr effects, strong spin-polarized transport and spin torques \cite{Smejkal2022_4,Yuanl2020,Samanta2020,Hernandez2021,Cuiq2023,Gomonay2024}. Up to now, several bulk material candidates of AMs have been theoretically predicted and experimentally verified such as RuO$_2$, CrSb, and MnTe \cite{Bose2022,Fengz2022,Lees2024,Krempasky2024,Reimers2024}. Although there has also been theoretically proposed of 2D altermagnetic materials (e.g., CrO and V$_2$Se$_2$O) \cite{Guop2023,Chenx2023,Sodequist2024,Bhattarai2025}, experimentally confirmed remain scarce. Furthermore, recent studies suggest the significant advantage of realizing 2D AMs in twisted bilayers of magnetic materials \cite{Her2023,Panb2024,Liuy2024,Sheoran2024,Zengs2024,Guos2024}, which presents a promising research direction for future research on moir\'e spintronics.
		\item \textbf{Twisted vdW hetersotructures and twist-angle dependent proximity-induced spin-orbit physics.} Beyond intrinsic 2D magnets, the integration of non-magnetic 2D materials (e.g., graphene, TMDs) with strong SOC or proximity-induced magnetism enables high efficient spin-charge interconversion and creates spin-momentum-locked states crucial for spintronics \cite{Gmitra2015,Avsar2020,Wangz2015,Zollner2023,Zhangg2024,Trier2022}. Recent breakthroughs in vdW heterostructures \cite{Huangb2020,Roche2024} demonstrate strong room-temperature ferromagnetism \cite{Burch2018,Bonilla2018}, significant DMI \cite{Yangh2018}, Field-Free SOTs \cite{Vojavek2024,Masseroni2024}, and gate-tunable spin Hall effects \cite{Duenas2024}. Furthermore, moir\'e potentials modulate spin-orbit coupling in graphene/2D-magnet heterostructures, which reveals the tremendous potential for twist-angle engineering of SOC strength and symmetry in twisted vdW hetersotructures \cite{Songk2018,Pezo2021,Liy2019,Naimer2021}. As TMDs are promising candidates for generating SOTs with both high SOC and charge-to-spin conversion ratios, it can be predicted that twist angle dependent SOTs generates energy-efficient magnetization reversal and tunable moir\'e skyrmions in TMDs/2D-magnet heterostructures \cite{Wuy2020,Stiehl2019,Husain2020,Zhaoz2025}. Howver, comprehensive roadmaps outline the vast potential and challenges of vdW materials for spintronics, including spin transport and relaxation, spin-orbit physics, magnetic proximity effects, scalable MRAM designs and SOT device integration \cite{Sierra2021,Yangh2022,Kurebayashi2022,Gish2024}. This burgeoning field of vdW spintronics provides versatile pathways for rapid expansion of moir\'e spintronics.
		\item \textbf{Role of defects and disorder.} While the current literature emphasizes idealized twist-angle engineering and pristine moir\'e superlattices, a significant disconnect remains between these theoretical models and the messy reality of experimental devices. These imperfections, ranging from atomic vacancies and adatoms to mesoscopic twist-angle inhomogeneity, crucially affect the robustness of spin transport, the stability of magnetic textures, and the coherence of magnon propagation. For example, disorder landscapes can act as pinning centers for moir\'e skyrmions, drastically altering their current-driven motions, or induce Anderson localization in flat bands, suppressing its unique transport phenomena one seeks to exploit. Addressing the role of defects and disorder in moir\'e spintronics is critical for the future development of the field. However, standard \textit{ab initio} methods, limited by cubic scaling ($O(N^3)$), cannot access the micrometer-scale dimensions required to statistically sample realistic disorder profiles. To bridge this gap, it is necessary to adopt linear-scaling ($O(N)$) quantum transport methodologies, as comprehensively reviewed by Ref. \cite{Fan2021}. These real-space numerical techniques enable the simulation of quantum transport in systems containing millions of atoms. They are uniquely capable of quantifying the impact of defects on spin diffusion lengths and topological invariants in device-sized systems. By transitioning from small, clean unit cells to massive, disordered supercells, these approaches provide the necessary theoretical framework to guide the engineering of disorder-tolerant moir\'e spintronics.
		\item \textbf{Closed-loop discovery platform for moir\'e spintronics.} The ultimate vision for accelerating progress is the establishment of a "closed-loop" autonomous discovery platform that seamlessly integrates AI, simulation, and eventually, experimental synthesis and characterization \cite{Tritsaris2021}. Such a platform would operate cyclically: (\romannumeral1) AI-driven hypothesis. An autonomous AI agent proposes a novel moir\'e structure designed to exhibit a target spintronic property, such as a large topological Hall effect. (\romannumeral2) Automated data generation via AI agents. To validate hypotheses and generate training data, researchers are increasingly adopting agentic workflows powered by Large Language Models (LLMs). A critical enabler in this space is the Model Context Protocol (MCP), a standardized interface that allows AI agents to directly interact with local or remote scientific software \cite{Li2025_survey}. In this framework, an MCP server wraps complex simulation codes (e.g., LAMMPS for molecular dynamics or VASP for DFT), allowing the AI agent to autonomously execute simulations, parse outputs, and error-check results. This automated pipeline efficiently produces the high-fidelity datasets required to train and refine surrogate models like \texttt{DPmoire} and \texttt{xDeepH}. (\romannumeral3) ML-accelerated simulation. Once trained on this agent-generated data, ML models like \texttt{DPmoire} and \texttt{xDeepH} rapidly perform virtual experiments, predicting the relaxed atomic structure and detailed spintronic properties of the proposed material, thereby validating the hypothesis \textit{in silico} \cite{Liu2024,Li2023}. (\romannumeral4) Robotic synthesis and characterization. Upon successful virtual screening, an automated robotic system could fabricate the vdW heterostructure with the specified twist angle. Automated probes would then characterize the real sample. (\romannumeral5) Iterative refinement. The AI agent refines its internal models based on the experimental feedback, and the discovery loop repeats.  	
\end{enumerate}

\renewcommand{\addcontentsline}[3]{}
\section*{Acknowledgements} 
F.Z. and H.Y. acknowledge the National Natural Science Foundation of China (Grant No. T2495212 and No. 12174405) and the supports from the National Key Research and Development Program of China (Grants No. 2022YFA1405100). Z.C. thanks Australia Research Council (Grant No. DP210101436) for support. K.C. acknowledges the support from the Strategic Priority Research Program of the Chinese Academy of Sciences (Grants No. XDB28000000 and No. XDB0460000), the NSFC under Grants No. 92265203 and No. 12488101, and the Innovation Program for Quantum Science and Technology under Grant No. 2024ZD0300104.

\section*{AUTHOR DECLARATIONS} 
\subsection*{Conflict of Interest}
The authors have no conflicts to disclose.

\subsection*{Author Contributions} 
Fengjun Zhuo and Zhenyu Dai contributed equally to this work.

\noindent \textbf{Fengjun Zhuo:} Conceptualization (lead); Visualization (lead); Writing - original draft (lead); Writing - review \& editing (lead). \textbf{Zhenyu Dai:} Conceptualization (equal); Visualization (equal); Writing - original draft (equal); Writing - review \& editing (equal). \textbf{Kai Chang:} Supervision (equal); Funding acquisition (equal); Project administration (supporting); Writing - review \& editing (supporting). \textbf{Hongxin Yang:} Funding acquisition (lead); Resources (equal); Supervision (equal); Project administration (equal); Writing – original draft (supporting); Writing - review \& editing (supporting). \textbf{Zhenxiang Cheng:} Supervision (lead); Project administration (lead); Resources (lead); Funding acquisition (equal); Conceptualization (equal); Writing – original draft (supporting); Writing - review \& editing (supporting).

\section*{DATA AVAILABILITY} 
Data sharing is not applicable to this article as no new data were created or analyzed in this study.


\begin{thebibliography}{999}%
\bibitem{Baibich1988} M. N. Baibich, J. M. Broto, A. Fert, F. Nguyen Van Dau, F. Petroff, P. Etienne, G. Creuzet, A. Friederich, and J. Chazelas, {\it Phys. Rev. Lett.} {\bf 61}, 2472 (1988).

\bibitem{Binasch1989} G. Binasch, P. Gr\"unberg, F. Saurenbach, and W. Zinn, {\it Phys. Rev. B} {\bf 39}, 4282 (1989).

\bibitem{Prinz1998} G.A. Prinz, {\it Science 282} {\bf 39}, 1660 (1998).

\bibitem{Wolf2001} S. A. Wolf, D. D. Awschalom, R. A. Buhrman, J. M. Daughton, S. von Moln\'er, M. L. Roukes, A. Y. Chtchelkanova, and D. M. Treger, {\it Science} {\bf 294}, 1488 (2001).

\bibitem{Slonczewski1996} J. Slonczewski, {\it J. Magn. Magn. Mater.} {\bf 159}, L1 (1996).

\bibitem{Berger1996} L. Berger, {\it Phys. Rev. B} {\bf 54}, 9353 (1996).

\bibitem{Manchon2015} A. Manchon, H. C. Koo, J. Nitta, S. M. Frolov, and R. A. Duine, {\it Nat. Mater.} {\bf 14}, 871 (2015).

\bibitem{Manchon2019} A. Manchon, J. \v{Z}elezn\'y, I. M. Miron, T. Jungwirth, J. Sinova, A. Thiaville, K. Garello, and P. Gambardella, {\it Rev. Mod. Phys.} {\bf 91}, 035004 (2019).

\bibitem{Dzyaloshinskii1958} I. Dzyaloshinskii, {\it J. Phys. Chem. Solids} {\bf 4}, 241 (1958).

\bibitem{Moriya1960} T. Moriya, {\it Phys. Rev. Lett.} {\bf 4}, 228 (1960).

\bibitem{Yangh2023} H. Yang, J. Liang, and Q. Cui, {\it Nat. Rev. Phys.} {\bf 5}, 43 (2023).

\bibitem{Bogdanov2001} A. N. Bogdanov and U. K. R\"{o}\ss{}ler, {\it Phys. Rev. Lett.} {\bf 87}, 037203 (2001).

\bibitem{Robler2006} U. K. R\"{o}\ss{}ler, A. N. Bogdanov, and C. Pfleiderer, {\it Nature} {\bf 442}, 797 (2006).

\bibitem{Muhlbauer2009} S. MR\"{u}hlbauer, B. Binz, F. Jonietz, C. Pfleiderer, A. Rosch, A. Neubauer, R. Georgii, P. B\"{o}n, and C. Pfleiderer, {\it Science} {\bf 323}, 915 (2009).

\bibitem{Yu2010} X. Z. Yu, Y. Onose, N. Kanazawa, J. H. Park, J. H. Han, Y. Matsui, N. Nagaosa, and Y. Tokura, {\it Nature} {\bf 465}, 901 (2010).

\bibitem{Yu2018} X. Z. Yu, W. Koshibae, Y. Tokunaga, K. Shibata, Y. Taguchi, N. Nagaosa, and Y. Tokura, {\it Nature} {\bf 564}, 95 (2018).

\bibitem{Jani2021} H. Jani, J.-C. Lin, J. Chen, J. Harrison, F. Maccherozzi, J. Schad, S. Prakash, C.-B. Eom, A. Ariando, T. Venkatesan, and P. G. Radaelli, {\it Nature} {\bf 590}, 74 (2021).

\bibitem{Zhuof2021} F. Zhuo, H. Li, and A. Manchon, {\it Phys. Rev. B} {\bf 104}, 144422 (2021).

\bibitem{Zhuof2022} F. Zhuo, H. Li, and A. Manchon, {\it New J. Phys.} {\bf 24}, 023033 (2022).

\bibitem{Zhuof2024} F. Zhuo, J. Kang, Z. Cheng, and A. Manchon, {\it Phys. Rev. B} {\bf 109}, 054412 (2024).

\bibitem{Dyakonov1971} M. I. Dyakonov and V. I. Perel,  {\it Phys. Lett. A} {\bf 35}, 459 (1971).

\bibitem{Hirsch1999} J. E. Hirsch, {\it Phys. Rev. Lett.} {\bf 83}, 1834 (1999).

\bibitem{Sinova2015} J. Sinova, S. O. Valenzuela, J. Wunderlich, C. H. Back, and T. Jungwirth, {\it Rev. Mod. Phys.} {\bf 87}, 1213 (2015).

\bibitem{Shindou2001} R. Shindou and N. Nagaosa, {\it Phys. Rev. Lett.} {\bf 87}, 116801 (2001).

\bibitem{Tokura2017} Y. Tokura,  M. Kawasaki, and  N. Nagaosa, {\it Nat. Phys.} {\bf 13}, 10561068 (2017).

\bibitem{Onose2010} Y. Onose, T. Ideue, H. Katsura, Y. Shiomi, N. Nagaosa, and Y. Tokura, {\it Science} {\bf 329}, 297 (2010).

\bibitem{Hirschberger2015} M. Hirschberger, R. Chisnell, Y. S. Lee, and N. P. Ong, {\it Phys. Rev. Lett.} {\bf 115}, 106603 (2015).

\bibitem{Zutic2004} I. \v{Z}uti\'{c}, J. Fabian, and S. D. Sarma, {\it Rev. Mod. Phys.} {\bf 76}, 323 (2004).

\bibitem{Chappert2007} C. Chappert, A. Fert, and F. Van Dau, {\it Nat. Mater.} {\bf 6}, 813 (2007).

\bibitem{Jungwirth2016} T. Jungwirth, X. Marti, P. Wadley, and J. Wunderlich, {\it Nat. Nanotechnol.} {\bf 11}, 231 (2016).

\bibitem{Baltz2018} V. Baltz, A. Manchon, M. Tsoi, T. Moriyama, T. Ono, and Y. Tserkovnyak, {\it Rev. Mod. Phys.} {\bf 90}, 015005 (2018).

\bibitem{Manipatruni2018} S. Manipatruni, D. E. Nikonov, and I. A. Young, {\it  Nat. Phys.} {\bf 14}, 338 (2018).

\bibitem{Lin2019} X. Lin, W. Yang, K. L. Wang, and W. Zhao, {\it Nat. Electron.} {\bf 2}, 274 (2019).

\bibitem{Fert2008} A. Fert, {\it Rev. Mod. Phys.} {\bf 1517}, 1517 (2008).

\bibitem{Mao2006} S. Mao, Y. Chen, F. Liu, X. Chen, B. Xu, P. Lu, M. Patwari, H. Xi, C. Chang, and B. Miller, {\it IEEE Trans. Magn.} {\bf 42}, 97 (2006). 

\bibitem{Dave2006} R. W. Dave, G. Steiner, J. Slaughter, J. Sun, B. Craigo, S. Pietambaram, K. Smith, G. Grynkewich, M. DeHerrera, and J. Akerman, {\it IEEE Trans. Magn.} {\bf 42}, 1935 (2006).

\bibitem{Fan2015} D. Fan, Y. Shim, A. Raghunathan, and K. Roy, {\it IEEE Trans. Nanotechnol.} {\bf 14}, 1013 (2015).

\bibitem{Zahedinejad2020} M. Zahedinejad, A. A. Awad, S. Muralidhar, R. Khymyn, H. Fulara, H. Mazraati, M. Dvornik, and J. \AA kerman, {\it Nat. Nanotechnol.} {\bf 15}, 47 (2020).

\bibitem{Kurenkov2019} A. Kurenkov, S. DuttaGupta, C. Zhang, S. Fukami, Y. Horio, and H. Ohno, {\it Adv. Mater.} {\bf 31}, 1900636 (2019).

\bibitem{Novoselov2004} K. S. Novoselov, A. K. Geim, S. V. Morozov, D. Jiang, Y. Zhang, S. V. Dubonos, I. V. Grigorieva, and A. A. Firsov, {\it Science} {\bf 306}, 666 (2004). 

\bibitem{Cheny2018} Y. Chen, Z. Fan, Z. Zhang, W. Niu, C. Li, N. Yang, B. Chen, and H. Zhang, {\it Chem. Rev.} {\bf 118}, 6409 (2018).

\bibitem{Sarkar2020} A. S. Sarkar and E. Stratakis, {\it Adv. Sci.} {\bf 7}, 2001655 (2020).

\bibitem{Wangt2021} T. Wang, M. Park, Q. Yu, J. Zhang, and Y. Yang, {\it Mater. Today Adv.} {\bf 8}, 100092 (2020).

\bibitem{Zhaob2021} B. Zhao, D. Y. Shen, Z. C. Zhang, P. Lu, M. Hossain, J. Li, B. Li, and X. D. Duan, {\it Adv. Funct. Mater.} {\bf 31}, 2105132 (2021).

\bibitem{Zhaos2021} S. Zhao, J. Zhang, and L. Fu, {\it Adv. Mater.} {\bf 33}, 2005544 (2021).

\bibitem{Zhouk2023} K. Zhou, G. Shang, H.-H. Hsu, S.-T. Han, V. A. L. Roy, and Y. Zhou, {\it Adv. Mater.} {\bf 35}, 2207774 (2023).

\bibitem{Koul2017} L. Kou, Y. Ma, Z. Sun, T. Heine, and C. Chen, {\it J. Phys. Chem. Lett.} {\bf 8}, 1905 (2017).

\bibitem{Rachel2018} S. Rachel, {\it Rep. Prog. Phys.} {\bf 81}, 116501 (2018).

\bibitem{Krishnamoorthy2023} H. N. S. Krishnamoorthy, A. M. Dubrovkin, G. Adamo, and C. Soci, {\it Chem. Rev.} {\bf 8}, 4416 (2023).

\bibitem{Maj2019} J. Ma, K. Deng, L. Zheng, S. Wu, Z. Liu, S. Zhou, and Dong Sun, {\it 2D Mater.} {\bf 6}, 032001 (2019).

\bibitem{Lix2025} X. Li, K. Liu, D. Wu, P. Lin, Z. Shi, X. Li, L. Zeng, Y. Chai, S. P. Lau, and Y. H. Tsang, {\it Adv. Mater.} {\bf } 2415717 (2025).

\bibitem{Yuh2025} H. Yu, H. Zeng, Y. Zhang, Y. Liu, W. ShangGuan, X. Zhang, Z. Zhang, and Y. Zhang, {\it Adv. Funct. Mater.} {\bf 35}, 2412913 (2025).

\bibitem{Qiu2021} D. Qiu, C. Gong, S. Wang, M. Zhang, C. Yang, X. Wang, and J. Xiong, {\it Adv. Mater.} {\bf 33}, 2006124 (2021).

\bibitem{Maggiora2024} J. Maggiora, X. Wang, and R. Zheng, {\it Phys. Rep.} {\bf 1076}, 1 (2024).

\bibitem{Gibertini2019} M. Gibertini, M. Koperski, A. F. Morpurgo, and K. S. Novoselov, {\it Nat. Nanotechnol.} {\bf 14}, 408 (2019).

\bibitem{Mak2019} K. F. Mak, J. Shan, and D. C. Ralph, {\it Nat. Rev. Phys.} {\bf 1}, 646 (2019).

\bibitem{Jiang2021} X. Jiang, Q. Liu, J. Xing, N. Liu, Y. Guo, Z. Liu, and J. Zhao, {\it Appl. Phys. Rev.} {\bf 8}, 031305 (2021).

\bibitem{Pesin2012} D. Pesin and A. H. MacDonald, {\it Nat. Mater.} {\bf 11}, 409 (2012).

\bibitem{Fiori2014} G. Fiori, F. Bonaccorso, G. Iannaccone, T. Palacios, D. Neumaier, A. Seabaugh, S. K. Banerjee, and Luigi Colombo, {\it Nat. Photon.} {\bf 8}, 899 (2014).

\bibitem{Chhowalla2016} M. Chhowalla, D. Jena, and H. Zhang, {\it Nat. Rev. Mater.} {\bf 1}, 16052 (2016).

\bibitem{Balandin2011} A. A. Balandin, {\it Nat. Mater.} {\bf 10}, 569 (2011).

\bibitem{Xiaf2014} F. Xia, H. Wang, D. Xiao, M. Dubey, and A. Ramasubramaniam, {\it Nat. Photon.} {\bf 8}, 899 (2014).

\bibitem{Mannix2017} A. J. Mannix, B. Kiraly, M. C. Hersam, and N. P. Guisinger, {\it Nat. Rev. Chem.} {\bf 1}, 0014 (2017)

\bibitem{Dongr2018} R. Dong, T. Zhang, and X. Feng, {\it Chem. Rev.} {\bf 118}, 6189 (2018).

\bibitem{Caix2018} X. Cai, Y. Luo, B. Liu, and H. Cheng , {\it 	Chem. Soc. Rev.} {\bf 47}, 6224 (2018).

\bibitem{Xiaox2018} X. Xiao, H. Wang, P. Urbankowski, and Y. Gogotsi, {\it Chem. Soc. Rev.} {\bf 47}, 8744 (2018).

\bibitem{Geim2007} A. K. Geim, and K. S. Novoselov, {\it Nat. Mater.} {\bf 6}, 183 (2007). 

\bibitem{Geim2009} A. K. Geim, {\it Science} {\bf 324}, 1530 (2009).

\bibitem{Liuh2015} H. Liu, Y. Du, Y. Denga, and P. D. Ye, {\it Chem. Soc. Rev.} {\bf 44}, 2732 (2015).

\bibitem{Gusmao2017} R. Gusm\~{a}o, Z. Sofer, and M. Pumera, {\it Angew. Chem. Int. Ed.} {\bf 56}, 8052 (2017).

\bibitem{Abate2018} Y. Abate, D. Akinwande, S. Gamage, H. Wang, M. Snure, N. Poudel, and S. B. Cronin, {\it Adv. Mater.} {\bf 30}, 1704749 (2018).

\bibitem{Wengq2016} Q. Weng, X. Wang, X. Wang, Y. Bando, and D. Golberg, {\it Chem. Soc. Rev.} {\bf 45}, 3989 (2016).

\bibitem{Zhangk2017} K. Zhang, Y. Feng, F. Wang, Z. Yanga, and J. Wang, {\it J. Mater. Chem. C} {\bf 5}, 11992 (2017).

\bibitem{Caldwell2019} J. D. Caldwell, I. Aharonovich, G. Cassabois, J. H. Edgar, B. Gil, and D. N. Basov, {\it Nat. Rev. Mater.} {\bf 4}, 552 (2019).

\bibitem{Roy2021} S. Roy, X. Zhang, A. B. Puthirath, A. Meiyazhagan, S. Bhattacharyya, M. M. Rahman, G. Babu, S. Susarla, S. K. Saju, M. K. Tran, L. M. Sassi, M. A. S. R. Saadi, J. Lai, O. Sahin, S. M. Sajadi, B. Dharmarajan, D. Salpekar, N. Chakingal, A. Baburaj, X. Shuai, A. Adumbumkulath, K. A. Miller, J. M. Gayle, A. Ajnsztajn, T. Prasankumar, V. V. J. Harikrishnan, V. Ojha, H. Kannan, A. Z. Khater, Z. Zhu, S. A. Iyengar, P. A. d. S. Autreto, E. F. Oliveira, G. Gao, A. G. Birdwell, M. R. Neupane, T. G. Ivanov, J. Taha-Tijerina, R. M. Yadav, S. Arepalli, R. Vajtai, and P. M. Ajayan, {\it Adv. Mater.} {\bf 33}, 2101589 (2021).

\bibitem{Wangq2012} Q. Wang, K. Kalantar-Zadeh, A. Kis, J. N. Coleman, and M. S. Strano, {\it Nat. Nanotechnol.} {\bf 7}, 699 (2012).

\bibitem{Voiry2015} D. Voiry, A. Mohiteb, and M. Chhowalla, {\it Chem. Soc. Rev.} {\bf 44}, 2702 (2015).

\bibitem{Manzeli2017} S. Manzeli, D. Ovchinnikov, D. Pasquier, O. V. Yazyev, and A. Kis, {\it Nat. Rev. Mater.} {\bf 2}, 17033 (2017).

\bibitem{Choi2017} W. Choi, N. Choudhary, G. H. Han, J. Park, D. Akinwande, and Y. H. Lee, {\it Mater. Today} {\bf 20}, 116 (2017).

\bibitem{Chowdhury2020} T. Chowdhury, E. C. Sadler, and T. J. Kempa, {\it Chem. Rev.} {\bf 120}, 12563 (2020).

\bibitem{Naguib2014} M. Naguib, V. N. Mochalin, M. W. Barsoum, and Y. Gogotsi, {\it Adv. Mater.} {\bf 26}, 992 (2014).

\bibitem{VahidMohammadi} A. VahidMohammadi, J. Rosen, and Y. Gogotsi, {\it Science} {\bf 372}, eabf1581 (2021).

\bibitem{Weip2021}Y. Wei, P. Zhang, R. A. Soomro, Q. Zhu, and B. Xu, {\it Adv. Mater.} {\bf 33}, 2103148 (2021).

\bibitem{Geim2013} A. K. Geim and I. V. Grigorieva, {\it Nature} {\bf 499}, 419 (2013).

\bibitem{Novoselov2016} K. S. Novoselov, A. Mishchenko, A. Carvalho, and A. H. Castro Neto, {\it Science} {\bf 353}, aac9439 (2016).

\bibitem{Liuy2016} Y. Liu, N. O. Weiss, X. Duan, H.-C. Cheng, Y. Huang, and X. Duan, {\it Nat. Rev. Mater.} {\bf 1}, 16042 (2016).

\bibitem{Haigh2012} S. J. Haigh, A. Gholinia, R. Jalil, S. Romani, L. Britnell, D. C. Elias, K. S. Novoselov, L. A. Ponomarenko, A. K. Geim, and R. Gorbachev , {\it Nat. Mater.} {\bf 11}, 764 (2012).

\bibitem{Butler2013} S. Z. Butler, S. M. Hollen, L. Cao, Y. Cui, J. A. Gupta, H. R. Guti\'{e}rrez, T. F. Heinz, S. S. Hong, J. Huang, A. F. Ismach, E. Johnston-Halperin, M. Kuno, V. V. Plashnitsa, R. D. Robinson, R. S. Ruoff, S. Salahuddin, J. Shan, L. Shi, M. G. Spencer, M. Terrones, W. Windl, and J. E. Goldberger, {\it ACS Nano} {\bf 7}, 2898 (2013).

\bibitem{Ponomarenko2011} L. A. Ponomarenko, A. K. Geim, A. A. Zhukov, R. Jalil, S. V. Morozov, K. S. Novoselov, I. V. Grigorieva, E. H. Hill, V. V. Cheianov, V. I. Fal’ko, K. Watanabe, T. Taniguchi, and R. V. Gorbachev, {\it Nat. Phys.} {\bf 7}, 958 (2011).

\bibitem{Georgiou2013} T. Georgiou, R. Jalil, B. D. Belle, L. Britnell, R. V. Gorbachev, S. V. Morozov, Y.-J. Kim, A. Gholinia, S. J. Haigh, O. Makarovsky, L. Eaves, L. A. Ponomarenko, A. K. Geim, K. S. Novoselov, and A. Mishchenko, {\it Nat. Nanotechnol.} {\bf 8}, 100 (2013).

\bibitem{Wangj2024} J. Wang, L. He, Y. Zhang, H. Nong, S. Li, Q. Wu, J. Tan, and B. Liu, {\it Adv. Mater.} {\bf 36}, 2314145 (2024).

\bibitem{Burch2018} K. S. Burch, D. Mandrus, and J.-G. Park, {\it Nature} {\bf 563}, 47 (2018).

\bibitem{Gong2019} C. Gong, and X. Zhang, {\it Science} {\bf 363}, eaav4450 (2019).

\bibitem{Lih2019} H. Li, S. Ruan, and Y.-J. Zeng, {\it Adv. Mater.} {\bf 31}, 1900065 (2019). 

\bibitem{Cortie2020} D. L. Cortie, G. L. Causer, K. C. Rule, H. Fritzsche, W. Kreuzpaintner, and F. Klose, {\it Adv. Funct. Mater.} {\bf 30}, 1901414 (2020). 

\bibitem{Wangq2022} Q. H. Wang, A. Bedoya-Pinto, M. Blei, A. H. Dismukes, A. Hamo, S. Jenkins, M. Koperski, Y. Liu, Q.-C. Sun, E. J. Telford, H. H. Kim, M. Augustin, U. Vool, J.-X. Yin, L. H. Li, A. Falin, C. R. Dean, F. Casanova, R. F. L. Evans, M. Chshiev, A. Mishchenko, C. Petrovic, R. He, L. Zhao, A. W. Tsen, B. D. Gerardot, M. Brotons-Gisbert, Z. Guguchia, X. Roy, S. Tongay, Z. Wang, M. Z. Hasan, J. Wrachtrup, A. Yacoby, A. Fert, S. Parkin, K. S. Novoselov, P. Dai, L. Balicas, and E. J. G. Santos, {\it ACS Nano} {\bf 16}, 6960 (2022). 

\bibitem{Gmitra2015} M. Gmitra and J. Fabian, {\it Phys. Rev. B} {\bf 92}, 155403 (2015).

\bibitem{Hanw2014} W. Han, R. K. Kawakami, M. Gmitra, and J. Fabian, {\it Nat. Nanotechnol.} {\bf 9}, 794 (2014). 

\bibitem{Guguchia2018} Z. Guguchia, A. Kerelsky, D. Edelberg, S. Banerjee, F. von Rohr, D. Scullion, M. Augustin, M. Scully, D. A. Rhodes, Z. Shermadini, H. Luetkens, A. Shengelaya, C. Baines, E. Morenzoni, A. Amato, J. C. Hone, R. Khasanov, S. J. L. Billinge, E. Santos, A. N. Pasupathy, and Y. J. Uemura, {\it Sci. Adv.} {\bf 4}, eaat3672 (2018). 

\bibitem{Castro2008} E. V. Castro, N. M. R. Peres, T. Stauber, and N. A. P. Silva, {\it Phys. Rev. Lett.} {\bf 100}, 186803  (2008).

\bibitem{Krasheninnikov2009} A. V. Krasheninnikov, P. O. Lehtinen, A. S. Foster, P. Pyykk\"o, and R. M. Nieminen, {\it Phys. Rev. Lett.} {\bf 102}, 126807 (2009).

\bibitem{Yangh2011} H. X. Yang, M. Chshiev, D. W. Boukhvalov, X. Waintal, and S. Roche, {\it Phys. Rev. B} {\bf 84}, 214404 (2011).

\bibitem{Yangh2013} H. X. Yang, A. Hallal, D. Terrade, X. Waintal, S. Roche, and M. Chshiev, {\it Phys. Rev. Lett.} {\bf 110}, 046603 (2013).

\bibitem{Caot2015} T. Cao, Z. Li, and S. G. Louie, {\it Phys. Rev. Lett.} {\bf 114}, 236602 (2015).
 
\bibitem{Huangb2017} B. Huang, G. Clark, E. Navarro-Moratalla, D. R. Klein, R. Cheng, K. L. Seyler, D. Zhong, E. Schmidgall, M. A. McGuire, D. H. Cobden, W. Yao, D. Xiao, P. Jarillo-Herrero, and Xiaodong Xu, {\it Nature} {\bf 546}, 270 (2017).

\bibitem{Gongc2017} C. Gong, L. Li, Z. Li, H. Ji, A. Stern, Y. Xia, T. Cao, W. Bao, C. Wang, Y. Wang, Z. Q. Qiu, R. J. Cava, S. G. Louie, J. Xia, and X. Zhang, {\it Nature} {\bf 546}, 265 (2017).

\bibitem{Mermin1966} N. D. Mermin and H. Wagner, {\it Phys. Rev. Lett.} {\bf 17}, 1133 (1966).

\bibitem{Halperin2019} B. I. Halperin, {\it J. Stat. Phys.} {\bf 175}, 521 (2019).

\bibitem{Jenkins2022} S. Jenkins, L. R\'{o}zsa, U. Atxitia, R. F. L. Evans, K. S. Novoselov, and E. J. G. Santos, {\it Nat. Commun.} {\bf 13}, 6917 (2022). 

\bibitem{Yangh2018} H. X. Yang, G. Chen, A. A. C. Cotta, A. T. N’Diaye, S. A. Nikolaev, E. A. Soares, W. A. A. Macedo, K. Liu, A. K. Schmid, A. Fert, and M. Chshiev, {\it Nat. Mater.} {\bf 17}, 605 (2018). 

\bibitem{Hellman2017} F. Hellman, A. Hoffmann, Y. Tserkovnyak, G. S. D. Beach, E. E. Fullerton, C. Leighton, A. H. MacDonald, D. C. Ralph, D. A. Arena, H. A. Dürr, P. Fischer, J. Grollier, J. P. Heremans, T. Jungwirth, A. V. Kimel, B. Koopmans, I. N. Krivorotov, S. J. May, A. K. Petford-Long, J. M. Rondinelli, N. Samarth, I. K. Schuller, A. N. Slavin, M. D. Stiles, O. Tchernyshyov, A. Thiaville, and B. L. Zink, {\it Rev. Mod. Phys.} {\bf 89}, 025006 (2017).

\bibitem{Kuepferling2023} M. Kuepferling, A. Casiraghi, G. Soares, G. Durin, F. Garcia-Sanchez, L. Chen, C. H. Back, C. H. Marrows, S. Tacchi, and G. Carlotti, {\it Rev. Mod. Phys.} {\bf 95}, 015003 (2023).

\bibitem{Yangh2025} H. Yang, M. Gobbi, F. Herling, V. T. Pham, F. Calavalle, B. Mart\'{i}n-Garc\'{i}a, A. Fert, L. E. Hueso, and F. Casanova, {\it Nat. Electron.} {\bf 8}, 15 (2025). 

\bibitem{Amiri2012} P. K. Amiri and K. L. Wang , {\it SPIN} {\bf 2}, 1240002 (2012).

\bibitem{Lix2017} X. Li, K. Fitzell, D. Wu, C. T. Karaba, A. Buditama, G. Yu, K. L. Wong, N. Altieri, C. Grezes, N. Kioussis, S. Tolbert, Z. Zhang, J. P. Chang, P. K. Amiri, K. L. Wang, {\it Appl. Phys. Lett.} {\bf 110}, 052401 (2017).

\bibitem{Dattas2012} S. Datta, S. Salahuddin, and B. Behin-Aein, {\it Appl. Phys. Lett.} {\bf 101}, 252411 (2012).

\bibitem{Weid2018} D. S. Wei, T. van der Sar, S. H. Lee, K. Watanabe, T. Taniguchi, Be. I. Halperin, and A. Yacoby, {\it Science} {\bf 362}, 229 (2018). 

\bibitem{Hanj2019} J. Han, P. Zhang, J. T. Hou, S. A. Siddiqui, and L. Liu, {\it Science} {\bf 366}, 1121 (2019).

\bibitem{Lum2015} J. M. Lu, O. Zheliuk, I. Leermakers, N. F. Q. Yuan, U. Zeitler, K. T. Law, and J. T. Ye, {\it Science} {\bf 350}, 1353 (2015).

\bibitem{Morales2019} A. Palacio-Morales, E. Mascot, S. Cocklin, H. Kim, S. Rachel, D. K. Morr, and R. Wiesendanger, {\it Sci. Adv.} {\bf 5}, eaav6600 (2019).

\bibitem{Saito2017} Y. Saito, T. Nojima, and Y. Iwasa, {\it Nat. Rev. Mater.} {\bf 2}, 16094 (2017).

\bibitem{Stier2018} A. V. Stier, N. P. Wilson, K. A. Velizhanin, J. Kono, X. Xu, and S. A. Crooker, {\it Phys. Rev. Lett.} {\bf }, 057405 (2018). 

\bibitem{Bae2022} Y. J. Bae, J. Wang, A. Scheie, J. Xu, D. G. Chica, G. M. Diederich, J. Cenker, M. E. Ziebel, Y. Bai, H. Ren, C. R. Dean, M. Delor, X. Xu, X. Roy, A. D. Kent, and X. Zhu, {\it Nature} {\bf 609}, 282 (2022). 

\bibitem{Xux2014} X. Xu, W. Yao, D. Xiao, and T. F. Heinz, {\it Nat. Phys.} {\bf 10}, 343 (2014).

\bibitem{Mak2018} K. F. Mak, D. Xiao, and J. Shan, {\it Nat. Photon.} {\bf 12}, 451 (2018).  

\bibitem{Liuj2023} J. Liu and T. Hesjedal, {\it Adv. Mater.} {\bf 35}, 2102427 (2023). 

\bibitem{Xingw2019} W. Xing, L. Qiu, X. Wang, Y. Yao, Y. Ma, R. Cai, S. Jia, X. C. Xie, and W. Han, {\it Phys. Rev. X} {\bf 9}, 011026 (2019). 

\bibitem{Jiangs2020} S. Jiang, H. Xie, J. Shan, and K. F. Mak, {\it Nat. Mater.} {\bf 19}, 1295 (2020). 

\bibitem{Rahman2021} S. Rahman, J. F. Torres, A. R. Khan, and Y. Lu, {\it ACS Nano} {\bf 15}, 17175 (2021).

\bibitem{Smejkal2022} L. \v{S}mejkal, J. Sinova, and T. Jungwirth, {\it Phys. Rev. X} {\bf 12}, 040501 (2022).

\bibitem{Songc2025} C. Song, H. Bai, Z. Zhou, L. Han, H. Reichlova, J. H. Dil, J. Liu, X. Chen, and F. Pan, {\it Nat. Rev. Mater.} {\bf 10}, 473 (2025).

\bibitem{Banerjee2016} A. Banerjee, C. A. Bridges, J.-Q. Yan, A. A. Aczel, L. Li, M. B. Stone, G. E. Granroth, M. D. Lumsden, Y. Yiu, J. Knolle, S. Bhattacharjee, D. L. Kovrizhin, R. Moessner, D. A. Tennant, D. G. Mandrus, and S. E. Nagler, {\it Nat. Mater.} {\bf 15}, 733 (2016).

\bibitem{Banerjee2017} A. Banerjee, J. Yan, J. Knolle, C. A. Bridges, M. B. Stone, M. D. Lumsden, D. G. Mandrus, D. A. Tennant, R. Moessner, and S. E. Nagler, {\it Science} {\bf 356}, 1055 (2017). 

\bibitem{Takagi2019} H. Takagi, T. Takayama, G. Jackeli, G. Khaliullin, and S. E. Nagler, {\it Nat. Rev. Phys.} {\bf 1}, 264 (2019). 

\bibitem{Zhangh2022} H. Zhang, D. Raftrey, Y.-T. Chan, Y.-T. Shao, R. Chen, X. Chen, X. Huang, J. T. Reichanadter, K. Dong, S. Susarla, L. Caretta, Z. Chen, J. Yao, P. Fischer, J. B. Neaton, W. Wu, D. A. Muller, R. J. Birgeneau, and R. Ramesh, {\it Sci. Adv.} {\bf 8}, eabm7103 (2022).

\bibitem{Tangj2021} J. Tang, Y. Wu, W. Wang, L. Kong, B. Lv, W. Wei, J. Zang, M. Tian, and H. Du, {\it Nat. Nanotechnol.} {\bf 16}, 1086 (2021).

\bibitem{Fragkos2022} S. Fragkos, P. Pappas, E. Symeonidou, Y. Panayiotatos, and A. Dimoulas, {\it Appl. Phys. Lett.} {\bf 120}, 182402 (2022).

\bibitem{Songt2024} T. Song and X. Xu, {\it Nat. Rev. Electr. Eng.} {\bf 1}, 696 (2024).

\bibitem{Mellnik2014} A. R. Mellnik, J. S. Lee, A. Richardella, J. L. Grab, P. J. Mintun, M. H. Fischer, A. Vaezi, A. Manchon, E.-A. Kim, N. Samarth, and D. C. Ralph, {\it Nature} {\bf 511}, 449 (2014).

\bibitem{Liw2020} W. Li, X. Lu, S. Dubey, L. Devenica, and A. Srivastava, {\it Nat. Mater.} {\bf 19}, 624 (2020).

\bibitem{Jinc2018} C. Jin, E. Y. Ma, O. Karni, E. C. Regan, F. Wang, and T. F. Heinz, {\it  Nat. Nanotechnol.} {\bf 13}, 994 (2018).

\bibitem{Liux2019} X. Liu, and M. C. Hersam, {\it Nat. Rev. Mater.} {\bf 4}, 669 (2019).

\bibitem{Carr2017} S. Carr, D. Massatt, S. Fang, P. Cazeaux, M. Luskin, and E. Kaxiras, {\it Phys. Rev. B} {\bf 95}, 075420 (2017).

\bibitem{Andrei2020} E. Y. Andrei and A. H. MacDonald, {\it Nat. Mater.} {\bf 19}, 1265 (2020).

\bibitem{Hennighausen2021} Z. Hennighausen and S. Kar, {\it Electron. Struct.} {\bf 3}, 014004 (2021).

\bibitem{Yangy2020} Y. Yang, J. Li, J. Yin, S. Xu, C. Mullan, T. Taniguchi, K. Watanabe, A. K. Geim, K. S. Novoselov, and A. Mishchenko, {\it Sci. Adv.} {\bf 6}, eabd3655 (2020).

\bibitem{Huangd2022} D. Huang, J. Choi, C. Shih, and X. Li, {\it Nat. Nanotechnol.} {\bf 17}, 227 (2022).

\bibitem{Adak2024} P. C. Adak, S. Sinha, A. Agarwal, and M. M. Deshmukh, {\it Nat. Rev. Mater.} {\bf 9}, 481 (2024).

\bibitem{Christos2022} M. Christos, S. Sachdev, and M. S. Scheurer, {\it Phys. Rev. X} {\bf 12}, 021018 (2022).

\bibitem{Liuy2021} Y. Liu, C. Zeng, J. Yu, J. Zhong, B. Li, Z. Zhang, Z. Liu, Z. M. Wang, A. Pan, and X. Duan, {\it Chem. Soc. Rev.} {\bf 50}, 6401 (2021).

\bibitem{Carr2020} S. Carr, S. Fang, and E. Kaxiras, {\it Nat. Rev. Mater.} {\bf 5}, 748 (2020).

\bibitem{Jorio2022} A. Jorio, {\it Nat. Mater.} {\bf 21}, 844 (2022).

\bibitem{Cui2024} Y. Cui, J. Wang, Y. Li, Y. Wu, E. Been, Z. Zhang, J. Zhou, W. Zhang, H. Y. Hwang, R. Sinclair1, and Y. Cui, {\it Science} {\bf 383}, 212 (2024).

\bibitem{Caoy2016} Y. Cao, J. Y. Luo, V. Fatemi, S. Fang, J. D. Sanchez-Yamagishi, K. Watanabe, T. Taniguchi, E. Kaxiras, and P. Jarillo-Herrero, {\it Phys. Rev. Lett.} {\bf 117}, 116804 (2016).

\bibitem{Ribeiro2018} R. Ribeiro-Palau, C. Zhang, K. Watanabe, T. Taniguchi, J. Hone, and C. R. Dean, {\it Science} {\bf 361}, 690 (2018).

\bibitem{Torma2022} P. T\"{o}rm\"{a}, S. Peotta, and B. A. Bernevig, {\it Nat. Rev. Phys.} {\bf 4}, 528 (2022).

\bibitem{Dul2023} L. Du, M. R. Molas, Z. Huang, G. Zhang, F. Wang, and Z. Sun, {\it Science} {\bf 379}, eadg0014 (2023).

\bibitem{Balents2020} L. Balents, C. R. Dean, D. K. Efetov, and A. F. Young, {\it Nat. Phys.} {\bf 16}, 725 (2020). 

\bibitem{Kim2022} J. Kim, E. Ko, J. Jo, M. Kim, H. Yoo, Y.-W. Son, and H. Cheong, {\it Nat. Mater.} {\bf 21}, 890 (2022).

\bibitem{Hejazi2020} K. Hejazi, Z. Luo, and L. Balents, {\it Proc. Natl. Acad. Sci. U.S.A.} {\bf 117}, 10721 (2020). 

\bibitem{Songt2021} T. Song, Q.-C. Sun, E. Anderson, C. Wang, J. Qian, T. Taniguchi, K. Watanabe , M. A. McGuire, R. St\"{o}hr, D. Xiao, T. Cao, J. Wrachtrup, and X. Xu, {\it Science} {\bf 374}, 1140 (2021).

\bibitem{Kennes2021} D. M. Kennes, M. Claassen, L. Xian, A. Georges, A. J. Millis, J. Hone, C. R. Dean, D. N. Basov, A. N. Pasupathy, and A. Rubio, {\it Nat. Phys.} {\bf 17}, 155 (2021).

\bibitem{Confalone2025} T. Confalone, S. Shokri, F. L. Sardo, V. M. Vinokur, K. Nielsch, H. Golam, and N. Poccia, {\it Nat. Rev. Electr. Eng.} {\bf 2}, 73 (2025).

\bibitem{Caiy2021} L. Cai and G. Yu, {\it Adv. Mater.} {\bf 33}, 2004974 (2021).

\bibitem{Wangj2019} J. Wang and X. Mu and L. Wang, and M. Sun, {\it Mater. Today Phys.} {\bf 9}, 100099 (2019). 

\bibitem{Sunx2024} X. Sun, M. Suriyage, A. R. Khan, M. Gao, J. Zhao, B. Liu, M. M. Hasan, S. Rahman, R. Chen, P. K. Lam, and Y. Lu, {\it Chem. Rev.} {\bf 124}, 1992 (2024). 

\bibitem{LeCun2015} Y. LeCun, Y. Bengio, and G. Hinton, {\it Nature} {\bf 521}, 436 (2015).

\bibitem{Vaswani2017} A. Vaswani, N. Shazeer, N. Parmar, J. Uszkoreit, L. Jones, A. N. Gomez, L. Kaiser, and I. Polosukhin, {\it Adv. Neural Inf. Process. Syst.} {\bf 30} (2017).

\bibitem{Jumper2021} J. Jumper, R. Evans, A. Pritzel, T. Green, M. Figurnov, O. Ronneberger, K. Tunyasuvunakool, R. Bates, A. \v{z}\'{i}dek, A. Potapenko, A. Bridgland, C. Meyer, S. A. A. Kohl, A. J. Ballard, A. Cowie, B. Romera-Paredes, S. Nikolov, R. Jain, J. Adler, T. Back, S. Petersen, D. Reiman, E. Clancy, M. Zielinski, M. Steinegger, M. Pacholska, T. Berghammer, S. Bodenstein, D. Silver, O. Vinyals, A. W. Senior, K. Kavukcuoglu, P. Kohli, and D. Hassabis, {\it Nature} {\bf 596}, 583 (2021).

\bibitem{Carleo2019} G. Carleo, I. Cirac, K. Cranmer, L. Daudet, M. Schuld, N. Tishby, L. Vogt-Maranto, and L. Zdeborov\'{a}, {\it Rev. Mod. Phys.} {\bf 91}, 045002 (2019).

\bibitem{Nelson2019} J. Nelson and S. Sanvito, {\it Phys. Rev. Materials} {\bf 3}, 104405 (2019).

\bibitem{Xia2022} W. Xia, M. Sakurai, B. Balasubramanian, T. Liao, R. Wang, C. Zhang, H. Sun, K. Ho, J. Chelikowsky, D. Sellmyer, and C.-Z. Wang, {\it Proc. Natl. Acad. Sci. U.S.A.} {\bf 119}, e2204485119 (2022).

\bibitem{Raissi2019} M. Raissi, P. Perdikaris, and G. E. Karniadakis, {\it J. Comput. Phys.} {\bf 378}, 686 (2019).

\bibitem{Karniadakis2021} G. E. Karniadakis, I. G. Kevrekidis, L. Lu, P. Perdikaris, S. Wang, and L. Yang, {\it Nat. Rev. Phys.} {\bf 3}, 422 (2021).

\bibitem{Torelli2019} D. Torelli, K. S. Thygesen, and T. Olsen, {\it 2D Mater.} {\bf 6}, 045018 (2019).

\bibitem{Lus2020} S. Lu, Q. Zhou, Y. Guo, Y. Zhang, Y. Wu, and J. Wang, {\it Adv. Mater.} {\bf 32}, 2002658 (2020).

\bibitem{Lub2024} B. Lu, Y. Xia, Y. Ren, M. Xie, L. Zhou, G. Vinai, S. A. Morton, A. T. S. Wee, W. G. van der Wiel, W. Zhang, and P. K. J. Wong, {\it  Adv. Sci.} {\bf 11}, 2305277 (2024).

\bibitem{Bhawsar2025} S. Bhawsar and E.-H. Yang, {\it J. Phys. D: Appl. Phys.} {\bf 48}, 073005 (2025).

\bibitem{Gouveia2025} J. D. Gouveia, T. L. P. Galv\~{a}o, K. I. Nassar, and J. R. B. Gomes, {\it npj 2D Mater. Appl.} {\bf 9}, 8 (2025).

\bibitem{Mounet2018} N. Mounet, M. Gibertini, P. Schwaller, D. Campi, A. Merkys, A. Marrazzo, T. Sohier, I. E. Castelli, A. Cepellotti, G. Pizzi and N. Marzari, {\it Nat. Nanotechnol.} {\bf 13}, 246 (2018).

\bibitem{Lu2024_review} B. Lu, Y. Xia, Y. Ren, M. Xie, L. Zhou, G. Vinai, S. A. Morton, A. T. S. Wee, W. G. van der Wiel, W. Zhang, and J. P. K. Wong, {\it Adv. Sci.} {\bf 11}, 2305277 (2024).

\bibitem{Lado2021} J. L. Lado, {\it Science} {\bf 374}, 1048 (2024). 

\bibitem{Nimbalkar2020} A. Nimbalkar and H. Kim, {\it Nano-Micro Lett.} {\bf 12}, 126 (2020). 

\bibitem{Lium2022} M. Liu, L. Wang, and G. Yu, {\it Adv. Sci.} {\bf 9}, 2103170 (2022). 

\bibitem{Bhowmik2024} S. Bhowmik, A. Ghosh, and U Chandni, {\it Rep. Prog. Phys.} {\bf 87}, 096401 (2024).

\bibitem{Mak2022} K. F. Mak and J. Shan, {\it  Nat. Nanotechnol.} {\bf 17}, 686 (2022). 

\bibitem{Fox2024} C. Fox, Y. Mao, X. Zhang, Y. Wang, and J. Xiao, {\it Chem. Rev.} {\bf 124}, 21862 (2024). 

\bibitem{Guoh2021} H.-W. Guo, Z. Hu, Z.-B. Liu, and J.-G. Tian, {\it  Adv. Funct. Mater.} {\bf 31}, 2007810 (2021).

\bibitem{Hef2021} F. He, Y. Zhou, Z. Ye, S.-H. Cho, J. Jeong, X. Meng, and Y. Wang, {\it ACS Nano} {\bf 15}, 5944 (2021).

\bibitem{Woodsc2014} C. R. Woods, L. Britnell, A. Eckmann, R. S. Ma, J. C. Lu, H. M. Guo, X. Lin, G. L. Yu, Y. Cao, R. V. Gorbachev, A. V. Kretinin, J. Park, L. A. Ponomarenko, M. I. Katsnelson, Yu. N. Gornostyrev, K. Watanabe, T. Taniguchi, C. Casiraghi, H-J. Gao, A. K. Geim, and K. S. Novoselov, {\it Nat. Phys.} {\bf 10}, 451 (2014). 

\bibitem{Liy2024} Y. Li, F. Zhang, V.-A. Ha, Y.-C. Lin, C. Dong, Q. Gao, Z. Liu, X. Liu, S. H. Ryu, H. Kim, C. Jozwiak, A. Bostwick, K. Watanabe, T. Taniguchi, B. Kousa, X. Li, E. Rotenberg, E. Khalaf, J. A. Robinson, F. Giustino, and C.-K. Shih, {\it Nature} {\bf 625}, 494 (2024).

\bibitem{Yaow2018} W. Yao, E. Wang, C. Bao, Y. Zhang, K. Zhang, K. Bao, C. K. Chan, C. Chen, J. Avila, M. C. Asensio, J. Zhu, and S. Zhou, {\it Proc. Natl. Acad. Sci. U.S.A.} {\bf 115}, 6928 (2018).

\bibitem{Ahn2018} S. J. Ahn, P. Moon, T.-H. Kim, H.-W. Kim, H.-C. Shin, E. H. Kim, H. W. Cha, S.-J. Kahng, P. Kim, M. Koshino, Y.-W. Son, C.-W. Yang, and J. R. Ahn, {\it Science} {\bf 361}, 782 (2018). 

\bibitem{Andrei2021} E. Y. Andrei, D. K. Efetov, P. Jarillo-Herrero, A. H. MacDonald, K. F. Mak, T. Senthil, E. Tutuc, A. Yazdani, and A. F. Young, {\it Nat. Rev. Mater.} {\bf 6}, 201 (2021).

\bibitem{Bistritzer2011} R. Bistritzer and A. H. MacDonald, {\it Proc. Natl. Acad. Sci. U.S.A.} {\bf 108}, 12233 (2011).

\bibitem{Kim2017} K. Kim, A. DaSilva, S. Huang, B. Fallahazad, S. Larentis, T. Taniguchi, K. Watanabe, B. J. LeRoy, A. H. MacDonald, and E. Tutuc, {\it Proc. Natl. Acad. Sci. U.S.A.} {\bf 114}, 3364 (2017).

\bibitem{Rosenberger2020} M. R. Rosenberger, H.-J. Chuang, M. Phillips, V. P. Oleshko, K. M. McCreary, S. V. Sivaram, C. S. Hellberg, and B. T. Jonker, {\it ACS Nano} {\bf 14}, 4550 (2020).

\bibitem{Mele2010} E. J. Mele, {\it Phys. Rev. B} {\bf 81}, 161405(R) (2010). 

\bibitem{Yankowitz2019} M. Yankowitz, S. Chen, H. Polshyn, Y. Zhang, K. Watanabe, T. Taniguchi, D. Graf, A. F. Young, and C. R. Dean, {\it Science} {\bf 363}, 1059 (2019).

\bibitem{Kangp2017} P. Kang, W.-T. Zhang, V. Michaud-Rioux, X.-H. Kong, C. Hu, G.-H. Yu, and H. Guo, {\it Phys. Rev. B} {\bf 96}, 195406 (2017). 

\bibitem{Srivastava2021} P. K. Srivastava, Y. Hassan, D. J. P. de Sousa, Y. Gebredingle, M. Joe, F. Ali, Y. Zheng, W. J. Yoo, S. Ghosh, J. T. Teherani, B. Singh, T. Low, and C. Lee, {\it Nat. Electron.} {\bf 4}, 269 (2021).

\bibitem{Huangs2024} S. Huang, B. Yu, Y. Ma, C. Pan, J. Ma, Y. Zhou, Y. Ma, K. Yang, H. Wu, Y. Lei, Q. Xing, L. Mu, J. Zhang, Y. Mou, and H. Yan, {\it Science} {\bf 386}, 526 (2024).

\bibitem{Xianl2019} L. Xian, D. M. Kennes, N. Tancogne-Dejean, M. Altarelli, and A. Rubio, {\it Nano Lett.} {\bf 19}, 4934 (2019).

\bibitem{Woods2021} C. R. Woods, P. Ares, H. Nevison-Andrews, M. J. Holwill, R. Fabregas, F. Guinea, A. K. Geim, K. S. Novoselov, N. R. Walet, and L. Fumagalli, {\it Nat. Commun.} {\bf 12}, 347 (2021).

\bibitem{Taboada2023} P. Roman-Taboada, E. Obregon-Castillo, A. R. Botello-Mendez, and C. Noguez, {\it Phys. Rev. B} {\bf 108}, 075109 (2023).

\bibitem{Bennett2023} D. Bennett, G. Chaudhary, Ro.-J. Slager, E. Bousquet, and P. Ghosez, {\it Nat. Commun.} {\bf 14}, 1629 (2023).

\bibitem{Zhangz2020} Z. Zhang, Y. Wang, K. Watanabe, T. Taniguchi, K. Ueno, E. Tutuc, and B. J. LeRoy, {\it Nat. Phys.} {\bf 16}, 1093 (2020). 

\bibitem{Naik2018} M. H. Naik and M. Jain, {\it Phys. Rev. Lett.} {\bf 121}, 266401 (2018). 

\bibitem{Wangl2020} L. Wang, E.-M. Shih, A. Ghiotto, L. Xian, D. A. Rhodes, C. Tan, M. Claassen, D. M. Kennes, Y. Bai, B. Kim, K. Watanabe, T. Taniguchi, X. Zhu, J. Hone, A. Rubio, A. N. Pasupathy, and C. R. Dean, {\it Nat. Mater.} {\bf 19}, 861 (2020). 

\bibitem{Weston2020} A. Weston, Y. Zou, V. Enaldiev, A. Summerfield, N. Clark, V. Z\'{o}lyomi, A. Graham, C. Yelgel, S. Magorrian, M. Zhou, J. Zultak, D. Hopkinson, A. Barinov, T. H. Bointon, A. Kretinin, N. R. Wilson, P. H. Beton, V. I. Fal'ko, S. J. Haigh, and R. Gorbachev, {\it Nat. Nanotechnol.} {\bf 15}, 592 (2020). 

\bibitem{Guoy2025} Y. Guo, J. Pack, J. Swann, L. Holtzman, M. Cothrine, K. Watanabe, T. Taniguchi, D. G. Mandrus, K. Barmak, J. Hone, A. J. Millis, A. Pasupathy, and C. R. Dean, {\it Nature} {\bf 637}, 839 (2025).

\bibitem{Xuy2022} Y. Xu, A. Ray, Y.-T. Shao, S. Jiang, K. Lee, D. Weber, J. E. Goldberger, K. Watanabe, T. Taniguchi, D. A. Muller, K. F. Mak, and J. Shan, {\it Nat. Nanotechnol.} {\bf 17}, 143 (2022). 

\bibitem{Xieh2022} H. Xie, X. Luo, G. Ye, Z. Ye, H. Ge, S. H. Sung, E. Rennich, S. Yan, Y. Fu, S. Tian, H. Lei, R. Hovden, K. Sun, R. He, and L. Zhao, {\it Nat. Phys.} {\bf 18}, 30 (2022).

\bibitem{Cheng2021} G. Chen and J. L. Lado, {\it Phys. Rev. Research} {\bf 3}, 033276 (2021).

\bibitem{Chend2022} D. Chen, W. Sun, W. Wang, X. Li, H. Li, and Z. Cheng, {\it J. Mater. Chem. C} {\bf 10}, 12741 (2022). 

\bibitem{Tarnopolsky2019} G. Tarnopolsky, A. J. Kruchkov, and A. Vishwanath, {\it Phys. Rev. Lett.} {\bf 122}, 106405 (2019).

\bibitem{Choi2021} Y. Choi, H. Kim, C. Lewandowski, Y. Peng, A. Thomson, R. Polski, Y. Zhang, K. Watanabe, T. Taniguchi, J. Alicea, and S. Nadj-Perge, {\it Nat. Phys.} {\bf 17}, 1375 (2021).

\bibitem{Kerelsky2019} A. Kerelsky, L. J. McGilly, D. M. Kennes, L. Xian, M. Yankowitz, S. Chen, K. Watanabe, T. Taniguchi, J.s Hone, C. Dean, A. Rubio, and A. N. Pasupathy, {\it Nature} {\bf 572}, 95 (2019).  

\bibitem{Brihuega2012} I. Brihuega, P. Mallet, H. Gonz\'alez-Herrero, G. T. de Laissardi\`ere, M. M. Ugeda, L. Magaud, J. M. G\'omez-Rodr\'{\i}guez, F. Yndur\'ain, and J.-Y. Veuillen, {\it Phys. Rev. Lett.} {\bf 109}, 196802 (2012). 

\bibitem{Sherkunov2018} Y. Sherkunov and J. J. Betouras, {\it Phys. Rev. B} {\bf 98}, 205151 (20).

\bibitem{Yinj2016} J. Yin, H. Wang, H. Peng, Z. Tan, L. Liao, L. Lin, X. Sun, A. L. Koh, Y. Chen, H. Peng, and Z. Liu, {\it Nat. Commun.} {\bf 7}, 10699 (2016).

\bibitem{Wus2021} S. Wu, Z. Zhang, K. Watanabe, T. Taniguchi, and E. Y. Andrei, {\it Nat. Mater.} {\bf 20}, 488 (2021).

\bibitem{Caoy2020} Y. Cao, D. Rodan-Legrain, O. Rubies-Bigorda, J. M. Park, K. Watanabe, T. Taniguchi, and P. Jarillo-Herrero, {\it Nature} {\bf 583}, 215 (2020).

\bibitem{Saito2020} Y. Saito, J. Ge, K. Watanabe, T. Taniguchi, and A. F. Young, {\it Nat. Phys.} {\bf 16}, 926 (2020).

\bibitem{Shenc2020} C. Shen, Y. Chu, Q. Wu, N. Li, S. Wang, Y. Zhao, J. Tang, J. Liu, J. Tian, K. Watanabe, T. Taniguchi, R. Yang, Z. Y. Meng, D. Shi, O. V. Yazyev, and G. Zhang, {\it Nat. Phys.} {\bf 16}, 520 (2020).

\bibitem{Yoo2019} H. Yoo, R. Engelke, S. Carr, S. Fang, K. Zhang, P. Cazeaux, S. H. Sung, R. Hovden, A. W. Tsen, T. Taniguchi, K. Watanabe, G.-C. Yi, M. Kim, M. Luskin, E. B. Tadmor, E. Kaxiras, and P. Kim, {\it Nat. Mater.} {\bf 18}, 448 (2019).

\bibitem{Shij2021} J. Shi, J. Zhu, and A. H. MacDonald, {\it Phys. Rev. B} {\bf 103}, 075122 (2021). 

\bibitem{Nuckolls2020} K. P. Nuckolls, M. Oh, D. Wong, B. Lian, K. Watanabe, T. Taniguchi, B. A. Bernevig, and A. Yazdani, {\it Nature} {\bf 588}, 610 (2020).

\bibitem{Oh2021} M. Oh, K. P. Nuckolls, D. Wong, R. L. Lee, X. Liu, K. Watanabe, T. Taniguchi, and A. Yazdani, {\it Nature} {\bf 600}, 240 (2021).

\bibitem{Lisi2021} S. Lisi, X. Lu, T. Benschop, T. A. de Jong, P. Stepanov, J. R. Duran, F. Margot, I. Cucchi, E. Cappelli, A. Hunter, A. Tamai, V. Kandyba, A. Giampietri, A. Barinov, J. Jobst, V. Stalman, M. Leeuwenhoek, K. Watanabe, T. Taniguchi, L. Rademaker, S. J. van der Molen, M. P. Allan, D. K. Efetov, and F. Baumberger, {\it Nat. Phys.} {\bf 17}, 189 (2021).

\bibitem{David2019} A. David, P. Rakyta, A. Korm\'{a}nyos, and G. Burkard, {\it Phys. Rev. B} {\bf 100}, 085412 (2019).

\bibitem{Naimer2021} T. Naimer, K. Zollner, M. Gmitra, and J. Fabian, {\it Phys. Rev. B} {\bf 104}, 195156 (2021). 

\bibitem{Arora2020} H. S. Arora, R. Polski, Y. Zhang, A. Thomson, Y. Choi, H. Kim, Z. Lin, I. Z. Wilson, X. Xu, J.-H. Chu, K. Watanabe, T. Taniguchi, J. Alicea, and S. Nadj-Perge, {\it Nature} {\bf 583}, 379 (2020). 

\bibitem{Seyler2019} K. L. Seyler, P. Rivera, H. Yu, N. P. Wilson, E. L. Ray, D. G. Mandrus, J. Yan, W. Yao, and X. Xu, {\it Nature} {\bf 576}, 66 (2019). 

\bibitem{Wangx2021} X. Wang, J. Zhu, K. L. Seyler, P. Rivera, H. Zheng, Y. Wang, M. He, T. Taniguchi, K. Watanabe, J. Yan, D. G. Mandrus, D. R. Gamelin, W. Yao, and X. Xu, {\it Nat. Nanotechnol. } {\bf 16}, 1208 (2021).

\bibitem{Zhangy2021} Y. Zhang,T. Devakul, and L. Fu, {\it Proc. Natl. Acad. Sci. U.S.A.} {\bf 118}, e2112673118 (2021).

\bibitem{Polovnikov2024} B. Polovnikov, J. Scherzer, S. Misra, X. Huang, C. Mohl, Z. Li, J. G\"{o}ser, J. F\"{o}rste, I. Bilgin, K. Watanabe, T. Taniguchi, A. H\"{o}gele, and A. S. Baimuratov, {\it Phys. Rev. Lett.} {\bf 132}, 076902 (2024).

\bibitem{Rafizadeh2025} N. Rafizadeh, G. Agunbiade, R. J. Scott, M. Vieux, and H. Zhao, {\it Appl. Phys. Lett.} {\bf 126}, 042103 (2025).

\bibitem{Caoy2018} Y. Cao, V. Fatemi, S. Fang, K. Watanabe, T. Taniguchi, E. Kaxiras, and P. Jarillo-Herrero, {\it Nature} {\bf 556}, 43 (2018).

\bibitem{Caoy2018_2} Y. Cao, V. Fatemi, A. Demir, S. Fang, S. L. Tomarken, J. Y. Luo, J. D. Sanchez-Yamagishi, K. Watanabe, T. Taniguchi, E. Kaxiras, R. C. Ashoori, and P. Jarillo-Herrero, {\it Nature} {\bf 556}, 80 (2018).

\bibitem{Cheng2019} G. Chen, L. Jiang, Sh. Wu, B. Lyu, H. Li, B. L. Chittari, K. Watanabe, T. Taniguchi, Z. Shi, J. Jung, Y. Zhang, and F. Wang, {\it Nat. Phys.} {\bf 15}, 237 (2019).

\bibitem{Burg2019} G. W. Burg, J. Zhu, T. Taniguchi, K. Watanabe, A. H. MacDonald, and E. Tutuc, {\it Phys. Rev. Lett.} {\bf 123}, 197702 (2019).

\bibitem{Hem2021} M. He, Y. Li, J. Cai, Y. Liu, K. Watanabe, T. Taniguchi, X. Xu, and M. Yankowitz, {\it Nat. Phys.} {\bf 17}, 26 (2021).

\bibitem{Sharpe2019} A. L. Sharpe, E. J. Fox, Arthur W. Barnard, Joe Finney, Kenji Watanabe, Takashi Taniguchi, M. A. Kastner, David Goldhaber-Gordon, {\it Science} {\bf 365}, 605 (2019).

\bibitem{Liux2020} X. Liu, Z. Hao, E. Khalaf, J. Y. Lee, Y. Ronen, H. Yoo, D. H. Najafabadi, K. Watanabe, T. Taniguchi, A. Vishwanath, and P. Kim, {\it Nature} {\bf 583}, 221 (2020).

\bibitem{Cheng2020} G. Chen, A. L. Sharpe, E. J. Fox, Y.-H. Zhang, S. Wang, L. Jiang, B. Lyu, H. Li, K. Watanabe, T. Taniguchi, Z. Shi, T. Senthil, D. Goldhaber-Gordon, Y. Zhang, and F. Wang, {\it Nature} {\bf 579}, 56 (2020).

\bibitem{Caij2023} J. Cai, E. Anderson, C. Wang, X. Zhang, X. Liu, W. Holtzmann, Y. Zhang, F. Fan, T. Taniguchi, K. Watanabe, Y. Ran, T. Cao, L. Fu, D. Xiao, W. Yao, and X. Xu, {\it Nature} {\bf 622}, 63 (2023).

\bibitem{Xuy2020} Y. Xu, S. Liu, D. A. Rhodes, K. Watanabe, T. Taniguchi, J. Hone, V. Elser, K. F. Mak, and J. Shan, {\it Nature} {\bf 587}, 214 (2020).

\bibitem{Devakul2021} T. Devakul, V. Cr\'{e}pel, Y. Zhang, and L. Fu, {\it Nat. Commun.} {\bf 12}, 6730 (2021).

\bibitem{Qiaoz2014} Z. Qiao, W. Ren, H. Chen, L. Bellaiche, Z. Zhang, A. H. MacDonald, and Q. Niu, {\it Phys. Rev. Lett.} {\bf 112}, 116404 (2014).

\bibitem{Zhaoc2017}  C. Zhao, T. Norden, P. Zhang, P. Zhao, Y. Cheng, F. Sun, J. P. Parry, P. Taheri, J. Wang, Y. Yang, T. Scrace, K. Kang, S. Yang, G.-X. Miao, R. Sabirianov, G. Kioseoglou, W. Huang, A. Petrou, and H. Zeng, {\it Nat. Nanotechnol.} {\bf 12}, 757 (2017).

\bibitem{Scharf2017} B. Scharf, G. Xu, A. Matos-Abiague, and I. \v{Z}uti\'{c}, {\it Phys. Rev. Lett.} {\bf 119}, 127403 (2017).

\bibitem{Zhongd2017} D. Zhong, K. L. Seyler, X. Linpeng, R. Cheng, N. Sivadas, B. Huang, E. Schmidgall, T. Taniguchi, K. Watanabe, M. A. McGuire, W. Yao, D. Xiao, K.-M. C. Fu, and X. Xu, {\it  Sci. Adv.} {\bf 3}, e1603113 (2017).

\bibitem{Zhongd2020}  D. Zhong, K. L. Seyler, X. Linpeng, N. P. Wilson, T. Taniguchi, K. Watanabe, M. A. McGuire, K.-M. C. Fu, D. Xiao, W. Yao, and X. Xu, {\it Nat. Nanotechnol.} {\bf 15}, 187 (2020).

\bibitem{Dengy2018} Y. Deng, Y. Yu, Y. Song, J. Zhang, N. Z. Wang, Z. Sun, Y. Yi, Y. Z. Wu, S. Wu, J. Zhu, J. Wang, X. H. Chen, and Y. Zhang, {\it Nature} {\bf 563}, 94 (2018).

\bibitem{Wangz2018} Z. Wang, T. Zhang, M. Ding, B. Dong, Y. Li, M. Chen, X. Li, J. Huang, H. Wang, X. Zhao, Y. Li, D. Li, C. Jia, L. Sun, H. Guo, Y.Ye, D. Sun, Y. Chen, T. Yang, J. Zhang, S. Ono, Z. Han, and Z. Zhang, {\it Nat. Nanotechnol.} {\bf 13}, 554 (2018).

\bibitem{Huangb2018} B. Huang, G. Clark, D. R. Klein, D. MacNeill, E. Navarro-Moratalla, K. L. Seyler, N. Wilson, M. A. McGuire, D. H. Cobden, D. Xiao, W. Yao, P. Jarillo-Herrero, and X. Xu, {\it Nat. Nanotechnol.} {\bf 13}, 554 (2018).

\bibitem{Jiangs2018} S. Jiang, J. Shan, and K. F. Mak, {\it Nat. Mater.} {\bf 17}, 406 (2018).

\bibitem{Songt2018} T. Song, X. Cai, M. W.-Y. Tu, X. Zhang, B. Huang, N. P. Wilson, K. L. Seyler, L. Zhu, T. Taniguchi, K. Watanabe, M. A. McGuire, D. H. Cobden, D. Xiao, W. Yao, and X. Xu, {\it Science} {\bf 360}, 1214 (2018). 

\bibitem{Wangz2018_2}  Z. Wang, I. Guti\'{e}rrez-Lezama, N. Ubrig, M. Kroner, M. Gibertini, T. Taniguchi, K. Watanabe, A. Imam\v{g}lu, E. Giannini, and A. F. Morpurgo, {\it Nat. Commun.} {\bf 9}, 2516 (2018). 

\bibitem{Sivadas2018} N. Sivadas, S. Okamoto, X. Xu, C. J. Fennie, and D. Xiao, {\it Nano Lett.} {\bf 18}, 7658 (2018). 

\bibitem{Jiangp2019} P. Jiang, C. Wang, D. Chen, Z. Zhong, Z. Yuan, Z.-Y. Lu, and W. Ji, {\it Phys. Rev. B} {\bf 99}, 144401 (2019).

\bibitem{Jangs2019} S. W. Jang, M. Y. Jeong, H. Yoon, S. Ryee, and M. J. Han, {\it Phys. Rev. Materials} {\bf 3}, 031001 (2019).

\bibitem{Songt2019} T. Song, Z. Fei, M. Yankowitz, Z. Lin, Q. Jiang, K. Hwangbo, Q. Zhang, B. Sun, T. Taniguchi, K. Watanabe, M. A. McGuire, D. Graf, T. Cao, J.-H. Chu, D. H. Cobden, C. R. Dean, D. Xiao, and X. Xu, {\it  Nat. Mater.} {\bf 18}, 1298 (2019).

\bibitem{Lit2019} T. Li, S. Jiang, N. Sivadas, Z. Wang, Y. Xu, D. Weber, J. E. Goldberger, K. Watanabe, T. Taniguchi, C. J. Fennie, K. F. Mak, and J. Shan, {\it  Nat. Mater.} {\bf 18}, 1303 (2019).

\bibitem{Chenw2019} W. Chen, Z. Sun, Z. Wang, L. Gu, X. Xu, S. Wu, and C. Gao, {\it  Science} {\bf 366}, 983 (2019).

\bibitem{Xieh2023} H. Xie, X. Luo, Z. Ye, Z. Sun, G. Ye, S. H. Sung, H. Ge, S. Yan, Y. Fu, S. Tian, H. Lei, K. Sun, R. Hovden, R. He, and L. Zhao, {\it Nat. Phys.} {\bf 19}, 1150 (2023).

\bibitem{Lis2024} S. Li, Z. Sun, N. J. McLaughlin, A. Sharmin, N. Agarwal, M. Huang, S. H. Sung, H. Lu, . Yan, H. Lei, R. Hovden, H. Wang, H. Chen, L. Zhao, and C. R. Du, {\it Nat. Commun.} {\bf 19}, 1150 (2023).

\bibitem{Chengg2023} G. Cheng, M. M. Rahman, A. L. Allcca, A. Rustagi, X. Liu, L. Liu, L. Fu, Y. Zhu, Z. Mao, K. Watanabe, T. Taniguchi, P. Upadhyaya, and Y. P. Chen, {\it Nat. Electron.} {\bf 6}, 434 (2023).

\bibitem{Zhuh2025} H. Zhu, H. Yu, W. Zhu, G. Yu, C. Xu, and H. Xiang, {\it Phys. Rev. Lett.} {\bf 135}, 196701 (2025).

\bibitem{Liang2023} S. Liang, J. Liang, J. C. Kotsakidis, H. S. Arachchige, D. Mandrus, A. L. Friedman, and C. Gong, {\it Phys. Rev. Materials} {\bf 7}, L061001 (2023).

\bibitem{Kartsev2020} A. Kartsev, M. Augustin, R. F. L. Evans, K. S. Novoselov, and E. J. G. Santos, {\it npj Comput. Mater.} {\bf 6}, 150 (2020).

\bibitem{Paul2020} S. Paul, S. Haldar, S. von Malottki, and S. Heinze, {\it Nat. Commun.} {\bf 11}, 4756 (2020).

\bibitem{Zhangs2023} S. Zhang, X. Li, H. Zhang, P. Cui, X.Xu, and Z. Zhang, {\it npj Comput. Mater.} {\bf 8}, 38 (2023).

\bibitem{Xuc2018} C. Xu, J. Feng, H. Xiang, and L. Bellaiche, {\it npj Comput. Mater.} {\bf 4}, 57 (2018).

\bibitem{Kims2022} S. Kim, B. Yuan, and Y.-J. Kim, {\it APL Mater.} {\bf 10}, 080903 (2022).

\bibitem{Wuhrer2024} D. Wuhrer, N. Rohling, W. Belzig, {\it Appl. Phys. Lett.} {\bf 125}, 022404 (2024).

\bibitem{DeBell2000} K. De'Bell, A. B. MacIsaac, and J. P. Whitehead, {\it Rev. Mod. Phys.} {\bf 72}, 225 (2000).

\bibitem{Lux2020} X. Lu, R. Fei, L. Zhu, and L. Yang, {\it Nat. Commun.} {\bf 11}, 4724 (2020).

\bibitem{Zhangc2017}  C. Zhang, C.-P. Chuu, X. Ren, M.-Y. Li, L.-J. Li, C. Jin, M.-Y. Chou, and C.-K. Shih, {\it Sci. Adv.} {\bf 3}, e1601459 (2017).

\bibitem{Huangm2023}  M. Huang, Z. Sun, G. Yan, H. Xie, N. Agarwal, G. Ye, S. H. Sung, H. Lu, J. Zhou, S. Yan, S. Tian, H. Lei, R. Hovden, R. He, H. Wang, L. Zhao, and C. R. Du, {\it Nat. Commun.} {\bf 14}, 5259 (2023).

\bibitem{Kapfer2023}  M. Kapfer, B. S. Jessen, M. E. Eisele, M. Fu, D. R. Danielsen, T. P. Darlington, S. L. Moore, N. R. Finney, A. Marchese, V. Hsieh, P. Majchrzak, Z. Jiang, D. Biswas, P. Dudin, J. Avila, K. Watanabe, T. Taniguchi, S. Ulstrup, P. B\o ggild, P. J. Schuck, D. N. Basov, J. Hone, and C. R. Dean, {\it Science} {\bf 381}, 667 (2023).

\bibitem{Wangc2020} C. Wang, Y. Gao, H. Lv, X. Xu, and D. Xiao, {\it Phys. Rev. Lett.} {\bf 125}, 247201 (2020).

\bibitem{Xiaof2021} F. Xiao, K. Chen, and Q. Tong, {\it Phys. Rev. Research} {\bf 3}, 013027 (2021).

\bibitem{Yangb2023} B. Yang, Y. Li, H. Xiang, H. Lin, and B. Huang, {\it Nat. Comput. Sci.} {\bf 3}, 314 (2023).

\bibitem{Kim2023} K.-M. Kim, D. H. Kiem, G. Bednik, M. J. Han, and M. J. Park, {\it Nano Lett.} {\bf 23}, 6088 (2023).

\bibitem{Kim2023_2} K.-M. Kim and M. J. Park, {\it Phys. Rev. B} {\bf 108}, L100401 (2023).

\bibitem{Smejkal2018} L. \u{S}mejkal, Y. Mokrousov, B. Yan, and A. H. MacDonald , {\it Nat. Phys.} {\bf 14}, 242 (2018).

\bibitem{Bonbien2021} V. Bonbien, F. Zhuo, A. Salimath, O. Ly, A. Abbout, and A. Manchon, {\it J. Phys. D: Appl. Phys.} {\bf 55}, 103002 (2021).

\bibitem{Heq2022} Q. L. He, T. L. Hughes, N. P. Armitage, Y. Tokura, and K. L. Wang, {\it Nat. Mater.} {\bf 21}, 15 (2022).

\bibitem{Nagaosa2013} N. Nagaosa, and Y. Tokura, {\it Nat. Nanotechnol.} {\bf 8}, 899 (2013).

\bibitem{Fert2017} A. Fert, N. Reyren, and V. Cros, {\it Nat. Rev. Mater.} {\bf 2}, 17031 (2017).

\bibitem{Tokura2021} Y. Tokura and N. Kanazawa, {\it Chem. Rev.} {\bf 121}, 2857 (2021).

\bibitem{Gobel2021} B. G\"{o}bel, I. Mertig, and O. A. Tretiakov, {\it Phys. Rep.} {\bf 895}, 1 (2021).

\bibitem{McClarty2021} P. A. McClarty, {\em Annu. Rev. Condens. Matter Phys.} {\bf 13}, 171 (2021).

\bibitem{Zhuo2025} F. Zhuo, J. Kang, A. Manchon, and Z. Cheng, {\em Adv. Phys. Res.} {\bf 4}, 2300054 (2025).

\bibitem{Smejkal2017} L. \u{S}mejkal, T. Jungwirth, and J. Sinova, {\it Phys. Status Solidi Rapid Res. Lett.} {\bf 11}, 1700044 (2017).

\bibitem{Tokura2019} Y. Tokura, K. Yasuda, and A. Tsukazaki, {\it Nat. Rev. Phys.} {\bf 1}, 126 (2019).

\bibitem{Changc2020} C.-Z. Chang, {\it Nat. Mater.} {\bf 19}, 484 (2020).

\bibitem{Romming2013} N. Romming, C. Hanneken, M. Menzel, J. E. Bickel, B. Wolter, K. von Bergmann, A. Kubetzka, and R. Wiesendanger, {\it Science} {\bf 341}, 636 (2013).

\bibitem{Luchaire2016} C. Moreau-Luchaire, C. Moutafis, N. Reyren, J. Sampaio, C. A. F. Vaz, N. Van Horne, K. Bouzehouane, K. Garcia, C. Deranlot, P. Warnicke, P. Wohlhüter, J.-M. George, M. Weigand, J. Raabe, V. Cros, and A. Fert, {\it Nat. Nanotechnol.} {\bf 11}, 444 (2016).

\bibitem{Wangl2018} L. Wang, Q. Feng, Y. Kim, R. Kim, K. H. Lee, S. D. Pollard, Y. J. Shin, H. Zhou, W. Peng, D. Lee, W. Meng, H. Yang, J. H. Han, M. Kim, Q. Lu, and T. W. Noh, {\it Nat. Mater.} {\bf 17}, 1807 (2018).

\bibitem{Tongq2018} Q. Tong, F. Liu, J. Xiao, and W. Yao, {\it Nano Lett.} {\bf 18}, 7194 (2018).

\bibitem{Akram2021} M. Akram and O. Erten, {\it Phys. Rev. B} {\bf 103}, L140406 (2021).

\bibitem{Akram2021_NL} M. Akram, H. LaBollita, D. Dey, J. Kapeghian, O. Erten, and A. S. Botana, {\it Nano Lett.} {\bf 21}, 6633 (2021).

\bibitem{Hejazi2021} K. Hejazi, Z.-X. Luo, and L. Balents, {\it Phys. Rev. B} {\bf 104}, L100406 (2021).

\bibitem{Fumega2023} A. O. Fumega, and J. L Lado, {\it 2D Mater.} {\bf 10}, 025026 (2023).

\bibitem{Ray2021} S. Ray and T. Das, {\it Phys. Rev. B} {\bf 104}, 014410 (2021).

\bibitem{Ghader2022} D. Ghader, B. Jabakhanji, and A. Stroppa, {\it Commun. Phys.} {\bf 5}, 192 (2022).

\bibitem{Shaban2023} P. S. Shaban, I. S. Lobanov, V. M. Uzdin, and I. V. Iorsh, {\it Phys. Rev. B} {\bf 108}, 174440 (2023).

\bibitem{Akram2024} M. Akram, J. Kapeghian, J. Das, R. Valent\'{i}, A. S. Botana, and O. Erten
, {\it Nano Lett.} {\bf 24}, 890 (2024).

\bibitem{Kim2024} K.-M. Kim, G. Go, M. J. Park, and S. K. Kim, {\it Nano Lett.} {\bf 24}, 74 (2024).

\bibitem{Kim2025} K.-M. Kim and S. K. Kim, {\it npj Quantum Mater.} {\bf 10}, 68 (2025).

\bibitem{Liy2020} Y. Li and R. Cheng, {\it Phys. Rev. B} {\bf 102}, 094404 (2020).

\bibitem{Wangh2023} H. Wang, M. Madami, J. Chen, H. Jia, Y. Zhang, R. Yuan, Y. Wang, W. He, L. Sheng, Y. Zhang, J. Wang, S. Liu, K. Shen, G. Yu, X. Han, D. Yu, J.-P. Ansermet, G. Gubbiotti, and H. Yu, {\it Phys. Rev. X} {\bf 13}, 021016 (2023).

\bibitem{Chenj2022} J. Chen, L. Zeng, H. Wang, M. Madami, G. Gubbiotti, S. Liu, J. Zhang, Z. Wang, W. Jiang, Y. Zhang, D. Yu, J.-P. Ansermet, and H. Yu, {\it Phys. Rev. B} {\bf 105}, 094445 (2022).

\bibitem{Chenj2024} J. Chen, J. M. Madami, G. Gubbiotti, and H. Yu, {\it  Appl. Phys. Lett.} {\bf 125}, 162403 (2024).

\bibitem{Liuj2025} J. Liu, X. Zhang, and G. Lu, {\it Proc. Natl. Acad. Sci. U.S.A} {\bf 122}, e2413326121 (2025).

\bibitem{Huac2023} C.-B. Hua, F. Xiao, Z.-R. Liu, J.-H. Sun, J.-H. Gao, C.-Z. Chen, Q. Tong, B. Zhou, and D.-H. Xu, {\it Phys. Rev. B} {\bf 107}, L020404 (2023).

\bibitem{Ganguli2023} S. C. Ganguli, M. Aapro, S. Kezilebieke, M. Amini, J. L. Lado, and P. Liljeroth, {\it Nano Lett.} {\bf 23}, 3412 (2023).

\bibitem{Ghader2020} D. Ghader, {\it Sci. Rep.} {\bf 10}, 15069 (2020).

\bibitem{Ferrer2003} A. V. Ferrer, P. F. Farinas, and A. O. Caldeira, {\it Phys. Rev. Lett.} {\bf 91}, 226803 (2003).

\bibitem{Sanchez2015} F. Garcia-Sanchez, P. Borys, R. Soucaille, J.-P. Adam, R. L. Stamps, and J.-V. Kim, {\it Phys. Rev. Lett.} {\bf 114}, 247206 (2015).

\bibitem{Demokritov2006} S. O. Demokritov, V. E. Demidov, O. Dzyapko, G. A. Melkov, A. A. Serga, B. Hillebrands, and A. N. Slavin, {\it Nature} {\bf 443}, 430 (2006).

\bibitem{Giamarchi2008} T. Giamarchi, C. R\"uegg, and O. Tchernyshyov, {\it Nat. Phys.} {\bf 4}, 198 (2008).

\bibitem{Tritsaris2021} G. A. Tritsaris, S. Carr, and G. R. Schleder, {\it Appl. Phys. Rev.} {\bf 8}, 031401 (2021).

\bibitem{Liu2024} J. Liu, Z. Fang, H. Weng, and Q. Wu, {\it npj Comput. Mater.} {\bf 11}, 248 (2025).

\bibitem{Li2023} H. Li, Z. Tang, X. Gong, N. Zou, W. Duan, and Y. Xu, {\it Nat. Comput. Sci.} {\bf 3}, 321 (2023).

\bibitem{Soriano2023} D. Soriano, {\it Nat. Comput. Sci.} {\bf 3}, 282 (2023).

\bibitem{Zheng2022_magnetic} F. Zheng, {\it Adv. Funct. Mater.} {\bf 33}, 2206923 (2023).

\bibitem{Liu2022_seeing} D. Liu, M. Luskin, and S. Carr, {\it Phys. Rev. Research} {\bf 4}, 043224 (2022).

\bibitem{Zhang2020_deep} L. Zhang, M. Chen, X. Wu, H. Wang, W. E, and R. Car, {\it Phys. Rev. B} {\bf 102}, 041121(R) (2020).

\bibitem{Zheng2025_machine} J.-D. Zheng, C.-S. Yao, S.-C. Zhou, Y.-K. Zhang, Z.-Q. Bao, W.-Y. Tong, J.-H. Chu, and C.-G. Duan, {\it Adv. Funct. Mater.} 2503011 (2025).

\bibitem{Sobral2023} J. A. Sobral, S. Obernauer, S. Turkel, A. N. Pasupathy, and M. S. Scheurer, {\it Nat. Commun.} {\bf 14}, 5012 (2023).

\bibitem{Labrie-Boulay2024_machine} I. Labrie-Boulay, T. B. Winkler, D. Franzen, A. Romanova, H. Fangohr, and M. Kl\"aui, {\it Phys. Rev. Appl.} {\bf 21}, 014014 (2024).

\bibitem{Antao2024} T. V. C. Ant\~ao, J. L. Lado, and A. O. Fumega, {\it Nano Lett.} {\bf 24}, 15767 (2024).

\bibitem{Yeh2025} H. Ye and S. Dong, {\it Phys. Rev. Lett.} {\bf 135}, 066701 (2025).

\bibitem{McGuire2017} M. A. McGuire, G. Clark, S. KC, W. M. Chance, G. E. Jellison, V. R. Cooper,  X. Xu, and B. C. Sales, {\it Phys. Rev. Mater.} {\bf 1}, 014001 (2017).

\bibitem{Dengy2020} Y. Deng, Y. Yu, M. Z. Shi, Z. Guo, Z. Xu, J. Wang, X. H. Chen, and Y. Zhang, {\it Science} {\bf 367}, 895 (2020).

\bibitem{Bonilla2018} M. Bonilla, S. Kolekar, Y. Ma, H. C. Diaz, V. Kalappattil, R. Das, T. Eggers, H. R. Gutierrez, M.-H. Phan, and M. Batzill, {\it Nat. Nanotechnol.} {\bf 13}, 289 (2018).

\bibitem{OHara2018} D. J. O’Hara, T. Zhu, A. H. Trout, A. S. Ahmed, Y. Kelly Luo,C. Hee Lee, M. R. Brenner, S. Rajan, J. A. Gupta, D. W. McComb, and R. K. Kawakami, {\it Nano Lett.} {\bf 18}, 3125 (2018).

\bibitem{Mengl2021} L. Meng, Z. Zhou, M. Xu, S. Yang, K. Si, L. Liu, X. Wang, H. Jiang, B. Li, P. Qin, P. Zhang, J. Wang, Z. Liu, P. Tang, Y. Ye, W. Zhou, L. Bao, H.-J. Gao, and Y. Gong, {\it Nat. Commun.} {\bf 12}, 809 (2021).

\bibitem{Yaoj2022} J. Yao, H. Wang, B. Yuan, Z. Hu, C. Wu, and A. Zhao, {\it Adv. Mater.} {\bf 34}, 2200236 (2022).

\bibitem{Telford2020} E. J. Telford, A. H. Dismukes, K. Lee, M. Cheng, A. Wieteska, A. M. Bartholomew, Y.-S. Chen, X. Xu, A. N. Pasupathy, X.-Y. Zhu, C. R. Dean, X. Roy, {\it Adv. Mater.} {\bf 32}, 2003240 (2020).

\bibitem{Parfenov2024} O. E. Parfenov, D. V. Averyanov, I. S. Sokolov, A. N. Mihalyuk, O. A. Kondratev, A. N. Taldenkov, A. M. Tokmachev, V. G. Storchak, {\it Adv. Mater.} {\bf 37}, 2412321 (2024).

\bibitem{Hul2023} L. Hu, H.-X. Wang, Y. Chen, K. Xu, M.-R. Li, H. Liu, P. Gu, Y. Wang, and M. Zhang, {\it Phys. Rev. B} {\bf 107}, L220407 (2023). 

\bibitem{Constant2024} C. Boix-Constant, S. Jenkins, R. Rama-Eiroa, E. J. G. Santos, S. Ma\~nas-Valero, and E. Coronado, {\it Nat. Mater.} {\bf 23}, 212 (2024).

\bibitem{Park2025} D. Park, C. Park, K. Yananose, E. Ko, B. Kim, R. Engelke, X. Zhang, K. Davydov, M. Green, H.-M. Kim, S. H. Park, J. H. Lee, S.-G. Kim, H. Kim, K. Watanabe, T. Taniguchi, S. M. Yang, K. Wang, P. Kim, Y.-W. Son, and H. Yoo, {\it Nature} {\bf 641}, 896 (2025).

\bibitem{Lih2021} H. Li, S. Li, M. H. Naik, J. Xie, X. Li, J. Wang, E. Regan, D. Wang, W. Zhao, S. Zhao, S. Kahn, K. Yumigeta, M. Blei, T. Taniguchi, K. Watanabe, S. Tongay, A. Zettl, S. G. Louie, F. Wang, and M. F. Crommie, {\it Nat. Mater.} {\bf 20}, 945 (2021).

\bibitem{Lie2021} E. Li, J.-X. Hu, X. Feng, Z. Zhou, L. An, K. T. Law, N. Wang, and N. Lin, {\it Nat. Commun.} {\bf 12}, 5601 (2021).

\bibitem{Sabani2025} D. \v{S}abani, C. Bacaksiz, and M. V. Milo\v{s}evi\'c, {\it Phys. Rev. Lett.} {\bf 135}, 036704 (2025).

\bibitem{Wangd2021} D. Wang and B. Sanyal, {\it J. Phys. Chem. C} {\bf 125}, 18467 (2021).

\bibitem{Tilak2023} N. Tilak, G. Li, T. Taniguchi, and K. Watanabe, and Eva Y. Andrei, {\it Nano Lett.} {\bf 23}, 73 (2023).

\bibitem{Agarwal2024} N. Agarwal, S. H. Sung, Z. Sun, L. Zhao, and R. Hovden, {\it Microsc. Microanal.} {\bf 30}, 1094 (2024).

\bibitem{Baih2022} H. Bai, Y. C. Zhang, L. Han, Y. J. Zhou, F. Pan, and C. Song, {\it Appl. Phys. Rev.} {\bf 9}, 041316 (2022).

\bibitem{Chuh2020} H. Chu, C. J. Roh, J. O. Island, C. Li, S. Lee, J. Chen, J.-G. Pak, A. F. Young, J. S. Lee, and D. Hsieh, {\it Phys. Rev. Lett.} {\bf 124}, 027601 (2020). 

\bibitem{Gaoy2023} Y. Gao, X. Jiang, Z. Qiu, and J. Zhao, {\it npj Comput. Mater.} {\bf 9}, 107 (2023). 

\bibitem{Mavani2025} H. Mavani, K. Huang, K. Samanta, and E. Y. Tsymbal, {\it Phys. Rev. B} {\bf 112}, L060401 (2025). 

\bibitem{Mah2021} H.-Y. Ma, M. Hu, N. Li, J. Liu, W. Yao, J.-F. Jia, and J. Liu, {\it Nat. Commun.} {\bf 12}, 2846 (2021).

\bibitem{Khan2025} I. Khan, D. Bezzerga, B. Marfoua, and J. Hong, {\it npj 2D Mater. Appl.} {\bf 9}, 18 (2025).

\bibitem{Gongj2024} J. Gong, Y. Wang, Y. Han, Z. Cheng, X. Wang, Z.-M. Yu, Y. Yao, {\it Adv. Mater.} {\bf 36}, 2402232 (2024).

\bibitem{Her2023} R. He, D. Wang, N. Luo, J. Zeng, K.-Q. Chen, and L.-M. Tang, {\it Phys. Rev. Lett.} {\bf 130}, 046401 (2023). 

\bibitem{Takahashi2016} R. Takahashi and N. Nagaosa, {\it Phys. Rev. Lett.} {\bf 117}, 217205 (2016).

\bibitem{Zhangx2019} X. Zhang, Y. Zhang, S. Okamoto, and D. Xiao, {\it Phys. Rev. Lett.} {\bf 123}, 167202 (2019).

\bibitem{Lius2021} S. Liu, A. Granados del \'Aguila, D. Bhowmick, C. K. Gan, T. Thu Ha Do, M. A.
Prosnikov, D. Sedmidubsk\'y, Z. Sofer, P. C. M. Christianen, P. Sengupta, and Q. Xiong, {\it Phys. Rev. Lett.} {\bf 127}, 097401 (2021). 

\bibitem{Pawbake2022} A. Pawbake, T. Pelini, A. Delhomme, D. Romanin, D. Vaclavkova, G. Martinez, M. Calandra, M.-A. Measson, M. Veis, M. Potemski, M. Orlita, and C. Faugeras, {\it ACS Nano} {\bf 16}, 12656 (2022).

\bibitem{Cuij2023} J. Cui, E. V. Bostr\"om, M. Ozerov, F. Wu, Q. Jiang, J.-H. Chu, C. Li, F. Liu, X. Xu, A. Rubio, and Q. Zhang, {\it Nat. Commun.} {\bf 14}, 3396 (2023). 

\bibitem{Linz2024} Z.-X. Lin and S. Zhang, {\it Appl. Phys. Lett.} {\bf 124}, 132402 (2024). 

\bibitem{Linm2018} M. Lin, Q. Tan, J. Wu, X. Chen, J. Wang, Y. Pan, X. Zhang, X. Cong, J. Zhang, W. Ji, P. Hu, K. Liu, and P. Tan, {\it ACS Nano} {\bf 12}, 8770 (2018).

\bibitem{Parzefall2021} P. Parzefall, J. H\"oller, M. Scheuck, A. Beer, K. Lin, B. Peng, B. Monserrat, P. Nagler, M. Kempf, T. Korn, and C. Sch\"uller, {\it 2D Mater.} {\bf 8}, 035030 (2021).

\bibitem{Chuangh2022} H. Chuang, M. Phillips, K. McCreary, D. Wickramaratne, M. Rosenberger, V. Oleshko, N. Proscia, M. Lohmann, D. O'Hara, P. Cunningham, C. Hellberg, and B. Jonker, {\it ACS Nano} {\bf 16}, 16260 (2022).

\bibitem{Diederich2023} G. M. Diederich, J. Cenker, Y. Ren, J. Fonseca, D. G. Chica, Y. J. Bae, X. Zhu, X. Roy, T. Cao, D. Xiao, and X. Xu, {\it Nat. Nanotechnol.} {\bf 18}, 23 (2023).

\bibitem{Wangz2023} Z. Wang, X.-X. Zhang, Y. Shiomi, T.-h. Arima, N. Nagaosa, Y. Tokura, and N. Ogawa, {\it Phys. Rev. Research} {\bf 5}, L042032 (2023).

\bibitem{Zhangy2022} Y. Zhang, H. Kim, W. Zhang, K. Watanabe, T. Taniguchi, Y. Gao, M. Maruyama, S. Okada, K. Shinokita, and K. Matsuda, {\it Adv. Mater.} {\bf 34}, 2200301 (2022).

\bibitem{Eerenstein2006} W. Eerenstein, N. D. Mathur, and J. F. Scott, {\it Nature} {\bf 442}, 759 (2006). 

\bibitem{Luc2015} C. Lu, W. Hu, Y. Tian, and T. Wu, {\it Appl. Phys. Rev.} {\bf 2}, 021304 (2015). 

\bibitem{Fiebig2016} M. Fiebig, T. Lottermoser, D. Meier, and M. Trassin, {\it Nat. Rev. Mater.} {\bf 1}, 16046 (2016).

\bibitem{Wangc2023} C. Wang, L. You, D. Cobden, and J. Wang, {\it Nat. Mater.} {\bf 22}, 542 (2023).

\bibitem{Qij2018} J. Qi, H. Wang, X. Chen, and X. Qian, {\it Appl. Phys. Lett.} {\bf 113}, 043102 (2018).

\bibitem{Tangw2023} W. Tang, D. Zhao, X. Weng, K. Wu, Z. Yang, C. Kang, Y. Sun, W.-C. Jiang, H. Liang, C. Wang, and Y.-J. Zeng, {\it Appl. Phys. Rev.} {\bf 10}, 031404 (2023). 

\bibitem{Wangx2023} X. Wang, Z. Shang, C. Zhang, J. Kang, T. Liu, X. Wang, S. Chen, H. Liu, W. Tang, Y.-J. Zeng, J. Guo, Z. Cheng, L. Liu, D. Pan, S. Tong, B. Wu, Y. Xie, G. Wang, J. Deng, T. Zhai, H.-X. Deng, J. Hong, and J. Zhao, {\it Nat. Commun.} {\bf 14}, 840 (2023). 

\bibitem{Songq2022} Q. Song, C.A. Occhialini, E. Ergeçen, B. Ilyas, D. Amoroso, P. Barone, J. Kapeghian, K. Watanabe, T. Taniguchi, A. S. Botana, S. Picozzi, N. Gedik, and R. Comin, {\it Nature} {\bf 602}, 601 (2022). 

\bibitem{Fumega2022} A. O. Fumega and J. L. Lado, {\it 2D Mater.} {\bf 9}, 025010 (2022).

\bibitem{Amini2024} M. Amini, A. O. Fumega, H. González-Herrero, V. Vaňo, S. Kezilebieke, J. L. Lado, P. Liljeroth, {\it Adv. Mater.} {\bf 36}, 2311342 (2024).

\bibitem{Tant2019} H. Tan, M. Li, H. Liu, Z. Liu, Y. Li and W. Duan, {\it Phys. Rev. B} {\bf 99}, 195434 (2019). 

\bibitem{Huangc2018} C. Huang, Y. Du, H. Wu, H. Xiang, K. Deng, and E. Kan, {\it Phys. Rev. Lett.} {\bf 120}, 147601 (2018).

\bibitem{Xum2020} M. Xu, C. Huang, Y. Li, S. Liu, X. Zhong, P. Jena, E. Kan, and Y. Wang, {\it Phys. Rev. Lett.} {\bf 124}, 067602 (2020).

\bibitem{Sunw2022} W. Sun, W. Wang, H. Li, X. Li, Z. Yu, Y. Bai, G. Zhang, and Z. Cheng, {\it npj Comput. Mater.} {\bf 8}, 159 (2022). 

\bibitem{Huangk2022} K. Huang, D.-F. Shao, and E. Y. Tsymbal, {\it Nano Lett.} {\bf 22}, 3349 (2022).

\bibitem{Sunw2023} W. Sun, W. Wang, C. Yang, X. Li, H. Li, S. Huang, and Z. Cheng, {\it Phys. Rev. B} {\bf 107}, 184439 (2023).

\bibitem{Smejkal2022_2} L. \v{S}mejkal, J. Sinova, and T. Jungwirth, {\it Phys. Rev. X} {\bf 12}, 031042 (2022).

\bibitem{Smejkal2022_3} L. \v{S}mejkal, A. H. MacDonald, J. Sinova, S. Nakatsuji, and T. Jungwirth, {\it Nat. Rev. Mater.} {\bf 7}, 482 (2022).

\bibitem{Jungwirth2024} T. Jungwirth, R. M. Fernandes, J. Sinova, and L. \v{S}mejkal, arXiv:2409.10034 (2024).

\bibitem{Jungwirth2025} T. Jungwirth, J. Sinova, P. Wadley, D. Kriegner, H. Reichlova, F. Krizek, H. Ohno, and L. \v{S}mejkal, arXiv:2508.09748 (2025).

\bibitem{Smejkal2022_4} L. \v{S}mejkal, A. B. Hellenes, R. Gonz\'alez-Hern\'andez, J. Sinova, and T. Jungwirth, {\it Phys. Rev. X} {\bf 12}, 011028 (2022). 

\bibitem{Yuanl2020} L.-D. Yuan, Z. Wang, J.-W. Luo, E. I. Rashba, and A. Zunger, {\it  Phys. Rev. B} {\bf 102}, 014422 (2022). 

\bibitem{Samanta2020} K. Samanta, M. Le\v{z}ai\'c, M. Merte, F. Freimuth, S. Bl\"ugel, and Y. Mokrousov, {\it J. Appl. Phys.} {\bf 127}, 213904 (2020). 

\bibitem{Hernandez2021} R. Gonz\'alez-Hern\'andez, L. \v{S}mejkal,  K. V\'yborn\'y, Y. Yahagi, J. Sinova, T. Jungwirth, and J. \v{Z}elezn\'y, {\it Phys. Rev. Lett.} {\bf 126}, 127701 (2021).

\bibitem{Cuiq2023} Q. Cui, B. Zeng, P. Cui, T. Yu, and H. Yang, {\it Phys. Rev. B} {\bf 108}, L180401 (2023).

\bibitem{Gomonay2024} O. Gomonay, V. P. Kravchuk, R. Jaeschke-Ubiergo, K. V. Yershov, T. Jungwirth, L. \v{S}mejkal, J. van den Brink, and J. Sinova, {\it npj Spintronics} {\bf 2}, 35 (2024). 

\bibitem{Bose2022}  A. Bose, N. J. Schreiber, R. Jain, D.-F. Shao, H. P. Nair, J. Sun, X. S. Zhang, D. A. Muller, E. Y. Tsymbal, D. G. Schlom, and D. C. Ralph, {\it Nat. Electron.} {\bf 5}, 267 (2022).

\bibitem{Fengz2022} Z. Feng, X. Zhou, L. \v{S}mejkal, L. Wu, Z. Zhu, H. Guo, R. Gonz\'alez-Hern\'andez, X. Wang, H. Yan, P. Qin,  X. Zhang, H. Wu, H. Chen, Z. Meng, L. Liu, Z. Xia, J. Sinova, T. Jungwirth, and Z. Liu, {\it Nat. Electron.} {\bf 5}, 735 (2022).
 
\bibitem{Lees2024} S. Lee, S. Lee, S. Jung, J. Jung, D. Kim, Y. Lee, B. Seok, J. Kim, B. G. Park, L. \v{S}mejkal, C.-J. Kang, and C. Kim, {\it Phys. Rev. Lett.} {\bf 132}, 036702 (2024). 

\bibitem{Krempasky2024} J. Krempask\'y, L. \v{S}mejkal, S. W. D'Souza, M. Hajlaoui, G. Springholz, K. Uhl\'i\v{r}ov\'a, F. Alarab, P. C. Constantinou, V. Strocov, D. Usanov, W. R. Pudelko, R. Gonz\'alez-Hern\'andez, A. B. Hellenes, Z. Jansa, H. Reichlov\'a, Z. \v{S}ob\'a\v{n}, R. D. G. Betancourt, P. Wadley, J. Sinova, D. Kriegner, J. Min\'ar, J. H. Dil, and T. Jungwirth, {\it Nature} {\bf 626}, 517 (2024). 

\bibitem{Reimers2024} S. Reimers, L. Odenbreit, L. \v{S}mejkal, V. N. Strocov, P. Constantinou, A. B. Hellenes, R. J. Ubiergo, W. H. Campos, V. K. Bharadwaj, A. Chakraborty, T. Denneulin, W. Shi, R. E. Dunin-Borkowski, S. Das, M. Kl\"aui, J. Sinova, and M. Jourdan, {\it Nat. Commun.} {\bf 15}, 2116 (2024). 

\bibitem{Guop2023} P.-J. Guo, Z.-X. Liu, and Z.-Y. Lu, {\it npj Comput. Mater.} {\bf 9}, 70 (2023).

\bibitem{Chenx2023} X. Chen, D. Wang, L. Li, and B. Sanyal, {\it Appl. Phys. Lett.} {\bf 123}, 022402 (2023).

\bibitem{Sodequist2024} J. S\o dequist and T. Olsen, {\it Appl. Phys. Lett.} {\bf 124}, 182409 (2024).

\bibitem{Bhattarai2025} R. Bhattarai, P. Minch, and T. D. Rhone, {\it Phys. Rev. Materials} {\bf 9}, 064403 (2025). 

\bibitem{Panb2024} B. Pan, P. Zhou, P. Lyu, H. Xiao, X. Yang, and L. Sun, {\it Phys. Rev. Lett.} {\bf 133}, 166701 (2024). 

\bibitem{Liuy2024} Y. Liu, J. Yu, and C. C. Liu, {\it Phys. Rev. Lett.} {\bf 133}, 206702 (2024).  

\bibitem{Sheoran2024} S. Sheoran and S. Bhattacharya, {\it Phys. Rev. Materials} {\bf 8}, L051401 (2024). 

\bibitem{Zengs2024} S. Zeng and Y.-J. Zhao, {\it Phys. Rev. B} {\bf 110}, 174410 (2024). 

\bibitem{Guos2024} S. D. Guo, Y. Liu, J. Yu, and C. C. Liu, {\it Phys. Rev. B} {\bf 110}, L220402 (2024).

\bibitem{Avsar2020}  A. Avsar, H. Ochoa, F. Guinea, B. \"{O}zyilmaz, B. J. van Wees, and I. J. Vera-Marun, {\it Rev. Mod. Phys.} {\bf 92}, 021003 (2020).

\bibitem{Wangz2015}  Z. Wang, C. Tang, R. Sachs, Y. Barlas, and J. Shi, {\it Phys. Rev. Lett.} {\bf 114}, 016603 (2015).

\bibitem{Zollner2023}  K. Zollner, S. M. Jo\~{a}o, B. K. Nikoli\'c, and J. Fabian, {\it Phys. Rev. B} {\bf 108}, 235166 (2023).

\bibitem{Zhangg2024}  G. Zhang, H. Wu, L. Yang, W. Jin, W. Zhang, and H. Chang, {\it Appl. Phys. Rev.} {\bf 11}, 021308 (2024).

\bibitem{Trier2022}  F. Trier, P. No\"{e}l, J.-V. Kim, J.-P. Attan\'e, L. Vila, and M. Bibes, {\it Nat. Rev. Mater.} {\bf 7}, 258 (2023).

\bibitem{Huangb2020}  B. Huang, M. A. McGuire, A. F. May, D. Xiao, P. Jarillo-Herrero, and X. Xu, {\it Nat. Mater.} {\bf 19}, 1276 (2020).

\bibitem{Roche2024}  S. Roche, B. v. Wees, K. Garello, and S. O. Valenzuela, {\it 2D Mater.} {\bf 11}, 043001 (2024).

\bibitem{Vojavek2024}  L. Voj\'{a}\v{c}ek, J. M. Due\~{n}as, J. Li, F. Ibrahim, A. Manchon, S. Roche, M. Chshiev, and J. H. Garc\'{i}a, {\it Nano Lett.} {\bf 24}, 11889 (2024).

\bibitem{Masseroni2024}  M. Masseroni, M. Gull, A. Panigrahi, N. Jacobsen, F. Fischer, C. Tong, J. D. Gerber, M. Niese, T. Taniguchi, K. Watanabe, L. Levitov, T. Ihn, K. Ensslin, and H. Duprez , {\it Nat. Commun.} {\bf 15}, 9251  (2024).

\bibitem{Duenas2024}  J. M. Due\~{n}as, J. H. Garc\'{i}a, and S. Roche, {\it Phys. Rev. Lett.} {\bf 132}, 266301 (2034).

\bibitem{Songk2018}  K. Song, D. Soriano, A. W. Cummings, R. Robles, P. Ordej\'{o}n, and S. Roche, {\it Nano Lett.} {\bf 18}, 2033 (2018).

\bibitem{Pezo2021}  A. Pezo, Z. Zanolli, N. Wittemeier, P. Ordej\'{o}n, A. Fazzio, S. Roche and J. H Garcia, {\it 2D Mater.} {\bf 9}, 015008 (2021).

\bibitem{Liy2019}  Y. Li, and M. Koshino, {\it Phys. Rev. B} {\bf 99}, 075438 (2019).

\bibitem{Wuy2020}  Y. Wu, S. Zhang, J. Zhang, W. Wang, Y. L. Zhu, J. Hu, G. Yin, K. Wong, C. Fang, C. Wan, X. Han, Q. Shao, T. Taniguchi, K. Watanabe, J. Zang, Z. Mao, X. Zhang, and K. L. Wang, {\it Nat. Commun.} {\bf 1}, 3680 (2020).

\bibitem{Stiehl2019}  G. M. Stiehl, R. Li, V. Gupta, I. E. Baggari, S. Jiang, H. Xie, L. F. Kourkoutis, K. F. Mak, J. Shan, R. A. Buhrman, and D. C. Ralph, {\it Phys. Rev. B} {\bf 100}, 184402 (2019).

\bibitem{Husain2020}  S. Husain, R. Gupta, A. Kumar, P. Kumar, N. Behera, R. Brucas, S. Chaudhary, and P. Svedlindh, {\it Appl. Phys. Rev.} {\bf 7}, 041312 (2020).

\bibitem{Zhaoz2025}  Z. Zhao, Y. Lin, and Ahmet Avsar, {\it npj 2D Mater. Appl.} {\bf 9}, 30 (2025).

\bibitem{Sierra2021}  J. F. Sierra, J. Fabian, R. K. Kawakami, S. Roche, and S. O. Valenzuela, {\it Nat. Nanotechnol.} {\bf 16}, 856 (2021).

\bibitem{Yangh2022}  H. Yang, S.O. Valenzuela, M. Chshiev, S. Couet, B. Dieny, B. Dlubak, A. Fert, K. Garello, M. Jamet, D.-E. Jeong, K. Lee, T. Lee, M.-B. Martin, G. S. Kar, P. S\'{e}n\'{e}or, H.-J. Shin, and S. Roche, {\it Nature} {\bf 606}, 663 (2022).

\bibitem{Kurebayashi2022}  H. Kurebayashi, J. H. Garcia, S. Khan, J. Sinova, and S. Roche, {\it Nat. Rev. Phys.} {\bf 4}, 150 (2022).

\bibitem{Gish2024}  J. T. Gish, D. Lebedev, T. W. Song, V. K. Sangwan, and M. C. Hersam, {\it Nat. Electron.} {\bf 7}, 336 (2024).

\bibitem{Fan2021}  Z. Fan, J. H. Garcia, A. W. Cummings, J. E. Barrios-Vargas, M. Panhans, A. Harju, F. Ortmann, and S. Roche, {\it Phys. Rep.} {\bf 903}, 1 (2021).

\bibitem{Li2025_survey}  X. Li, et al., {\it Agentic AI for Scientific Discovery: A Survey of Progress, Challenges, and Future Directions}, arXiv:2503.08979 (2025).

\bibitem{Lil2024}  L. Li, X. Li, L. Lin, D. Zhang, M. Chen, D. Wu, and Y. Yang, {\it Phys. Rev. B} {\bf 110}, 205119 (2024).


\end{thebibliography}
\end{document}